\documentclass[sigconf,nonacm]{acmart}

\AtBeginDocument{%
  \providecommand\BibTeX{{%
    \normalfont B\kern-0.5em{\scshape i\kern-0.25em b}\kern-0.8em\TeX}}}
\usepackage{bbding}

\usepackage{bm}
\usepackage{siunitx}
\usepackage{colortbl} 
\usepackage{multirow} 
\usepackage{makecell} 
\usepackage{booktabs} 
\usepackage{utfsym}
\usepackage{bbm}
\usepackage{pifont}
\usepackage[most]{tcolorbox}
\usepackage{graphicx}
\usepackage{float} 
\usepackage{subfigure}
\usepackage{multirow}
\usepackage{tabularx}
\newcolumntype{Y}{>{\raggedleft\arraybackslash}X}
\usepackage{makecell}
\usepackage{multirow}
\usepackage{mdframed}
\usepackage{balance}
\usepackage{xcolor} %
\usepackage{color}
\usepackage{tikz}
\usepackage{tikz-qtree}
\usepackage{pgfplots}
\usepackage{varwidth}
\usepackage{url}
\usepackage{mathtools}
\usepackage{rotating} 
\usepackage{hyperref}
\usepackage{graphicx}
\usepackage{todonotes}
\usepackage{cleveref}

\usepackage{threeparttable}

\newtheorem{example}{Example}
\usepackage[ruled,linesnumbered,vlined]{algorithm2e}
\usetikzlibrary{patterns}

\usepackage[noend]{algpseudocode} 
\usepackage{hyperref}
\hypersetup{
    colorlinks=true, 
    urlcolor=blue,  
}

\newcommand{\tabincell}[2]{\begin{tabular}{@{}#1@{}}#2\end{tabular}}
\definecolor{lightergray}{RGB}{242,242,255}

\newtcolorbox{mybox}{
  colback=lightergray,    
  coltext=black,    
  boxrule=1.2pt,    
  arc=4pt           
  width=0.3\textwidth, 
  center            
}

\usepackage[most]{tcolorbox}

\newtheorem{problem}{Problem}

\newcommand{\Yes}{\usym{2713}}
\newcommand{\No}{\usym{2717}}

\newcommand\vldbdoi{XX.XX/XXX.XX}
\newcommand\vldbpages{XXX-XXX}
\newcommand\vldbvolume{18}
\newcommand\vldbissue{1}
\newcommand\vldbyear{2025}
\newcommand\vldbauthors{\authors}
\newcommand\vldbtitle{\shorttitle} 
\newcommand\vldbpagestyle{plain} 
\title{In-depth Analysis of Densest Subgraph Discovery in~a~Unified~Framework}

\newcommand{\zhou}[1]{\textcolor{magenta}{{#1}}}

\newcommand{\eat}[1]{}

\author{Yingli Zhou}
\affiliation{%
  \institution{The Chinese University of Hong Kong, Shenzhen}
    \country{China}
}
\email{yinglizhou@link.cuhk.edu.cn}

\author{Qingshuo Guo$^{\#}$}
\affiliation{%
  \institution{The Chinese University of Hong Kong, Shenzhen}
  \country{China}
}
\email{qingshuoguo@link.cuhk.edu.cn}

\author{Yi Yang$^{\#}$}
\affiliation{%
  \institution{The Chinese University of Hong Kong, Shenzhen}
  \country{China}
}
\email{yiyang3@link.cuhk.edu.cn}

\author{Yixiang Fang$^{*}$}
\affiliation{%
  \institution{The Chinese University of Hong Kong, Shenzhen}
  \country{China}
}
\email{fangyixiang@cuhk.edu.cn}

\author{Chenhao Ma$^{*}$}
\affiliation{%
  \institution{The Chinese University of Hong Kong, Shenzhen}
  \country{China}
}
\email{machenhao@cuhk.edu.cn}

\author{Laks V.S. Lakshmanan}
\affiliation{%
  \institution{The University of
British Columbia}
  \country{Canada}
}
\email{laks@cs.ubc.ca}

\begin{abstract}

As a fundamental topic in graph mining, {\it Densest Subgraph Discovery (DSD)} has found a wide spectrum of real applications.
Several DSD algorithms, including exact and approximation algorithms, have been proposed in the literature. 
However, these algorithms have not been systematically and comprehensively compared under the same experimental settings.
In this paper, we first propose a unified framework to incorporate all DSD algorithms from a high-level perspective.
We then extensively compare representative DSD algorithms over a range of graphs -- from small to billion-scale -- and examine the effectiveness of all methods, which provide a thorough analysis of DSD algorithms.
From our experimental analysis,  we identify new variants of the DSD algorithms over undirected graphs, by combining existing techniques, which are up to $10\times$ faster than the state-of-the-art algorithm with the same accuracy guarantee.
Finally, based on the findings, we offer promising research opportunities.
We believe that a deeper understanding of the behavior of existing algorithms can provide new valuable insights for future research.
The codes are released at \zhou{\url{https://anonymous.4open.science/r/DensestSubgraph-245A }}
\end{abstract} 
\begin{document}




\maketitle

\pagestyle{\vldbpagestyle}
\begingroup\small\noindent\raggedright\textbf{PVLDB Reference Format:}\\
\vldbauthors. \vldbtitle. PVLDB, \vldbvolume(\vldbissue): \vldbpages, \vldbyear.\\
\href{https://doi.org/\vldbdoi}{doi:\vldbdoi}
\endgroup
\begingroup
\renewcommand\thefootnote{}\footnote{\noindent
This work is licensed under the Creative Commons BY-NC-ND 4.0 International License. Visit \url{https://creativecommons.org/licenses/by-nc-nd/4.0/} to view a copy of this license. For any use beyond those covered by this license, obtain permission by emailing \href{mailto:info@vldb.org}{info@vldb.org}. Copyright is held by the owner/author(s). Publication rights licensed to the VLDB Endowment. \\
\raggedright Proceedings of the VLDB Endowment, Vol. \vldbvolume, No. \vldbissue\ %
ISSN 2150-8097. \\
\href{https://doi.org/\vldbdoi}{doi:\vldbdoi} \\
}\addtocounter{footnote}{-1}\endgroup

\ifdefempty{\vldbavailabilityurl}{}{
}

\section{Introduction}
\label{sec:intro}

\begin{figure}[t]
        \centering
	\includegraphics[width=0.8\linewidth]{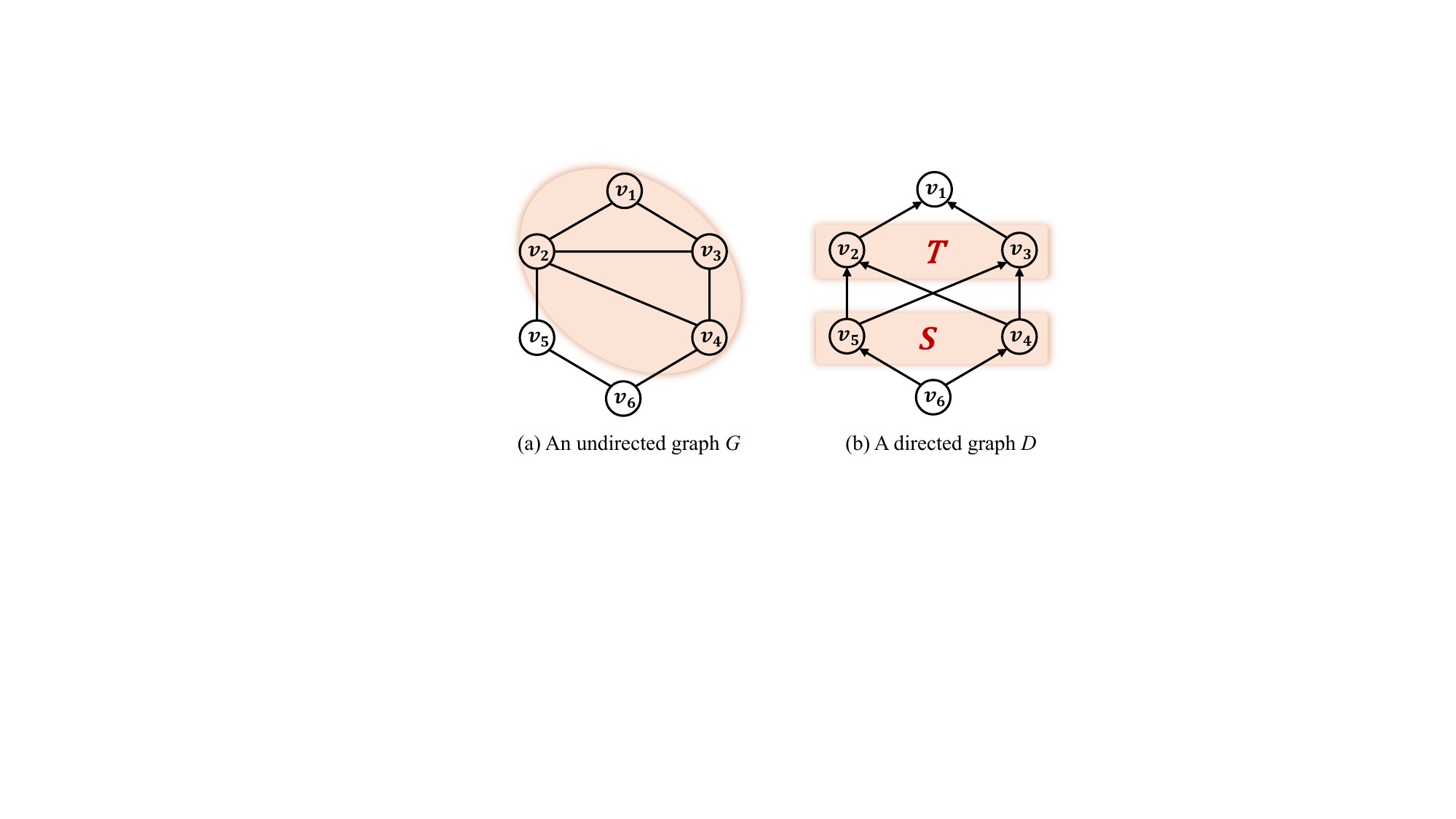}
\caption{Examples of undirected and directed graphs.}
\label{fig:graphs}
\end{figure}

Graph data are often used to model relationships between objects in various real-world applications \cite{ching2015one,java2007we,karlebach2008modelling,ma2019linc,albert1999diameter}. For example, the Facebook friendship network can be modelled as an undirected graph by mapping users to vertices and friendships to edges \cite{ching2015one}; In X (formerly known as   Twitter), a directed edge can represent the ``following'' relationship between two users \cite{java2007we}; The Web network itself can also be modelled as a vast directed graph \cite{albert1999diameter}.
Figures \ref{fig:graphs} (a) and (b) depict an undirected graph and a directed graph respectively.

As a fundamental problem in graph mining, the {\it densest subgraph discovery (DSD)} problem aims to discover a very ``dense'' subgraph from a given graph \cite{lanciano2023survey,luo2023survey}. More precisely, given an undirected graph, the DSD problem \cite{goldberg1984finding} asks for a subgraph with the highest {\em density}, defined as the number of edges over the number of vertices in the subgraph, and it is often termed the densest subgraph (DS).
The DSD problem lies at the core of graph mining \cite{bahmani2012densest,gionis2015dense}, and is widely used in many areas.
For instance, in social networks, the DS discovered can be used to detect communities \cite{chen2010dense,tsourakakis2013denser}, reveal fake followers~\cite{beutel2013copycatch}, and identify echo chambers and groups of actors engaged in spreading misinformation~\cite{lakshmanan2022quest}.
In e-commerce networks~\cite{beutel2013copycatch}, the DS is useful for detecting fake accounts.
In graph databases, the DSD  is a building block for solving many graph problems, such as reachability queries~\cite{cohen2003reachability} and motif detection~\cite{fratkin2006motifcut,saha2010dense}.
In biological data analysis, DSD solutions have been shown to be useful in identifying regulatory motifs in genomic DNA \cite{fratkin2006motifcut} and gene annotation graphs \cite{saha2010dense}.
Besides, the DSD problem is closely related to other fundamental graph problems, such as network flow and bipartite matching \cite{sawlani2020near}.
Due to the theoretical and practical importance, researchers from the database, data mining, computer science theory, and network communities have designed efficient and effective solutions to the DSD problem.

\begin{table*}[h]
  \centering
  \small
  \caption{Classification of existing DSD works.}
    \setlength{\tabcolsep}{3.5pt}
  \label{tab:overall_table}
  \resizebox{\textwidth}{!}{
  \begin{threeparttable}
  \begin{tabular}{c|l|l|c|c|c|c|c|c|c}
     \toprule
     	\multirow{3}{*}{{\bf \tabincell{c}{Graph\\Type} }}& \multirow{3}{*}{{\bf \tabincell{c}{Algorithm} }} & \multirow{3}{*}{{\bf \tabincell{c}{Key Technique} }} &
            \multicolumn{2}{c|}{\bf Complexity} &  \multirow{3}{*}{{\bf \tabincell{c}{Approx. \\ Ratio } }} & \multirow{3}{*}{\bf \# Iteration}  & \multicolumn{3}{c}{\bf Optimization}\\
     		\cline{4-5}  \cline{8-10}
       & & & \multirow{2}{*}{{\bf \tabincell{c}{Time \\ Complexity} }} & \multirow{2}{*}{{\bf \tabincell{c}{Space \\ Complexity} }} &  &    &\multirow{2}{*}{{\bf \tabincell{c}{Early \\ Termination} }} & \multirow{2}{*}{{\bf \tabincell{c}{Graph\\ Reduction} }}  & \multirow{2}{*}{{\bf \tabincell{c}{Parallel \\ Firendly} }}\\
        & & & &&&&&&
       \\

     \hline\hline
     \multirow{13}{*}{\tabincell{c}{Undirected\\graphs}}     & \texttt{FlowExact~}\cite{goldberg1984finding}& Network flow & $\mathcal{O}(\log n \cdot t_{\text{Flow}})$ & $\mathcal{O}(m)$ & 1  & N/A  & \No & \No  & \No\\ \cline{2-10}

   & \texttt{CoreExact~}\cite{fang2019efficient} & Network flow & $\mathcal{O}(\log n \cdot t_{\text{Flow}})$ & $\mathcal{O}(m)$  & 1  & N/A  & \Yes & \Yes  & \No\\  \cline{2-10}

     & \texttt{FWExact~}\cite{danisch2017large} & Convex Programming & $\mathcal{O}(T \cdot m + \log T \cdot t_{\text{Flow}})$ & $\mathcal{O}(m)$ & 1  & $\Omega(T)$  & \Yes & \No  & \Yes      

                        \\     \cline{2-10}

       & \texttt{MWUExact~}\cite{harb2022faster} & Convex Programming & $\mathcal{O}(T \cdot m + \log T \cdot t_{\text{Flow}})$ & $\mathcal{O}(m)$ & 1  & $\Omega(T)$  & \Yes & \No  & \Yes      

                        \\     \cline{2-10}   

         & \texttt{FISTAExact~}\cite{harb2022faster} & Convex Programming & $\mathcal{O}(T \cdot m + \log T \cdot t_{\text{Flow}})$ & $\mathcal{O}(m)$ & 1  & $\Omega(T)$  & \Yes & \No  & \Yes      

                        \\     \cline{2-10}                   

      & \texttt{Greedy~}\cite{charikar2000greedy} & Peeling & $\mathcal{O}(m + n)$ &$\mathcal{O}(m)$ & 2  & N/A & \No & \No & \No \\ \cline{2-10} 





      & \texttt{CoreApp~}\cite{fang2019efficient}& Peeling & $\mathcal{O}(m + n)$ & $\mathcal{O}(m)$ & 2  & N/A & \No & \No & \No       

                        \\ \cline{2-10}                  

        & \texttt{Greedy++~}\cite{boob2020flowless} & Peeling & $\mathcal{O}(T \cdot m \log n)$ & $\mathcal{O}(m)$ &($1 + \epsilon$)  &  $\Omega(\frac{\Delta(G)}{\rho^{*}_G \epsilon^2})$ & \No & \No & \No       

                    \\ \cline{2-10}

       & \texttt{FWApp~}\cite{danisch2017large} & Convex Programming & $\mathcal{O}(T \cdot m)$ &$\mathcal{O}(m)$ & ($1 + \epsilon$)  & $\Omega(\frac{mn \Delta(G))}{\epsilon^2})$ & \Yes & \No & \No       

                        \\ \cline{2-10}     

      & \texttt{MWUApp~}\cite{harb2022faster} & Convex Programming & $\mathcal{O}(T \cdot m)$ & $\mathcal{O}(m)$ & ($1 + \epsilon$)  &$\Omega(x)$ & \Yes & \No & \No       

                        \\ \cline{2-10}   

      & \texttt{FISTAApp~}\cite{harb2022faster} & Convex Programming & $\mathcal{O}(T \cdot m)$ &$\mathcal{O}(m)$ & ($1 + \epsilon$)  &$\Omega(\frac{\sqrt{m n \Delta(G)}}{\epsilon^2})$ & \Yes & \No & \No       

                        \\ \cline{2-10}   

      &  \texttt{FlowApp*}\cite{xu2023efficient} & Network flow & $\mathcal{O}(T \cdot m \log m)$ & $\mathcal{O}(m)$ & ($1 + \epsilon$)  & $\Omega(\frac{\log m}{\epsilon})$ & \Yes & \Yes & \No       

                        \\

     \bottomrule

      \multirow{7}{*}{\tabincell{c}{Directed\\graphs}}          & \texttt{DFlowExact~}\cite{khuller2009finding} & Network flow & $\mathcal{O}(n^2 \cdot t_{\text{Flow}})$ & $\mathcal{O}(m)$ & 1  & N/A & \No & \No & \No       

                        \\ \cline{2-10}        

    & \texttt{DCExact~}\cite{ma2020efficient} &Network flow & $\mathcal{O}(k \cdot t_{\text{Flow}})$ & $\mathcal{O}(m)$ & 1 & N/A & \Yes & \Yes & \No       

                        \\ \cline{2-10}  

    & \texttt{DFWExact~}\cite{ma2021efficient} & Convex Programming & $\mathcal{O}(T \cdot t_{\text{Fw}})$ &$\mathcal{O}(m)$ & 1  & $\Omega(T)$ & \Yes & \Yes & \No       

\\ \cline{2-10}  

    & \texttt{DGreedy~}\cite{charikar2000greedy} & Peeling & $\mathcal{O}(n^2 \cdot (n + m))$ & $\mathcal{O}(m)$ & 2  & N/A & \No & \No & \No       

                        \\ \cline{2-10}  





   & \texttt{XYCoreApp~}\cite{ma2020efficient} &Peeling & $\mathcal{O}(\sqrt{m} \cdot (n + m))$ & $\mathcal{O}(m)$ & 2  & N/A & \No & \No & \No       

                        \\ \cline{2-10}   

  & \texttt{WCoreApp~}\cite{luo2023scalable} & Peeling & $\mathcal{O}(\Delta(G) \cdot m)$ & $\mathcal{O}(m)$ & 2 &  N/A & \No & \No & \Yes       

                        \\ \cline{2-10}                            

     & \texttt{DFWApp~}\cite{ma2021efficient} & Convex Programming &$\mathcal{O}(T \cdot \log_{1+\epsilon} n m)$& $\mathcal{O}(m)$ & ($1 + \epsilon$)  & $\Omega( \frac{m \cdot \kappa}{\epsilon^2} )$ & \Yes & \Yes & \Yes
    \\           
     \bottomrule
  \end{tabular}
  {\raggedright $\star$ Note: $n$ and $m$ denote the numbers of vertices and edges in the graph respectively; $\epsilon>0$ is a real value.\par}
  

  {\raggedright $\star$ Note: $\Delta(G)$ denotes the highest degree of $G$; $\kappa$ is an integer that is proportional to the maximum value of the highest out-degree and in-degree in $D$;  \par}
  
  \end{threeparttable}
  }
\end{table*}

In Table \ref{tab:overall_table}, we categorize representative DSD algorithms by their computation models, main methods, and approximation ratio guarantee.
The exact algorithms include maximum flow and convex programming-based algorithms; the approximation algorithms include peeling-based, convex programming-based, and network flow-based algorithms.
After a careful literature review, we make the following observations.
First, no prior work has proposed a unified framework to abstract the DSD solutions and identify key performance factors. 
Second, existing works focus on evaluating the overall performance, but not individual components. 
Third, there is no existing comprehensive comparison between all these algorithms in terms of efficiency and accuracy. 
%

\textbf{Our work.} To address the above issues, in this paper we conduct an in-depth study on sequential DSD algorithms\footnote{We refer readers to \cite{sukprasert2024practical} for a 
comparison of parallel DSD algorithms.}.  
We first propose a unified framework with three modules, namely {\it graph reduction}, {\it vertex weight update} ({\tt VWU}) and {\it candidate subgraph extract and verify} ({\tt CSV}), which capture the core ideas of all existing algorithms.
Given a graph $G$ and an error threshold $\epsilon$, {\it graph reduction} aims to locate the DS in a small subgraph; {\tt VWU} aims to update vertex weights over $T$ iterations; and  {\tt CSV} extracts a candidate subgraph based on vertex weights and verifies if it satisfies the $\epsilon$ error requirement. 
Under this framework, we systematically compare 12 and 7 representative algorithms for undirected and directed graphs, respectively.
We conduct comprehensive experiments on both real-world and synthetic datasets and provide an in-depth analysis. 
%
%
%

In summary, our principal contributions are as follows.

\begin{itemize}
    \item Provide a unified framework for DSD solutions from a high-level perspective (Section \ref{sec:generic}); 
    
    \item Comprehensively examine DSD algorithms for both undirected and directed graphs respectively (Sections \ref{sec:compare} and \ref{sec:extended_to_DG}); 
    
    \item Conduct extensive experiments from different angles using various datasets which provide a thorough analysis of DSD algorithms. Based on our analysis, we identify new variants of the DSD algorithms over undirected graphs, by combining existing techniques, that significantly outperform state-of-the-art. (Section \ref{sec:experiments}); 
    
    \item  Summarize lessons learned and propose practical research opportunities that can facilitate future studies (Section \ref{sec:lessons_opp}).
\end{itemize}


In Section \ref{sec:pre}, we present the preliminaries and 
 introduce a unified DSD framework in Section \ref{sec:generic}.
\eat{
In Sections \ref{sec:compare} and \ref{sec:extended_to_DG}, we compare the DSD algorithms under the framework for undirected and directed graphs, respectively.
The comprehensive experimental results and analysis are reported in Section \ref{sec:experiments}. 
We present the learned lessons and a list of research opportunities in Section \ref{sec:lessons_opp},
} 
Section \ref{sec:related}  reviews related work while   Section \ref{sec:conclusions} summarizes the paper.

\section{PRELIMINARIES}
\label{sec:pre}


We first provide the definitions of DSD problems over both undirected and directed graphs, i.e., UDS and DDS problems respectively, and then present their convex program (CP) formulations.

\subsection{Problem definitions}
\label{sec:problem}

\begin{table}[h]
  \caption{Notations and meanings.}
  \label{tab:commands}
  \centering
  \small
  \begin{tabular}{c|l}
    \toprule
    \textbf{Notation} & \textbf{Meaning}\\
    \hline
        $G=(V, E)$ & An undirected graph with vertex set $V$ and edge set $E$\\
    \hline
         $D=(V, E)$  &  A directed graph with vertex set $V$ and edge set $E$ \\
    \hline
       $N(v, G)$ & The set of neighbors of a vertex $v$ in $G$ \\ 
    \hline
       $d_G(v)$ & The degree of $v$ in $G$, i.e., $d_G(v)$=$|N(v, G)|$\\ 
    \hline
       $d^+_D(v)$, $d^-_D(v)$ & The out-degree and in-degree of $v$ in $D$, respectively\\ 
    \hline
        $G[S]$ & The subgraph of $G$ induced by vertices in $S$ \\
      \hline
    $D[S, T]$ & The subgraph of $D$ induced by vertices in $S$ and $T$ \\
    \hline
        $\mathcal{D}(G)$ & The densest subgraph of $G$ \\
    \hline
        $\rho(S, T)$ & The density of subgraph $D[S,T]$ \\
    \hline
    $k^*$ & The largest $k$ such that the $k$-core in $G$ exists.\\
 \hline
    $\Delta(G)$ & The highest degree of $G$. \\
    \bottomrule
  \end{tabular}
\end{table}

We denote an undirected graph by $G = (V,E)$, where $|V|$ = $n$ and $|E|$ = $m$ are the numbers of vertices and edges of $G$, respectively.
The set of neighbors of a vertex $u$ in $G$ is denoted by $N(u, G)$, and the degree of $u$ is $d_G(u)$ = $|N(u, G)|$. 
Given a vertex set $S$, we use $G[S]$ = $(S, E(S))$ to denote the subgraph of $G$ induced by $S$, where $E(S)$= $\{(u, v) \in E \mid  u, v \in S\}$ denotes the set of edges in $G$ contained in $S$.
For a given undirected graph $H$, we denote its sets of vertices and edges by $V(H)$ and $E(H)$, respectively.

\begin{definition}[\textbf{Density of undirected graph }~\cite{goldberg1984finding}]
\label{def:un-edge-udensity}
	Given an undirected graph $G=(V, E)$, its density $\rho(G)$ is defined as the number of edges over the number of vertices, i.e., 
     $\rho(G)= \frac{{|E|}}{{|V|}}$.
\end{definition}

\begin{problem}[\textbf{UDS problem}~\cite{goldberg1984finding,tsourakakis2015k,fang2019efficient}]
\label{def:densest}
Given an undirected graph $G$, find the subgraph $\mathcal{D}(G)$ whose density is the highest among all the possible subgraphs, which is also called the undirected densest subgraph (UDS).
\end{problem}

Let $D$ = $(V,E)$ be a directed graph.
For each vertex $v \in V$, denote by $N_D^+(v)$
(resp. $N_D^-(v)$) the out-neighbors (resp. in-neighbors) of $v$, and correspondingly denote by $d_D^+(v) := |N_D^+(v)|$  (resp. $d_D^-(v) := |N_D^-(v)|$) the out-degree (resp. in-degree) of $v$.
Given two vertex subsets $S, T \subseteq V$ that are not necessarily disjoint, $E(S, T)$ = $E \cap (S \times T)$ denotes the set of all edges from $S$ to $T$ in the graph $D$.
The $(S, T)$-induced subgraph of $D$ contains the vertex sets $S$, $T$ and the edge set $E(S, T)$.

\begin{definition}[\textbf{Directed graph density}\cite{kannan1999analyzing,khuller2009finding,ma2020efficient,ma2021directed}]
\label{def:di-edge-ddensity}
Given a directed graph $D$=$(V,E)$ and two vertex sets $S$ and $T$, the density of an ($S$, $T$)-induced subgraph is defined as $\rho(S, T)= \frac{{|E(S, T)|}}{{\sqrt{|S||T|}}}$.
\end{definition}

\begin{problem}[{\bf DDS problem }\cite{kannan1999analyzing,gionis2015dense,charikar2000greedy,khuller2009finding,bahmani2012densest}]
\label{def:densest}
Given a directed graph $D$, find the subgraph $\mathcal{D}(D)$=$D[S^{*}, T^{*}]$ whose corresponding density is the highest among all the possible ($S$, $T$)-induced subgraphs, also called the directed densest subgraph (DDS).
\end{problem}

Denote the density of $\mathcal{D}(G)$ and $\mathcal{D}(D)$ by $\rho^{*}_G$ and $\rho^{*}_D$ respectively.
E.g., in Figures~\ref{fig:graphs} (a) and (b), the subgraphs in the dashed ellipses are the UDS and DDS respectively, with $\rho^{*}_G$=5/4 and $\rho^{*}_D$=2.

\subsection{The $\mathsf{CP}$ formulations of UDS problems}
\label{sec:lp_formulation}

A well-known CP formulation of the UDS problem \cite{danisch2017large} is as follows:
\begin{equation}
\label{equ:danisch-cp}
\small
\begin{aligned}
\mathsf{CP}(G) \quad \min & \sum_{u \in V}\mathbf{w}^2(u) \\
\text{s.t.} \quad &\mathbf{w}(u) = \sum_{(u,v) \in E } \alpha_{u,v}, && \forall u \in V   \\
&\alpha_{u,v} + \alpha_{v,u} = 1, && \forall  (u,v) \in E \\
&\alpha_{u,v} \geq 0,  \alpha_{v,u} \geq 0 && \forall (u,v) \in E
\end{aligned}
\end{equation}
    This $\mathsf{CP}(G)$ can be visualized as follows. Each edge $(u, v) \in E$ has a weight of 1, which it wants to assign to its endpoints: $u$ and $v$ such that the weight sum received by the vertices is as even as possible. 
Indeed, in the DS $\mathcal{D}(G)$ of $G$, it is possible to distribute all edge weights such that the weight sum received by each vertex in $\mathcal{D}(G)$ is exactly $\rho(\mathcal{D}(G))$ = $\frac{E(\mathcal{D}(G))}{|V(\mathcal{D}(G))|}$.
Following this intuition, $\alpha_{u,v}$ in $\mathsf{CP}(G)$  indicates the weight assigned to $u$ from edge ($u$,$v$), and $\mathbf{w}(u)$ is the weight sum received by $u$ from its adjacent edges.

For the DDS problem, Ma et al. \cite{ma2022convex} derived its CP formulation by a series of linear programs w.r.t. the possible values of $c=|S|/|T|$.
Since the ratio $c$ is unknown in advance, there are $O(n^{2})$ possible values, which result in $O(n^{2})$ different CPs.
For each $c$, the corresponding CP is formulated by Equation (\ref{equ:ma-dual-lp}).
\begin{equation}
\label{equ:ma-dual-lp}
\small
\begin{aligned}
\mathsf{CP}(c)&&  \min  &&   \max_{u \in V} \{|\mathbf{w}_{\alpha}(u)|, & |\mathbf{w}_{\beta}(u)|\}		 \\
	&& \text{s.t.}	&& \alpha_{u,v}  + \beta_{v,u} &=1,					&& \forall (u,v) \in E  \\
	&& && 2\sqrt{c} \sum_{(u,v) \in E}\alpha_{u,v}&=\mathbf{w}_{\alpha}(u),				&& \forall u \in V \\
 	&& && \frac{2}{\sqrt{c}} \sum_{(u,v) \in E}\beta_{v,u}&=\mathbf{w}_{\beta}(u),				&& \forall u \in V \\
 &&	&& \alpha_{u,v} , \beta_{v,u} &\geq 0,					&& \forall (u,v) \in E  
\end{aligned}
\end{equation}

\section{A Unified Framework}
\label{sec:generic}
\begin{table*}[h]
\centering
\small
\begin{tabular}{l|l|l|l}
\toprule
\textbf{Method} 
& \textbf{Stage (1):} {\tt ReduceGraph}
& \textbf{Stage (2):} {\tt VWU}
& \textbf{Stage (3):} {\tt CSV}
\\ \hline
\begin{tabular}[c]{@{}l@{}} {\tt FlowExact} \\   {\tt CoreExact}\end{tabular} & \begin{tabular}[c]{@{}l@{}} no reduction  \\ locate graph into $\lceil \underline{\rho} \rceil$-core \end{tabular} & compute the maximum flow & \begin{tabular}[c]{@{}l@{}} {\Large\ding{182}} extract the minimum cut;\\ {\Large\ding{183}}  verify if it is optimal;\end{tabular} \\ \hline
{\tt FlowApp} & locate graph into $\lceil \underline{\rho} \rceil$-core & perform the blocking flow & \begin{tabular}[c]{@{}l@{}}{\Large\ding{182}}  extract the residual graph;\\ {\Large\ding{183}} verify the approximation ratio;\end{tabular} \\ \hline
CP-based & no reduction & optimize \(\mathsf{CP}(G)\); & \begin{tabular}[c]{@{}l@{}}{\Large\ding{182}}  extract the maximum prefix sum set;\\ {\Large\ding{183}} verify if it is exact or satisfies the approximation ratio criteria;\end{tabular} \\ \hline
Peeling-based & no reduction & iteratively remove vertices &  {\Large\ding{182}} extract the subgraph with the highest density during the peeling process; \\ 
\bottomrule
\end{tabular}
\caption{Overview of the three stages of the existing UDS algorithms.}
\label{tab:brief_overview}
\end{table*}
In this section, we develop a unified framework, consisting of three stages: {\it Graph reduction}, {\it Vertex Weight Update} ({\tt VWU}), and {\it Candidate subgraph Extract and Verify} ({\tt CSV}), which can cover all existing DSD algorithms, as shown in Algorithm \ref{alg:framework}.

\begin{algorithm}[h]
  \caption{A unified framework of DSD} 
  \label{alg:framework}
  \small
   \SetKwInOut{Input}{input}\SetKwInOut{Output}{output}
    \Input{$G=(V,E)$, $\epsilon$,  $T$}
    \Output{An exact / approximation densest subgraph $\mathcal{D}(G)$}
    $f \gets$ {\tt False}; $\underline{\rho} \gets k^* / 2$; $\overline{\rho} \gets k^*$\; 
    \Repeat{$f$={\tt True}}{
        \tcp{\textcolor{teal}{(1) The graph reduction method.}}
        $G\gets${\tt  ReduceGraph}($G$, $\underline{\rho}$)\;
        
        \tcp{\textcolor{teal}{(2) The vertex weight update method.}}
        $\mathbf{w} \gets$ {\tt VWU}($G$, $\mathbf{w}$, $T$, $\underline{\rho}$, $\overline{\rho}$, $\epsilon$)\;
        
        \tcp{\textcolor{teal}{(3) The candidate subgraph extract and verify.}}
        $(f, \mathcal{D}(G)) \gets$ {\tt CSV}($G,\underline{\rho}, \overline{\rho}, \mathbf{w}$, $\epsilon$);
        
    
    }
    
    \Return{$\mathcal{D}(G)$;}

\end{algorithm}

Specifically, given a graph $G$, an error threshold $\epsilon$, and the number of iterations $T$, we initialize the upper and lower bounds of $\rho^{*}_G$ (line 1), where $k^*$ is the maximum core number which will be introduced later (refer to Lemma \ref{lemma:uds:kcore}).
%
%
We then iteratively execute operations in the following three stages (lines 2-6):
\begin{enumerate}
      \item Locate the graph into a smaller subgraph (i.e., $\lceil \underline{\rho} \rceil$-core) utilizing the lower bound $\underline{\rho}$ (Section \ref{sec:graph_reduction});
    
    \item Update the vertex weight vector $\mathbf{w}$ for each vertex over $T$ iterations (Section \ref{sec:vw_update});
    
    \item Extract the candidate subgraph using vertex weight vector $\mathbf{w}$, and verify if the candidate subgraph meets the requirements  (Section \ref{sec:candidate_extra}). Terminate the process if it does; otherwise, update the parameters and repeat the above steps.
\end{enumerate}

In Table \ref{tab:brief_overview}, we illustrate the details of these three stages for each category of DSD algorithms for undirected graphs.
We note that the vertex weight vector $\mathbf{w}$ holds different meanings for and serves different purposes in different algorithms:
in the network flow-based algorithms, $\mathbf{w}$ denotes the flows from the vertices to the target node;
in the CP-based algorithms, $\mathbf{w}$ represents the weight sum received by each vertex;
in the peeling-based algorithms, $\mathbf{w}$ is used to select which vertex should be removed. Our abstraction unifies the different uses of weights in different algorithms.

\section{Comparison and Analysis Of DSD Algorithms for Undirected Graphs}
\label{sec:compare}

In this section, we systematically compare and analyze all the UDS algorithms in terms of the three stages of our unified framework.

\subsection{Graph reduction}
\label{sec:graph_reduction}

We first review the notion of \textit{$k$-core}.

\begin{definition}[\textbf{$k$-core }~\cite{goldberg1984finding,tsourakakis2015k,fang2019efficient}]
\label{def:densest}
Given an undirected graph $G$ and an integer $k$ ($k \geq 0$), its $k$-core, denoted  $\mathcal{H}_k$, is the largest subgraph of $G$, such that $\forall v \in \mathcal{H}_k$, $deg_{\mathcal{H}_k} (v) \geq k$.
\end{definition}

The \textit{core number} of a vertex $v \in V$ is the largest $k$ for which a $k$-core contains $v$;  the maximum core number among all vertices is denoted $k^*$.
A $k$-core has an interesting ``nested'' property \cite{batagelj2003m}:  for any two non-negative integers $i$ and $j$ s.t. $i < j$,  $\mathcal{H}_j \subseteq \mathcal{H}_i$.

\texttt{CoreExact}~\cite{fang2019efficient} locates the UDS into some $k$-cores by leveraging following lemma. 
%

\begin{lemma}[\cite{fang2019efficient}]
\label{lemma:core_reduce}
Given an undirected graph $G$ let its UDS $\mathcal{D}(G)$ have density $\rho:=\rho^{*}_G$.  Then $\mathcal{D}(G)$ is contained in the $\lceil \rho \rceil$-core.
\end{lemma}

Since $\rho^{*}_G$ is unknown in advance, we cannot use it directly.
Instead, we locate the UDS into some $k$-cores using  a lower bound on $\rho^{*}_G$, thanks to the nested property of $k$-cores.
\begin{figure}[h]
	\centering
	\includegraphics[width=0.9\linewidth]{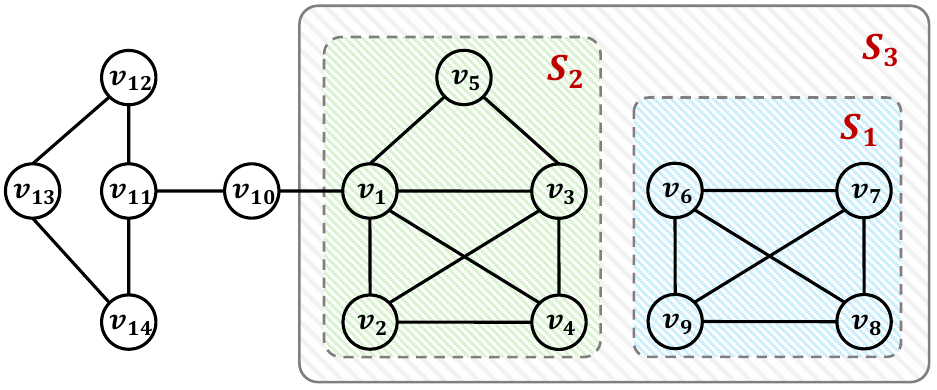}
	\caption{An example of the core-based graph reduction.}
	\label{fig:coreexact}
\end{figure}

\begin{example}
For example, for the undirected graph in Figure \ref{fig:coreexact}, suppose the lower bound of $\rho^{*}_G$ is 3. Then, the UDS can be located in the 3-core, i.e., $S_1$ and $S_2$, which is smaller than the entire graph.
\end{example}

Due to the nested property of $k$-core,  vertices with smaller core numbers in the remaining graph can be successively removed in the search process thus  gradually reducing the size of the graph, since as we shall see, the lower bound of $\rho^{*}_G$ is progressively increasing.
We next present the lower and upper bounds on the density of a $k$-core.

\begin{lemma}[\cite{fang2019efficient}]
\label{lemma:uds:kcore}
Given an undirected graph $G$, let $\mathcal{H}_k$ be a $k$-core of $G$. Then, the density of  $\mathcal{H}_k$ satisfies: $k^* / 2 \leq \rho(\mathcal{H}_k) \leq k^*$.
\end{lemma}

The lemma says that $\rho_G^*$ cannot be larger than $k^*$ and cannot be smaller than $k^* / 2$.

\subsection{Vertex weight updating}
\label{sec:vw_update}

We show that the key components in various UDS algorithms can be considered as the process of vertex weight updating.

\begin{algorithm}[h]
  \caption{The template of {\tt VWU}} 
  \small
  \label{alg:templat_vwu}
      {\Large\ding{182}} initialize $\mathbf{w}$ and auxiliary variables\; 
    \ForEach {$t$ = $1, \cdots, T$} {
 \tcp{\textcolor{teal}{Algorithms have varied stop conditions.}}
        \While{the stop condition is not met} {
             {\Large\ding{183}} update the auxiliary variables\;
            {\Large\ding{184}} update $\mathbf{w}$ via auxiliary variables\;   
        }
    }
\end{algorithm}

We outline the {\tt VWU} template in Algorithm \ref{alg:templat_vwu}, which begins by initializing the vertex weight vector $\mathbf{w}$ along with auxiliary variables facilitating the update process of $\mathbf{w}$.
We update the variables over $T$ iterations, where each iteration updates the auxiliary variables and $\mathbf{w}$
until the stop conditions are met (lines 3-5).

\begin{figure}[h]
	\centering
	\includegraphics[width=0.8\linewidth]{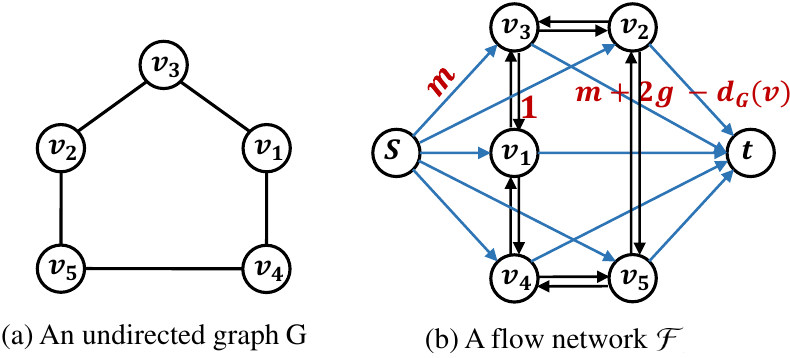}
	\caption{An undirected Graph $G$ and its flow network $\mathcal{F}$.}
	\label{unflow}
\end{figure}

$\bullet$ \textbf{Network flow-based algorithms.}
For the exact algorithms, {\tt FlowExact} and {\tt CoreExact} need to {\Large\ding{182}} build a flow network based on the guessed maximum density $g$ (i.e., $g$ = ($\underline{\rho}$ +$\overline{\rho}$)$ / 2$).
For lack of space, we omit the detailed steps of building the flow network \cite{goldberg1984finding}.
%
%
%
For example, Figure \ref{unflow} shows a flow network of an undirected graph.
We use auxiliary variables to denote the capacity and flow of each edge, and $\mathbf{w}(v)$ represents the value of flow from vertex $v$ to the sink node $t$. 
{\Large\ding{183}} After constructing the flow network, they try to update the flows of some edges in $\mathcal{F}$, and {\Large\ding{184}} increase $\mathbf{w}$.
The above process is repeated until the maximum flow is reached.
%
%
{\tt CoreExact} follows the same steps as {\tt FlowExact}, but it utilizes $k$-core for graph reduction (refer Section \ref{sec:graph_reduction}).
%
%
Unlike exact algorithms, {\tt FlowApp} does not need to calculate the exact maximum flow, it only needs to perform partial maximum flow computations. 

%

$\bullet$ \textbf{CP-based algorithms.} All CP-based algorithms aim to solve $\mathsf{CP(G)}$ in Eq. (\ref{equ:danisch-cp}).
There are three widely used types of CP solvers for the DSD problem: {\tt Frank-wolfe}, {\tt MWU}, and {\tt FISTA}-based algorithms.
All these algorithms, no matter whether they are designed for approximation or exact solutions, share the same {\tt VWU} procedure.

\begin{algorithm}[h]
  \caption{{\tt VWU}  of {\tt Frank-wolfe} and {\tt MWU}} 
  \label{alg:vwu:fw-mwu}
   \SetKwInOut{Input}{input}\SetKwInOut{Output}{output}
     {\Large\ding{182}} \textcolor{teal}{initialize $\mathbf{w}$ and auxiliary variables}\; 
  \textbf{foreach} $(u,v) \in E$ \textbf{do} $\alpha^{(0)}_{u,v} \gets 1 / 2$; $\alpha^{(0)}_{v,u} \gets 1 / 2$ \;
   \textbf{foreach} $u \in V$ \textbf{do} $\mathbf{w}^{(0)}(u) \gets \sum_{(u,v) \in E} \alpha^{(0)}_{u,v}$ \;
   \ForEach{$t$ = $1, \cdots, T$}{
        {\Large\ding{183}} \textcolor{teal}{update the auxiliary variables}\; 
        \textbf{foreach}{ $(u,v) \in E$} \textbf{do}{ update $\widehat{\!\alpha}_{u,v}$ and $\widehat{\!\alpha}_{v,u}$\;}
        $\alpha^{(t)} \gets (1 - \gamma_t) \cdot \alpha^{(t - 1)} + \gamma_t \cdot \widehat{\alpha}$, with $\gamma_t = \frac{2}{t + 2}$ or $\frac{1}{t+1}$\;      
        {\Large\ding{184}} \textcolor{teal}{update $\mathbf{w}$ via $\alpha$}\; 
        \textbf{foreach} {$v \in V$} \textbf{do} {
           $\mathbf{w}^{(t)}(u) \gets \sum_{(u,v) \in E} \alpha^{(t)}_{u,v}$\;
       }
    }
\end{algorithm}

The {\tt Frank-wolfe} and {\tt MWU}-based algorithms solve $\mathsf{CP}(G)$ in an iterative manner, where $\gamma_t = \frac{2}{t+2}$ for \texttt{Frank-Wolfe} algorithms and $\gamma_t = \frac{1}{t+1}$ for \texttt{MWU}-based algorithms.
In each iteration, algorithms linearize the objective function at the current point and move towards minimizing it \cite{jaggi2013revisiting,danisch2017large,chekuri2022densest}.
Algorithm \ref{alg:vwu:fw-mwu} shows the process, {\Large\ding{182}} starting with initializing $ \alpha$ and $\mathbf{w}$ (lines 2-3).
Next, in the $t$-th  iteration, {\Large\ding{183}} each edge $(u,v) \in E$ attempts to distribute its weight, i.e., 1, to the endpoint with a smaller $\mathbf{w}^{(t-1)}$ value, and  the $\alpha^{(t)}$
values of all vertices are computed as a convex combination by $\alpha^{(t-1)}$ and $\widehat{\alpha}$  (lines 5–7).
Then, {\Large\ding{184}} $\mathbf{w}^{(t)}$ is updated by the weight sum received by each vertex in $V$ (line 9). 

{\tt FISTA}-based algorithms \cite{harb2022faster} also adopt the iterative paradigm, but leverage the {\em projections} along with Nesterov-like momentum terms~\cite{nesterov1983method} to address quadratic objective functions with linear constraints.
Harb et al. \cite{harb2022faster} further proved that they converge faster in theory, but experimentally there is a gap between the theoretical conclusion and practical results, i.e., they are slower than other CP-based algorithms as the {\it projection} is very time-consuming.

\begin{algorithm}[h]
  \caption{{\tt VWU}  of {\tt Greedy} and {\tt Greedy++}} 
  \label{alg:vwu:peeling}
    {\large \ding{182}} \textcolor{teal}{initialize $\mathbf{w}$ and auxiliary variables}\; 
   \textbf{foreach} $v \in V$ \textbf{do} $\mathbf{w}^{(0)}(v) \gets 0$;
   $H \gets \emptyset$\;
    \ForEach{$t$ = $1 \cdots T$}{
    $H \gets G$\;
    \While{$V(H) \neq \emptyset$}{
        Select vertex $v$ minimizing $\mathbf{w}^{(t - 1)}(v)$ + $d_H(v)$\;
         {\large \ding{183}}  \textcolor{teal}{update the auxiliary variables}\; 
         \textbf{foreach} {$u \in N(v, H)$} {{\bf do}} {
            $d_H(u) \gets d_H(u) - 1$\;
       }
          {\large \ding{184}}  \textcolor{teal}{update $\mathbf{w}$ via $d_H$}\; 
     
        $\mathbf{w}^{(t)}(v) \gets \mathbf{w}^{(t - 1)}(v) + d_H(v)$\;
        Remove $v$ and all its adjacent edges $(u,v)$ from $H$; 
    }
    }
\end{algorithm}

$\bullet$ \textbf{Peeling-based algorithms.} The key idea of peeling-based algorithms is to find a vertex with the minimum weight, remove it from the graph and update the weights of the remaining vertices.
Algorithm \ref{alg:vwu:peeling} presents the details of {\tt Greedy} and {\tt Greedy++}.
Specifically, when $T$ = 1, this algorithm is {\tt Greedy}, removing the vertex with the minimum degree one by one.
{\tt Greedy++} algorithm is based on {\tt Greedy} and iteratively removes the vertices via $T$ iterations.
In each iteration, it iteratively removes the vertex with the smallest weight, where the weight of vertex $v$ in each iteration is the sum of its induced degree (w.r.t. the remaining vertices) and the weight of $v$ in the previous iteration (line 6).
{\tt CoreApp} reveals that $k^*$-core can serve as a 2-approximation solution, where the $k^*$-core can be computed by following a peeling-based  process.
\subsection{Candidate subgraph extraction and verification}
\label{sec:candidate_extra}

In this section, we explore the {\tt CSV} stage across various UDS algorithms and examine how to utilize upper and lower bounds of the optimal density $\rho^{*}_G$ for verifying results.
Let $g$ represent the guessed density, and set $g$ = ($\underline{\rho}$ +$\overline{\rho}$)$ / 2$.

$\bullet$ \textbf{FlowExact and CoreExact.} {\tt FlowExact} and CoreExact need to compute the minimum cut ($\mathcal{S}$, $\mathcal{T}$)  via the maximum flow.
If $\mathcal{S}$ contains only the source node $\{s\}$, $\overline{\rho}$ is updated to $g$; otherwise, $\underline{\rho}$ is set to $g$ and $\mathcal{S} \setminus \{s\}$ is returned as the candidate subgraph.

To verify the quality of the candidate subgraph, {\tt FlowExact} checks if the difference between $\overline{\rho}$ and $\underline{\rho}$ is less than $ \frac{1}{n \cdot (n - 1)}$, as the density difference between any two subgraphs must be larger than $\frac{1}{n \cdot (n - 1)}$.
When the condition is satisfied, the exact solution has been found.
In contrast, {\tt CoreExact} utilizes less stringent stop conditions for verifying results:
it checks if the density difference is $< \frac{1}{|V_C| \cdot (|V_C| - 1)}$,
where $V_C$ is the largest connected component in the graph.

$\bullet$ \textbf{FlowApp.} {\tt FlowApp} searches for an augmenting path in $\mathcal{F}$ after $h$ blocking flows.
If such a path exists, $\underline{\rho}$ is updated to $g$ and the residual graph of $\mathcal{F}$ is returned as the candidate subgraph; otherwise, $\overline{\rho}$ is updated to $g$ and no candidate subgraph is obtained.

{\tt FlowApp} verifies whether a  candidate subgraph is a ($1+\epsilon$)-approximation solution by checking if  $\frac{g - \underline{\rho}}{2g} < \frac{\epsilon}{3-2\epsilon}$~\cite{xu2023efficient}.

$\bullet$ \textbf{CP-based algorithms.} All CP-based algorithms use the vertex weight vector $\mathbf{w}$ and the {\tt PAVA} algorithm \cite{danisch2017large} to extract the candidate subgraph. 

\begin{algorithm}[h]
  \caption{\tt PAVA} 
  \label{alg:cse:lp_based}
   \SetKwInOut{Input}{input}\SetKwInOut{Output}{output}
    \Input{$G=(V,E)$, $\mathbf{w}$}
    \Output{The candidate subgraph $\mathcal{S}(G)$}
      \ForEach{$1 \leq i \leq |V(G)|$}{
       $u_i \gets$ the vertex with the $i$-th highest weight in $V(G)$\;
       $G_i \gets$ the induced subgraph of top-$i$ weight vertices\;
       $y_i \gets d_{G_i}(u_i)$\; 
    }
    
    $s^* \gets \arg \max_{1 \leq s \leq n} \frac{1}{s} \sum_{i=1}^s y_i$\;
   $S \gets$ the subgraph induced by the first $s^*$ vertices\;
    \Return {  $S$;}
\end{algorithm}

Algorithm \ref{alg:cse:lp_based} presents the details.
Intuitively, the vertices with higher weights are more likely to appear in the densest subgraph $\mathcal{D}(G)$, since they are linked by more edges.
Thus, the subgraph induced by the first $s^*$ vertices with the largest weights is returned as the candidate subgraph.
Here, $\underline{\rho}$ is updated to $\rho(S)$, and  $\overline{\rho}$ is updated to  $\max_{1 \leq i \leq n} \min \left \{\frac{1}{i} {i \choose 2}, \frac{1}{i} \sum_{j = 1}^i \mathbf{w}(u_j) \right\}$ \cite{danisch2017large,sun2020kclist++}.

After a sufficient number of iterations, CP-based algorithms converge to the optimal solution, which can be verified by using the concept of stable set with theoretical guarantee \cite{danisch2017large}. We omit the details for lack of space.
For the approximation algorithms, we can use the ratio of $\overline{\rho}$ over $\rho(S)$ to estimate the empirical approximation ratio, as the optimal density is not known in advance \cite{danisch2017large,sun2020kclist++}.

$\bullet$ \textbf{Peeling-based algorithms.} {\tt Greedy} and {\tt Greedy++} extract the highest density subgraph during the process of vertex peeling. 
In addition, \texttt{CoreApp} utilizes the $k^*$-core as their returned candidate subgraph. 
These algorithms do not require updating $\underline{\rho}$ and $\overline{\rho}$ as they do not need to verify results.
\subsection{Optimizations}
\label{sec:optimize}

There are two optimization strategies used in existing UDS works.

$\bullet$ \textbf{Locating the UDS in a connected component.}
{\tt CoreExact} locates the UDS in a connected $\lceil \rho \rceil$-core; notice that the $\lceil \rho \rceil$-core may be disconnected. Connected cores tend to be smaller than the disconnected cores they are part of.
For example, in Figure \ref{fig:coreexact}, the two connected cores $S_1$ and $S_2$ can be processed one by one by {\tt CoreExact} algorithm.

$\bullet$  \textbf{Simultaneous update strategy.}
Danisch \cite{danisch2017large} employs a simultaneous weight update strategy for {\tt FWExact} and {\tt FWApp}.
Specifically, within each iteration, if a vertex $v$'s weight $\mathbf{w}(v)$ changes, then the updated vertex weight is promptly visible to subsequent updates of other vertices in the same iteration.
The simultaneous weight update strategy enables a more balanced weight distribution among vertices, making the algorithm converge faster, as all the vertices in the UDS have the same weight upon convergence.
%


\textbf{}
\section{Comparison and Analysis of DSD Algorithms for Directed Graphs}
\label{sec:extended_to_DG}


The DDS problem is more complicated than the UDS problem because it is an induced subgraph of two vertex sets $S$ and $T$, which leads to a search space of $n^2$ possible values of the ratio between the size of the two vertex sets (i.e., $c = |S|/|T|$) which must be examined when computing the maximum density.
Next, we show how to adapt our framework (Algorithm \ref{alg:framework}) for the DDS problem.

$\bullet$ \textbf{Exact algorithms.} The exact algorithms need to enumerate all possible values of $c$ and compute the DS for each $c$.
For each fixed $c$, the exact algorithms share the same paradigm with UDS algorithms, indicating that they can be easily incorporated into our framework.    
Similar to {\tt CoreExact}, {\tt DC-Exact}~\cite{ma2020efficient} reduces the size of the graph by locating the DDS into some $[x, y]$-core \cite{ma2020efficient}.
\begin{definition}[ {$[ x, y ]$}-core \cite{ma2020efficient}]
\label{def:x-y-core}
Given a directed graph $D$=($V$, $E$), the {\bf $[x$, $y]$-core} is the largest ($S$, $T$)-induced subgraph $D[S,T]$, which satisfies:
\begin{enumerate}
	\item $\forall u \in S, d^{+}_{D[S,T]}(u) \geq x$ and $\forall v \in T, d^{- }_{D[S,T]}(v) \geq y$;
	\item $\nexists D[S', T']\neq D[S,T]$, such that $D[S,T]$ is a subgraph of $D[S',T']$, i.e., $S \subseteq S'$, $T \subseteq T'$, and $D[S',T']$ satisfies (1);
\end{enumerate}
The $[x^{*}, y^{*}]$-core is the $[x, y]$-core with the largest $x \cdot y$ values.
\end{definition}

\begin{theorem}[\cite{ma2020efficient}]
\label{theorem:divide_cp}
Given a directed graph $D$ = $(V, E)$, its DDS $D[S^*,T^*]$ is contained in the $ \left[  \frac{\rho_D^*}{2\sqrt{c}}, \frac{\sqrt{c}\rho_D^{*}}{2}\right ]$-core, where c = $\frac{|S^*|}{|T*|}$.    
\end{theorem}

By Theorem \ref{theorem:divide_cp}, we only need to discover DDS in the $ \left[  \frac{\underline{\rho}}{2\sqrt{c_r}}, \frac{\sqrt{c_l}\underline{\rho}}{2}\right ]$-core, where ($c_l$, $c_r$) specifies the interval of $c$ values under consideration during a particular stage of the divide-and-conquer approach and $\underline{\rho}$ is the lower bound of the optimal density.

$\bullet$ \textbf{Approximation algorithms}. There are two groups of approximation algorithms, i.e., peeling-based~\cite{charikar2000greedy,khuller2009finding,bahmani2012densest,ma2020efficient,luo2023scalable} and CP-based~\cite{ma2022convex} algorithms.
{\tt DGreedy} needs to enumerate all $c$ values to obtain the approximation solution, and when a specific $c$ is fixed, it is analogous to {\tt Greedy}.
{\tt XYCoreApp} focuses on identifying the $[x^{*}, y^{*}]$-core.
Each time it fixes one dimension and optimizes the other dimension via iteratively peeling vertices in the other dimension to find the $[x^{*}, y^{*}]$-core, where the peeling process is similar to {\tt Greedy}.
\texttt{WCoreApp} sequentially eliminates vertices based on the lowest weight, where a vertex's weight is determined by the product of its out-degree and in-degree.
The key difference between \texttt{WCoreApp} and \texttt{Greedy} is in how they calculate a vertex's weight.
The CP-based algorithms need to update $\mathbf{w}_{\alpha}$ and $\mathbf{w}_{\beta}$ via two auxiliary variables $\alpha$ and $\beta$, whose process is similar to Algorithm~\ref{alg:vwu:fw-mwu}.

Moreover, {\tt DCExact} \cite{ma2020efficient} first introduced a divide and conquer method to reduce the number of  $\frac{|S|}{|T|}$ values examined from $n^2$ to $k$, where theoretically $k \leq n^2$, but practically $k \ll n^2$.
%
{\tt DFWExact} \cite{ma2022convex} introduced a new divide-and-conquer method, utilizing the relationship between the DDS and the $c$-biased DS to skip searches for certain $c$ values in the DSD process.

In addition, Sawlani and Wang \cite{sawlani2020near} showed that the DDS problem can be transformed into $O(\log_{1+\epsilon}n)$ vertex-weighted UDS problems, where the vertex weights are assigned according to $O(\log_{1+\epsilon}n)$ different guesses of $\frac{|S|}{|T|}$.
In our study, we test its performance by adapting the SOTA algorithm for the UDS problem.
%



\section{Experiments}
\label{sec:experiments}

We now present the experimental results.
Section \ref{sec:exp:setup} discusses
the setup.
The experimental results of the UDS algorithms and DDS algorithms are reported in Sections \ref{sec:exp:uds} and \ref{sec:exp:dds}, respectively.

\subsection{Setup}
\label{sec:exp:setup}
\input{sections/6_Experiment/experiment_plot}
\begin{table}[h]
\small
  \caption{Undirected graphs used in our experiments.}
  \label{tab:un_dataset}
  \begin{tabular}{l|l|r|r}
    \toprule
     Dataset &Category & $|V|$ & $|E|$  \\ \hline
     bio-SC-GT (BG)& Biological & 1,716 & 31,564  \\
    econ-beacxc  (EB) &  Economic  &507 &42,176\\
    DBLP (DP)& Collaboration & 317,080 & 1,049,866  \\
    Youtube  (YT)& Multimedia & 3,223,589&  9,375,374 \\
    LiveJournal (LJ) & Social & 4,036,538 & 34,681,189  \\
    UK-2002 (UK)  & Road & 18,483,186 &261,787,258\\
   WebBase (WB)& Web & 118,142,155 & 881,868,060  \\
    Friendster (FS)& Social & 124,836,180 & 1,806,067,135  \\
   \bottomrule
  \end{tabular}
\end{table}
\begin{table}[h]
\small
  \caption{Directed graphs used in our experiments.}
  \label{tab:dd_dataset}
  \begin{tabular}{l|l|r|r}
    \toprule
     Dataset &Category & $|V|$ & $|E|$   \\
    \hline
  maayan-lake (ML) & Foodweb & 183 & 2,494 \\
    maayan-figeys (MF)   &  Metabolic  &2,239 &6,452 \\
    Openflights (OF)   & Infrastructure &2,939 & 30,501  \\
    Advogato (AD)   & Social &6,541 & 51,127   \\
   Amazon (AM) & E-commerce & 403,394 & 3,387,388   \\
   Baidu-zhishi (BA)& Hyperlink & 2,141,300 & 17,794,839   \\
   Wiki-en (WE) & Hyperlink & 13,593,032 & 437,217,424 \\ 
    SK-2005 (SK) & Web & 50,636,154 & 1,949,412,601    \\
   \bottomrule
  \end{tabular}
\end{table}

\begin{table*}[h]
\renewcommand\arraystretch{0.892}
\caption{Comparison of 2-approx. UDS algorithms (\textcolor{red}{Red} denotes the most efficient algorithm, and \textcolor{t3}{Blue} denotes the best accuracy).}
\renewcommand\arraystretch{0.89}
\label{tab:uds:2_approx}
\begin{tabular}{l|rr|rr|rr|rr|rr|rr}
\toprule
\multirow{2}{*}{Dataset}  & \multicolumn{2}{c|}{EB} & \multicolumn{2}{c|}{DP} & \multicolumn{2}{c|}{YT} & \multicolumn{2}{c|}{LJ}  & \multicolumn{2}{c|}{WB} & \multicolumn{2}{c}{FS} \\ \cline{2-13}
~  &time & ratio &time & ratio &time & ratio &time & ratio &time & ratio &time & ratio \\ \hline
\texttt{FlowApp*}    & 43.8 ms& 1.556 &  126.3 ms &\textcolor{t3}{1.001} & 9.1 s &1.761 & 2.4 s  &1.075   & 41.0 s &1.085 & 2095.6s & 1.799    \\ \hline
\texttt{CoreApp}   &  1.6 ms & 1.182 &  98.0 ms  & \textcolor{t3}{1.001} & 0.7 s  & 1.139 & 2.0 s  & 1.033  & 41.1 s  & 1.085 & 156.0 s & 1.065 \\ \hline
\texttt{PKMC}    & 4.3 ms  & 1.182 &  104.5 ms  & \textcolor{t3}{1.001} & 11.2 s & 1.139 & 8.5 s & 1.033  & 37.8 s  & 1.085 & 5546.4 s & 1.065 \\ \hline
\texttt{FWApp}         &     11.8 ms  & 1.072     &      371.6 ms & 1.111  &  9.1 s & 1.004   &     16.4 s & 1.078    &  383.6 s &1.214   &  1468.2 s &1.200 \\ \hline
\texttt{MWUApp}       &     2.7 ms & 1.001     &      213.6 ms & 1.420  &  24.8 s & 1.001    &     9.8 s & 1.115     &  281.6 s & 1.197   &   723.7 s & 1.029  \\ \hline
\texttt{FISTAApp}     &     3.4 ms & \textcolor{t3}{1.000}     &      390.3 ms & 1.152 &  49.1 s & 1.016   &     28.7 s & 1.480   &  2377.0 s & 1.105   &  2053.8 s & 1.026 \\ \hline
\texttt{Greedy} &    \textcolor{t4}{0.8 ms} & \textcolor{t3}{1.000} & \textcolor{t4}{72.8 ms} & \textcolor{t3}{1.001} & \textcolor{t4}{0.6 s} & \textcolor{t3}{1.000} & \textcolor{t4}{1.8 s} & \textcolor{t3}{1.013} & \textcolor{t4}{37.3 s} & \textcolor{t3}{1.016} & \textcolor{t4}{130.6 s} & \textcolor{t3}{1.000}\\ 
\bottomrule
\end{tabular}
\end{table*}

\begin{figure*}[h]
 \small
 \setlength{\abovecaptionskip}{-0.1cm}
  \setlength{\belowcaptionskip}{-0.1cm}
	\centering		
 \ref{poltk}\\
	%
\subfigure[EB]{
		\begin{tikzpicture}[scale=0.36]
			\begin{axis}[
         legend to name=poltk,
			   legend style = {
				    legend columns=-1,
				    draw=none,
				},
                 xtick = {2, 4, 6, 8},
                xticklabels={0.1,  0.01,  0.001, 0.0001},
				ymode = log,
				mark size=6.0pt, 
				width=0.43\textwidth,
    			height=.33\textwidth,
				ylabel={\Huge \bf time (ms)},
				xlabel={\Huge \bf $\epsilon$}, 
				ticklabel style={font=\Huge},
				every axis plot/.append style={line width = 2.5pt},
				every axis/.append style={line width = 2.5pt},
				]
                       \addplot [mark=o,color=c10] table[x=k,y=Greedy]{\approxEB};
            \addplot [mark=square,color=c5] table[x=k,y=Flow]{\approxEB};
           \addplot [mark=diamond,color=c3] table[x=k,y=FW]{\approxEB};
            \addplot [mark=star,color=c8] table[x=k,y=MWU]{\approxEB};
           \addplot [mark=triangle,color=c4] table[x=k,y=FISTA]{\approxEB};
			  \legend{ {\tt Greedy++},  {\tt FlowApp}*,   {\tt FWApp},  {\tt MWUApp}, {\tt FISTAApp}}
			\end{axis}
		\end{tikzpicture}
	}
	%
	\subfigure[DP]{
		\begin{tikzpicture}[scale=0.36]
			\begin{axis}[
				width=0.43\textwidth,
    			height=.33\textwidth,
                       xtick = {2, 4, 6, 8},
                xticklabels={0.1,  0.01, 0.001, 0.0001},
				ymode = log,
				mark size=6.0pt, 
				xlabel={\Huge \bf $\epsilon$}, 
				ticklabel style={font=\Huge},
				every axis plot/.append style={line width = 2.5pt},
				every axis/.append style={line width = 2.5pt},
				]
             \addplot [mark=o,color=c10] table[x=k,y=Greedy]{\approxDP};
            \addplot [mark=square,color=c5] table[x=k,y=Flow]{\approxDP};
           \addplot [mark=diamond,color=c3] table[x=k,y=FW]{\approxDP};
            \addplot [mark=star,color=c8] table[x=k,y=MWU]{\approxDP};
           \addplot [mark=triangle,color=c4] table[x=k,y=FISTA]{\approxDP};
			\end{axis}
		\end{tikzpicture}
	}
	%
	\subfigure[YT]{
		\begin{tikzpicture}[scale=0.36]
			\begin{axis}[ 
				width=0.43\textwidth,
    			height=.33\textwidth,
			    legend style = {
				    legend columns=2,
				    draw=none,
				},
			          xtick = {2, 4, 6, 8},
                xticklabels={0.1, 0.01, 0.001, 0.0001},
				ymode = log,
				mark size=6.0pt, 
				xlabel={\Huge \bf $\epsilon$}, 
				ticklabel style={font=\Huge},
				every axis plot/.append style={line width = 2.5pt},
				every axis/.append style={line width = 2.5pt},
				]
                       \addplot [mark=o,color=c10] table[x=k,y=Greedy]{\approxYT};
            \addplot [mark=square,color=c5] table[x=k,y=Flow]{\approxYT};
           \addplot [mark=diamond,color=c3] table[x=k,y=FW]{\approxYT};
            \addplot [mark=star,color=c8] table[x=k,y=MWU]{\approxYT};
           \addplot [mark=triangle,color=c4] table[x=k,y=FISTA]{\approxYT};
			\end{axis}
		\end{tikzpicture}
	}
	\subfigure[LJ]{
		\begin{tikzpicture}[scale=0.36]
			\begin{axis}[
			legend style = {
				    legend columns=-1,
				    draw=none,
				},
			width=0.43\textwidth,
			height=.33\textwidth,
	          xtick = {2, 4, 6, 8},
                xticklabels={0.1,  0.01, 0.001, 0.0001},
				ymode = log,
				mark size=6.0pt, 
				xlabel={\Huge \bf $\epsilon$}, 
				ticklabel style={font=\Huge},
			every axis plot/.append style={line width = 2.5pt},
			every axis/.append style={line width = 2.5pt},
			]
           \addplot [mark=o,color=c10] table[x=k,y=Greedy]{\approxLJ};
            \addplot [mark=square,color=c5] table[x=k,y=Flow]{\approxLJ};
           \addplot [mark=diamond,color=c3] table[x=k,y=FW]{\approxLJ};
            \addplot [mark=star,color=c8] table[x=k,y=MWU]{\approxLJ};
           \addplot [mark=triangle,color=c4] table[x=k,y=FISTA]{\approxLJ};
			\end{axis}
		\end{tikzpicture}
	}
	%
	%
		\subfigure[WB]{
		\begin{tikzpicture}[scale=0.36]
			\begin{axis}[
				 legend style = {
				    legend columns=-1,
				    draw=none,
				},
              xtick = {2, 4, 6, 8},
                xticklabels={0.1,  0.01, 0.001, 0.0001},
				ymode = log,
				mark size=6.0pt, 
				width=0.43\textwidth,
    			height=.33\textwidth,
				xlabel={\Huge \bf $\epsilon$}, 
				ticklabel style={font=\Huge},
				every axis plot/.append style={line width = 2.5pt},
				every axis/.append style={line width = 2.5pt},
				]
           \addplot [mark=o,color=c10] table[x=k,y=Greedy]{\approxWB};
            \addplot [mark=square,color=c5] table[x=k,y=Flow]{\approxWB};
           \addplot [mark=diamond,color=c3] table[x=k,y=FW]{\approxWB};
            \addplot [mark=star,color=c8] table[x=k,y=MWU]{\approxWB};
           \addplot [mark=triangle,color=c4] table[x=k,y=FISTA]{\approxWB};
			\end{axis}
		\end{tikzpicture}
	}
	%
		\subfigure[FS]{
		\begin{tikzpicture}[scale=0.36]
			\begin{axis}[
				 legend style = {
				    legend columns=-1,
				    draw=none,
				},
              xtick = {2, 4, 6, 8},
                xticklabels={0.1,  0.01, 0.001, 0.0001},
				ymode = log,
				width=0.43\textwidth,
    			height=.33\textwidth,
				mark size=6.0pt, 
				xlabel={\Huge \bf $\epsilon$}, 
				ticklabel style={font=\Huge},
				every axis plot/.append style={line width = 2.5pt},
				every axis/.append style={line width = 2.5pt},
				]
           \addplot [mark=o,color=c10] table[x=k,y=Greedy]{\approxFS};
            \addplot [mark=square,color=c5] table[x=k,y=Flow]{\approxFS};
           \addplot [mark=diamond,color=c3] table[x=k,y=FW]{\approxFS};
            \addplot [mark=star,color=c8] table[x=k,y=MWU]{\approxFS};
           \addplot [mark=triangle,color=c4] table[x=k,y=FISTA]{\approxFS};
			\end{axis}
		\end{tikzpicture}
	}
	\caption{Efficiency results of ($1+\epsilon$)-approximation algorithms on undirected graph.}
	\label{fig:uds:time_approx}
\end{figure*}
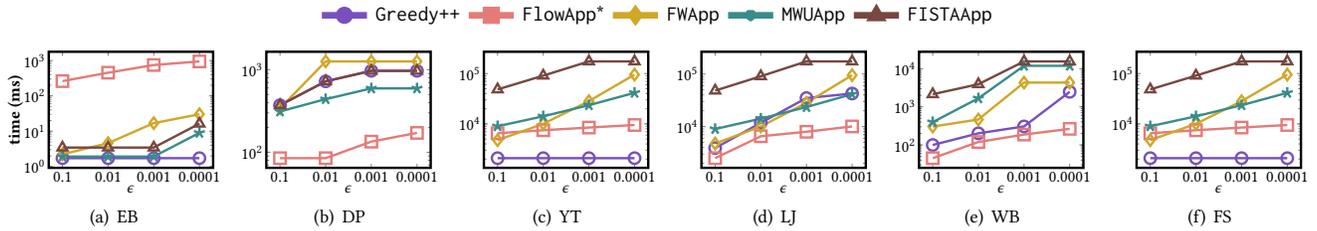

We use sixteen real datasets from different domains including 8 undirected graphs and 8 directed graphs, which are available on the Stanford Network Analysis Platform \footnote{http://snap.stanford.edu/data/}, Laboratory of Web Algorithmics \footnote{http://law.di.unimi.it/datasets.php}, Network Repository \footnote{https://networkrepository.com/network-data.php}, and Konect \footnote{http://konect.cc/networks/}.
%
%
\Cref{tab:un_dataset} and \ref{tab:dd_dataset} report the statistics of these graphs.
Due to space limitations, we present results for twelve datasets here and include additional data and results in our supplementary material.
We implement all the algorithms in C++ and run experiments on a machine having an Intel(R) Xeon(R) Gold  6338R 2.0GHz CPU and 512GB of memory, with Ubuntu installed.
If an exact algorithm cannot finish in three days or an approximation algorithm cannot finish in one day, we mark its running time as \textbf{INF} in the figures and ``---'' in the tables.

\subsection{Evaluation of UDS algorithms}
\label{sec:exp:uds}

\noindent \textbf{1. Overall performance.}
%
We first report the running time of all 2-approximation algorithms and the actual approximation ratios of approximation solutions returned by each algorithm in Table \ref{tab:uds:2_approx}.
Here, we set $\epsilon$=1 for ($1+\epsilon$)-approximation algorithms.
We observe that {\tt Greedy} always achieves the best performance when $\epsilon$=1 in terms of accuracy and efficiency.

We then examine the efficiency of ($1+\epsilon$)-approximation algorithms by varying $\epsilon$ from $0.1$ to $0.0001$, and report their running time in Figure \ref{fig:uds:time_approx}.
Specifically, we report the running time needed by  different ($1+\epsilon$)-approximation algorithms to find an approximation solution whose density is at least $(1+\epsilon)\times\rho_G^*$.
%
%
For a fair comparison, we stop algorithms when they achieve a density of $(1+\epsilon)\times\rho_G^*$, rather than the theoretical number of iterations, since it often takes far fewer iterations in practice than the theoretical value.
We observe that {\tt FlowApp}* and {\tt Greedy++} always perform faster than the others, since {\tt FlowApp}* utilizes the $k$-core-based graph reduction to reduce the search space; {\tt Greedy++} uses fewer iterations to obtain higher accuracy, and for each iteration, {\tt Greedy++} requires just a straightforward operation: iteratively remove the vertex with the lowest weight.
As for CP-based algorithms, {\tt FISTAApp} often takes more time to obtain solutions with the same accuracy as {\tt FWApp} and {\tt MWUApp}, despite theoretically needing fewer iterations. The is because it requires an extra projection operation in each iteration, which is very time-consuming, especially for large graphs.
Besides, {\tt FWApp} and {\tt MWUApp} achieve the comparable performance.
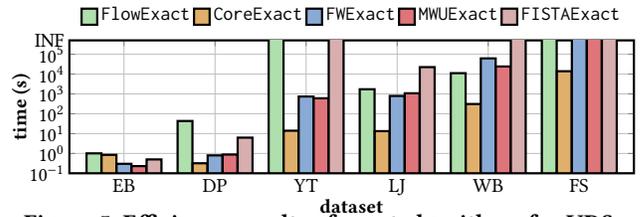
\begin{figure}[h]
    \centering
        \setlength{\abovecaptionskip}{-0.1cm}
  \setlength{\belowcaptionskip}{-0.1cm}
       	\begin{tikzpicture}[scale=0.52]
        		\begin{axis}[
        			ybar=0.11pt,
                        grid = major,
        			bar width=0.38cm,
        			width=0.9\textwidth,
    				height=0.28\textwidth,
        			xlabel={\huge \bf dataset}, 
        			xtick=data,	xticklabels={EB,DP,YT,LJ,WB,FS},
                     legend style={at={(0.5,1.30)}, anchor=north,legend columns=-1,draw=none},
                           legend image code/.code={
            \draw [#1] (0cm,-0.263cm) rectangle (0.4cm,0.15cm); },
        			xmin=0.8,xmax=13.2,
    			ymin=0.1,ymax=500000,
                    ytick = {0.1, 1, 10, 100, 1000, 10000, 100000,500105},
    	        yticklabels = {$10^{-1}$, $10^0$,$10^1$, $10^2$, $10^3$, $10^4$, $10^5$, INF},
                    ymode = log,    
                log origin=infty,
        			tick align=inside,
        			ticklabel style={font=\huge},
        			every axis plot/.append style={line width = 1.6pt},
        			every axis/.append style={line width = 1.6pt},
                        ylabel={\textbf{\huge time (s) }}
        			]
        			\addplot[fill=p1] table[x=datasets,y=Flow]{\exactUDS};
        			\addplot[fill=p2] table[x=datasets,y=Core]{\exactUDS};
           			\addplot[fill=p3] table[x=datasets,y=FW]{\exactUDS};
              \addplot[fill=p4] table[x=datasets,y=MWU]{\exactUDS};
                \addplot[fill=p5] table[x=datasets,y=FISTA]{\exactUDS};
                \legend{\huge {\tt FlowExact},\huge {\tt CoreExact}, \huge {\tt FWExact}, \huge {\tt MWUExact},\huge {\tt FISTAExact}};
        		\end{axis}
        	\end{tikzpicture}
    	\caption{Efficiency results of exact algorithms for UDS.}
    \label{fig:uds:exact_time}
\end{figure}
Finally, we present the running time of all exact algorithms in Figure \ref{fig:uds:exact_time}.
To be specific, {\tt CoreExact} performs best for the most part, since it can compute the minimum cut on the smaller flow network, and these CP-based algorithms achieve comparable performance.


\begin{figure*}[h]
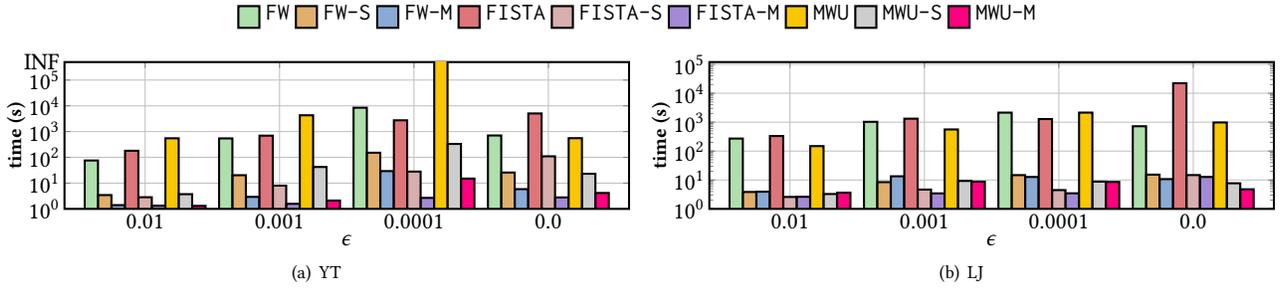

 \ref{poltReduce}\\ 
    \centering
     \setlength{\abovecaptionskip}{-0.1cm}
  \setlength{\belowcaptionskip}{-0.2cm}
    	\subfigure[YT]{

         }
    	\caption{The effect of graph reduction on undirected graphs.}
    \label{fig:uds:time_reduction}
\end{figure*}
\begin{figure*}[h]
 \small
      \setlength{\abovecaptionskip}{-0.1cm}
  \setlength{\belowcaptionskip}{-0.3cm}
	\centering		
    \begin{minipage}[h]{0.49\textwidth}
	\subfigure[YT]{
		%
	}
     \caption{The number of edges in graphs.}
    \label{fig:uds:reduction_edges}
\end{minipage}
    \begin{minipage}[h]{0.49\textwidth}
 \vspace{3pt}
      \setlength{\abovecaptionskip}{-0.1cm}
  \setlength{\belowcaptionskip}{-0.3cm}
	\subfigure[YT]{
		%
	}
        \vspace{2pt}
	\caption{The effect of graph reduction for {\tt Greedy++}.}
	\label{fig:reduce_greedy}
 \end{minipage}
\end{figure*}

\begin{table}[h]
\renewcommand\arraystretch{0.892}
\fontsize{6.5pt}{1em}
\selectfont
 \caption{Speed up of graph reduction.}
%
	}
	\caption{The number of iterations w.r.t $\epsilon$ on undirected graphs.}
	\label{fig:iter_appro}
\end{figure*}
\begin{figure*}[]
 \setlength{\abovecaptionskip}{-0.1cm}
  \setlength{\belowcaptionskip}{-0.2cm}
 \ref{poltUpdate}\\ 
    \centering
    	\subfigure[YT]{
       	%
         }
    	\caption{The effect of update strategies on undirected graphs.}
    \label{fig:uds:update_strategy}
\end{figure*}

\begin{figure*}[h]
 \small
 \setlength{\abovecaptionskip}{-0.1cm}
  \setlength{\belowcaptionskip}{-0.2cm}
	\centering		
 \ref{poltNew}\\
\subfigure[EB]{
		%
	}
	\caption{Efficiency results of new ($1+\epsilon$)-approximation algorithms on undirected graphs.}
	\label{fig:uds:new_approx}
\end{figure*}

\noindent \textbf{2. Evaluation of graph reduction.} We evaluate the effectiveness of two graph reduction strategies on all algorithms, except  2-approximation algorithms:
1) \textit{Single-round reduction}, which involves reducing the entire graph to a  $k^*/2$-core once and then executing all algorithms on this smaller graph, instead of the original large graph;
and 2) \textit{Multi-round reduction}, which is applied whenever a tighter lower bound of $\rho_G*$ is identified, as it locates the UDS within a smaller subgraph with a higher core number, resulting in a smaller graph.
We present the efficiency of all algorithms and their variants in Figure \ref{fig:uds:time_reduction} on two datasets, where the original algorithm names appended with ``S'' and ``M'' indicate the adoption of a single-round and multi-round reduction, respectively.
We also present the number of remaining edges after each of the first five rounds of graph reductions for all CP-based exact algorithms, on two datasets (Figure \ref{fig:uds:reduction_edges}), wherein the x-axis, ``0'' denotes the original graph before any reduction, and we record the speedup ratio of graph reduction over different algorithms in Table \ref{tab:uds:reduction_speedup}. 


We make the following analysis:
1) Graph reduction significantly enhances the efficiency of all algorithms.
2) Using multi-round reduction can result in a much greater performance increase than just using single-round reduction in most cases, yet in some cases, the multi-round reduction might bring extra time consumption due to the time needed to perform the reduction itself.
For example, applying multi-round reduction can speed up \texttt{FISTA} 1097 times, more than 98 $\times$ speedup achieved with a single-round reduction on the YT dataset with $\epsilon$ = 0.0001.
3) \texttt{FISTA}, employing multi-round graph reduction outperforms MWU and FW-based algorithms in terms of efficiency while using similar reduction strategies, with the same accuracy.
This efficiency boost stems from less projection time on smaller graphs and a reduction in the number of iterations needed.
4) The first graph reduction drastically reduces the graph size, pruning over 98\% edges after five rounds of reductions.
5) As shown in Figure \ref{fig:reduce_greedy}, graph reduction significantly enhances the performance of \texttt{Greedy++}, achieving an improvement of up to one order of magnitude faster than \texttt{Greedy++} without graph reduction.

\noindent \textbf{3. Accuracy of CP-based algorithms.} In this experiment, we report the number of iterations each algorithm needed to achieve different levels of accuracy in Figure \ref{fig:iter_appro}.
%
%
We make the following observations:
1) The actual numbers of iterations needed for all CP-based algorithms are much less than their theoretical numbers.
For example, on the YT dataset with $\epsilon$ = 0.1, {\tt FWApp}, {\tt MWUApp}, and {\tt FISTAApp} theoretically require at least $10^{16}$, $10^{16}$, and $10^{8}$ iterations on the original graph to obtain the approximated solution, while in practice they only need 16, 256, and 256 iterations, respectively; similarly, for the reduced graph, {\tt FWAPP-M}, {\tt MWUApp-M}, and {\tt FISTAApp-M} theoretically require $10^{14}$, $10^{14}$, and $10^{6}$ iterations, but practically they also only need 16, 8, and 16 iterations.
%
2) Graph reduction techniques significantly decrease the number of iterations required by CP-based algorithms, and this phenomenon is especially marked in {\tt FISTA}-based algorithms because it uses Nesterov-like momentum~\cite{harb2022faster} in its projection phase and adopts $\frac{1}{2 \cdot \Delta(G)}$ as the learning rate, since graph reduction can reduce $\Delta(G)$.

\begin{figure}[]
 \small
 \setlength{\abovecaptionskip}{-0.1cm}
  \setlength{\belowcaptionskip}{-0.1cm}
 \ref{poltRealEps}\\
	\centering		
	\subfigure[LJ]{
		\begin{tikzpicture}[scale=0.35]
			\begin{axis}[
            legend to name=poltRealEps,
			legend style = {
				    legend columns=-1,
				    draw=none,
				},
			width=0.43\textwidth,
			height=.35\textwidth,
	          xtick = {2, 4, 6, 8, 10},
                xticklabels={0.1, 0.05, 0.01, 0.001, 0.0001},
				ymode = log,
                ymin = 1500,
                ytick = {1500, 10000},
                yticklabels = {$10^3$, $10^4$},
				mark size=6.0pt, 
				ylabel={\Huge \bf time (ms)},
				xlabel={\Huge \bf $\epsilon$}, 
				ticklabel style={font=\Huge},
			every axis plot/.append style={line width = 2.5pt},
			every axis/.append style={line width = 2.5pt},
			]
            \addplot [mark=square,color=c5] table[x=k,y=Flow]{\approxLJ};
           \addplot [mark=diamond,color=c3] table[x=k,y=FWReal]{\approxLJ};
            \addplot [mark=star,color=c8] table[x=k,y=MWUReal]{\approxLJ};
           \addplot [mark=triangle,color=c4] table[x=k,y=FISTAReal]{\approxLJ};
		  \legend{ \texttt{FlowApp}*,   \texttt{FWApp}*,  \texttt{MWUApp}*,  \texttt{FISTAApp}*}
			\end{axis}
		\end{tikzpicture}
	}
	%
	%
		\subfigure[WB]{
		\begin{tikzpicture}[scale=0.35]
			\begin{axis}[
				 legend style = {
				    legend columns=-1,
				    draw=none,
				},
              xtick = {2, 4, 6, 8, 10},
                xticklabels={0.1, 0.05, 0.01, 0.001, 0.0001},
				ymode = log,
				mark size=6.0pt, 
				width=0.43\textwidth,
    			height=.35\textwidth,
                ymin = 30, ymax = 1000, 
				xlabel={\Huge \bf $\epsilon$}, 
				ticklabel style={font=\Huge},
				every axis plot/.append style={line width = 2.5pt},
				every axis/.append style={line width = 2.5pt},
				]
            \addplot [mark=square,color=c5] table[x=k,y=Flow]{\approxWB};
           \addplot [mark=diamond,color=c3] table[x=k,y=FWReal]{\approxWB};
            \addplot [mark=star,color=c8] table[x=k,y=MWUReal]{\approxWB};
           \addplot [mark=triangle,color=c4] table[x=k,y=FISTAReal]{\approxWB};
			\end{axis}
		\end{tikzpicture}
	}
	%
		\subfigure[FS]{
		\begin{tikzpicture}[scale=0.35]
			\begin{axis}[
				 legend style = {
				    legend columns=-1,
				    draw=none,
				},
              xtick = {2, 4, 6, 8, 10},
                xticklabels={0.1, 0.05, 0.01, 0.001, 0.0001},
				ymode = log,
				width=0.43\textwidth,
    			height=.35\textwidth,
				mark size=6.0pt, 
				xlabel={\Huge \bf $\epsilon$}, 
				ticklabel style={font=\Huge},
				every axis plot/.append style={line width = 2.5pt},
				every axis/.append style={line width = 2.5pt},
				]
            \addplot [mark=square,color=c5] table[x=k,y=Flow]{\approxFS};
           \addplot [mark=diamond,color=c3] table[x=k,y=FWReal]{\approxFS};
            \addplot [mark=star,color=c8] table[x=k,y=MWUReal]{\approxFS};
           \addplot [mark=triangle,color=c4] table[x=k,y=FISTAReal]{\approxFS};
			\end{axis}
		\end{tikzpicture}
	}
	\caption{Efficiency results of ($1+\epsilon$)-approximation algorithms for a given $\epsilon$  on undirected graphs.}
	\label{fig:uds:real_new_approx}
\end{figure}
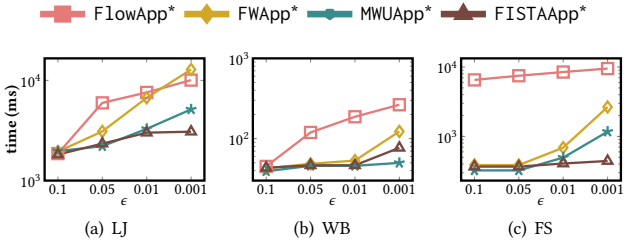

\noindent \textbf{4. Evaluation of weight update strategies.} In this experiment, we test the effect of the vertex weight update strategies, i.e., sequential and simultaneous update strategies, in all CP-based algorithms.
As shown in Figure \ref{fig:uds:update_strategy}, we observe that the algorithms using the simultaneous update strategy almost always take less time to achieve the same approximation ratios than the algorithms using the sequential weight update strategy.
This is because the simultaneous update strategy makes a more balanced weight distribution.

Based on the above results and discussions, we obtain two major conclusions:
1) Graph reduction is highly beneficial for all algorithms, and employing multi-round reduction often yields better outcomes than single-round reduction.
2) For CP-based algorithms, the simultaneous update strategy typically outperforms the sequential one.
By combining existing techniques, we introduce new algorithms which we present and evaluate in the next section.

\begin{figure}[h]
\renewcommand\arraystretch{0.95}
    \centering
     \setlength{\abovecaptionskip}{-0.1cm}
  \setlength{\belowcaptionskip}{-0.1cm}
       	\begin{tikzpicture}[scale=0.52]
        		\begin{axis}[
                        grid = major,
        			ybar=0.11pt,
        			bar width=0.45cm,
        			width=0.9\textwidth,
    				height=0.27\textwidth,
        			xlabel={\huge \bf dataset}, 
        			xtick=data,	xticklabels={EB,DP,YT,LJ,WB,FS},
                     legend style={at={(0.5,1.30)}, anchor=north,legend columns=-1,draw=none},
                           legend image code/.code={
            \draw [#1] (0cm,-0.263cm) rectangle (0.4cm,0.15cm); },
        			xmin=0.8,xmax=13.2,
                   	ymin=0.001,ymax=800000,
                         ytick = {0.01, 0.1, 1, 10, 100, 1000, 10000, 100000,800105},
    	        yticklabels = {$10^{-2}$,$10^{-1}$, $10^0$,$10^1$, $10^2$, $10^3$, $10^4$, $10^5$, \textbf{INF}},
                    ymode = log,    
                log origin=infty,
        			tick align=inside,
        			ticklabel style={font=\huge},
        			every axis plot/.append style={line width = 1.6pt},
        			every axis/.append style={line width = 1.6pt},
                        ylabel={\textbf{\huge time (s) }}
        			]
        			\addplot[fill=p1] table[x=datasets,y=Core]{\exactUDS};
           			\addplot[fill=p2] table[x=datasets,y=FWI]{\exactUDS};
              \addplot[fill=p3] table[x=datasets,y=MWUI]{\exactUDS};
                \addplot[fill=p4] table[x=datasets,y=FISTAI]{\exactUDS};
                \legend{\huge {\tt CoreExact}, \huge {\tt FWExact}*, \huge {\tt MWUExact}*,\huge {\tt FISTAExact}*};
        		\end{axis}
        	\end{tikzpicture}
        	\caption{Efficiency of exact algorithms for UDS.}
    \label{fig:uds:new_exact_time}
    
\end{figure}
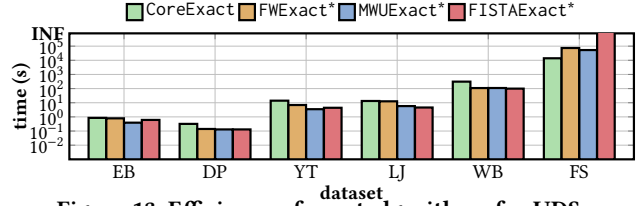

\noindent \textbf{5. New algorithms.} 
%
%
Specifically, we apply a simultaneous update strategy and multi-round graph reductions for all CP-based algorithms. We denote modified versions of original algorithms with an asterisk (*).
For \texttt{Greedy++}, we enhance efficiency through graph reduction, and denote this variant by \texttt{Greedy-M}.
We first evaluate the efficiency of various ($1+\epsilon$)-approximation algorithms with the optimal density used in the early stop check.
Figure \ref{fig:uds:new_approx} indicates that for smaller datasets, graph reduction is not very useful.
However, for large graphs, {\tt Greedy-M} is always the best one, and those new CP-based algorithms achieve comparable performance. This is remarkable given that {\tt Greedy-M} is a much simpler algorithm than the CP-based ones.

We test all CP-based and {\tt FlowApp}* algorithms by reporting their running time on the four largest datasets.
In real situations, the optimal density is not known in advance, so we terminate an algorithm when it meets the early stop conditions that are set according to the specified $\epsilon$, and report the running time in Figure \ref{fig:uds:real_new_approx}.
Our findings lead to two main insights: firstly, all CP-based algorithms perform similarly well. Secondly, CP-based algorithms are up to one order of magnitude faster than {\tt FlowApp}*. 

In addition, we compare the efficiency of the newly designed CP-based exact algorithms and {\tt CoreExact}.
As shown in Figure \ref{fig:uds:new_exact_time}, the new algorithms generally require less time than \texttt{CoreExact} except on the FS dataset.
%
After we deeply delve into the FS dataset, we discover a subgraph with a density of 273.518, nearly matching the optimal density of 273.519. 
We conjecture that if some subgraphs have densities very close to the optimal density, the exact CP-based algorithms often require a larger number of iterations to obtain the optimal solution.
To verify it, we synthesise four datasets as follows: 1) we generate two cliques, each with 1,000 vertices, and connect these two cliques by a single edge; 2) we then randomly remove $0.001\%$, $0.01\%$, $0.1\%$, and $1\%$ of the edges from the second clique to simulate if there exists a subgraph with a density close to that of the densest subgraph (i.e., the first clique).

\begin{table}[h]
\renewcommand\arraystretch{0.98}
    \centering
    \small
    \caption{The number of iterations.}
    \begin{tabular}{l|r|r|r|r}
    \toprule
        Method & 1\% & 0.1\% & 0.01 \%  & 0.001\%  \\ \hline
        FW & 128 & 2,048 & 32,768  & 131,072 \\ \hline
        MWU & 128 & 1,024 & 16,384 & 65,536 \\ \hline
        FISTA & 1 & 128 & 512 & 1,024 \\ \bottomrule
    \end{tabular}
    \label{tab:uds:create_graph}
\end{table}

Table~\ref{tab:uds:create_graph} reports the number of iterations required by these algorithms on these datasets.
We observe that as the density of the second graph approaches that of a clique, CP-based exact algorithms require a significantly higher number of iterations.

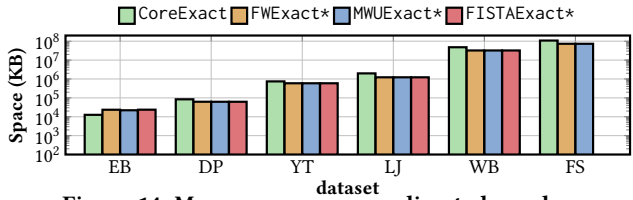
\begin{figure}[h]
    \centering
     \setlength{\abovecaptionskip}{-0.1cm}
  \setlength{\belowcaptionskip}{-0.1cm}
       	\begin{tikzpicture}[scale=0.52]
        		\begin{axis}[
                        grid = major,
        			ybar=0.11pt,
        			bar width=0.45cm,
        			width=0.9\textwidth,
    				height=0.26\textwidth,
        			xlabel={\huge \bf dataset}, 
        			xtick={2, 4, 6, 8, 10, 12},	
                        xticklabels={EB,DP,YT,LJ,WB,FS},
                     legend style={at={(0.5,1.30)}, anchor=north,legend columns=-1,draw=none},
                           legend image code/.code={
            \draw [#1] (0cm,-0.263cm) rectangle (0.4cm,0.15cm); },
        			xmin=0.8,xmax=13.2,
    					ymin=100,ymax=200000050,
                    ytick = {100, 1000, 10000, 100000, 1000000, 10000000, 100000000},
                    yticklabels = {$10^2$,$10^3$, $10^4$, $10^5$, $10^6$, $10^7$, $10^8$},
                    ymode = log,    
                log origin=infty,
        			tick align=inside,
        			ticklabel style={font=\huge},
        			every axis plot/.append style={line width = 1.6pt},
        			every axis/.append style={line width = 1.6pt},
                        ylabel={\textbf{\huge Space (KB)}}
        			]	
                \addplot[fill=p1] table[x=datasets,y=CORE]{\UDSmem};
                \addplot[fill=p2] table[x=datasets,y=FW]{\UDSmem};
                \addplot[fill=p3] table[x=datasets,y=MWU]{\UDSmem};
                \addplot[fill=p4] table[x=datasets,y=FISTA]{\UDSmem};
                \legend{\huge {\tt CoreExact}, \huge {\tt FWExact*}, \huge {\tt MWUExact*}, \huge {\tt FISTAExact*}};
        		\end{axis}
        	\end{tikzpicture}
    	\caption{Memory usage on undirected graphs.}
    \label{fig:uds:exact_memory}
\end{figure}

\noindent \textbf{6. Memory usage.} We report the memory usage of all exact UDS algorithms in Figure \ref{fig:uds:exact_memory}.
We observe that the memory costs of all algorithms are almost at the same scale because all algorithms incur linear memory usage w.r.t. the graph size (i.e., $O(m)$).
For approximation algorithms, their theoretical space costs are also the same, so we omit the evaluation.

\begin{table*}[h]
\renewcommand\arraystretch{0.892}
\caption{Comparison of 2-approx. DDS algorithms (\textcolor{red}{Red} denotes the most efficient algorithm, and \textcolor{t3}{Blue} denotes the best accuracy).}
\begin{tabular}{l|rr|rr|rr|rr|rr|rr}
\toprule
\multirow{2}{*}{Dataset} &  \multicolumn{2}{c|}{OF} & \multicolumn{2}{c|}{AD} & \multicolumn{2}{c|}{AM} & \multicolumn{2}{c|}{BA} & \multicolumn{2}{c|}{WE} & \multicolumn{2}{c}{SK} \\ \cline{2-13}
~ & time & ratio &time & ratio &time & ratio &time & ratio &time & ratio &time & ratio  \\ \hline
\texttt{DGreedy} & ---     &     --- &---     &     --- &---     &     --- &---     &     --- &---     &     --- & ---     &     ---  \\ \hline
\texttt{DFWApp}   &  87.8 ms &\textcolor{t3}{1.001} & 175.3 ms &\textcolor{t3}{1.009} & 822.2 ms &\textcolor{t3}{1.004} & 22.0 s &\textcolor{t3}{1.000} & 432.3 s &\textcolor{t3}{1} & 1256.6 s & \textcolor{t3}{1.000}    \\ \hline
\texttt{FWVW}   &  139.0 ms & 1.135 & 391.1 ms & 1.122 & 6623.3 s & \textcolor{t3}{1.004} & 557.1 s & 1.226 & 7855.1 s & \textcolor{t3}{1} & 19585.2 s & 1.010   \\ \hline
\texttt{MWUVW}   &  204.7 ms & 1.024 & 600.0 ms & 1.122 & --- & --- & 480.0 s & 1.226 & 7739.7 s & \textcolor{t3}{1} & 18417.0 s & \textcolor{t3}{1.000}
   \\ \hline
    \texttt{FISTAVW}   &  251.7 ms & 1.135 & 480.1 ms & 1.122 & 846.6 s & \textcolor{t3}{1.004} & 408.1 s & \textcolor{t3}{1.000} & 7349.1 s & \textcolor{t3}{1} & 22076.3 s & \textcolor{t3}{1.000}   \\ \hline
\texttt{XYCoreApp}  &   \textcolor{t4}{27.9 ms} &1.422 & 17.2 ms &1.130 & 751.7 ms &\textcolor{t3}{1.004} & 4.6 s &\textcolor{t3}{1.000} & 131.0 s &1.007 & 1423.9 s &\textcolor{t3}{1.000}  \\ \hline
\texttt{WCoreApp}  &   333.7 ms & 1.396& \textcolor{t4}{6.0 ms} & 1.130& \textcolor{t4}{253.9 ms} & \textcolor{t3}{1.004}& \textcolor{t4}{1.4 s} & \textcolor{t3}{1.000}& \textcolor{t4}{12.6 s} & 1.007& \textcolor{t4}{100.3 s} & \textcolor{t3}{1.000} \\ 
\bottomrule
\end{tabular}
\label{tab:dds:2_approx}
\end{table*}

\begin{figure*}[h]
\small
\setlength{\abovecaptionskip}{-0.1cm}
\setlength{\belowcaptionskip}{-0.3cm}
\centering		
\ref{polt_appro_dds}\\
	\subfigure[OF]{
		\begin{tikzpicture}[scale=0.35]
			\begin{axis}[
				width=0.43\textwidth,
    			height=.33\textwidth,
    			 legend style = {
				    legend columns=-1,
				    draw=none,
				},
                ymax=100050,
                xtick = {2, 4, 6, 8},
                legend to name=polt_appro_dds,
                xticklabels={0.1, 0.01, 0.001, 0.0001},
                ytick = {0.01 ,0.1, 1, 10, 100, 1000, 10000, 100000},
                yticklabels={$10^{-1}$, $10^{0}$, $10^{1}$, $10^{2}$, $10^{3}$, $10^{4}$, \textbf{INF}},
				ymode = log,
				mark size=6.0pt, 
				ylabel={\Huge \bf time (s)},
				xlabel={\Huge \bf $\epsilon$}, 
				ticklabel style={font=\Huge},
				every axis plot/.append style={line width = 2.5pt},
				every axis/.append style={line width = 2.5pt},
				]
            \addplot [mark=o,color=c8] table[x=k,y=DFW]{\approxOF};
            \addplot [mark=square,color=c3] table[x=k,y=FW]{\approxOF};
            \addplot [mark=star,color=c1] table[x=k,y=MWU]{\approxOF};
            \addplot [mark=triangle,color=c2] table[x=k,y=FISTA]{\approxOF};
            \legend{\texttt{DFWApp}, \texttt{FWVW}, \texttt{MWUVW}, \texttt{FISTAVW}};
			\end{axis}
		\end{tikzpicture}
	}
	%
 \hspace{-1pt}
	\subfigure[AD]{
		\begin{tikzpicture}[scale=0.35]
			\begin{axis}[ 
				width=0.43\textwidth,
    			height=.33\textwidth,
			    legend style = {
				    legend columns=-1,
				    draw=none,
				},
			    xtick = {2, 4, 6, 8},
                xticklabels={0.1, 0.01, 0.001, 0.0001},
                ytick = {0.1, 1, 10, 100, 1000, 10000, 100000},
                yticklabels={$10^{-1}$, $10^{0}$, $10^{1}$, $10^{2}$, $10^{3}$, $10^{4}$, \textbf{INF}},
				ymode = log,
                    ymin=0.1, ymax=100050,
				mark size=6.0pt, 
				xlabel={\Huge \bf $\epsilon$}, 
				ticklabel style={font=\Huge},
				every axis plot/.append style={line width = 2.5pt},
				every axis/.append style={line width = 2.5pt},
				]
            \addplot [mark=o,color=c8] table[x=k,y=DFW]{\approxAD};
            \addplot [mark=square,color=c3] table[x=k,y=FW]{\approxAD};
            \addplot [mark=star,color=c1] table[x=k,y=MWU]{\approxAD};
            \addplot [mark=triangle,color=c2] table[x=k,y=FISTA]{\approxAD};
			\end{axis}
		\end{tikzpicture}
	}
\hspace{-1pt}
	\subfigure[AM]{
		\begin{tikzpicture}[scale=0.35]
			\begin{axis}[
			legend style = {
				    legend columns=-1,
				    draw=none,
				},
			width=0.43\textwidth,
			height=.33\textwidth,
	          xtick = {2, 4, 6, 8},
                xticklabels={0.1, 0.01, 0.001, 0.0001},
				ymode = log,
				mark size=6.0pt, 
                    ymin=0.1, ymax=100050,
                    ytick = {0.1, 1, 10, 100, 1000, 10000, 100000},
                    yticklabels={$10^{-1}$,$10^{0}$, $10^{1}$, $10^{2}$, $10^{3}$, $10^{4}$, \textbf{INF}},
				xlabel={\Huge \bf $\epsilon$}, 
				ticklabel style={font=\Huge},
			every axis plot/.append style={line width = 2.5pt},
			every axis/.append style={line width = 2.5pt},
			]
            \addplot [mark=o,color=c8] table[x=k,y=DFW]{\approxAM};
            \addplot [mark=square,color=c3] table[x=k,y=FW]{\approxAM};
            \addplot [mark=star,color=c1] table[x=k,y=MWU]{\approxAM};
            \addplot [mark=triangle,color=c2] table[x=k,y=FISTA]{\approxAM};
			\end{axis}
		\end{tikzpicture}
	}
	%
\hspace{-1pt}
	\subfigure[BA]{
		\begin{tikzpicture}[scale=0.35]
			\begin{axis}[
		        legend style = {
				    legend columns=-1,
				    draw=none,
				},
                xtick = {2, 4, 6, 8},
                xticklabels={0.1, 0.01, 0.001, 0.0001},
                ytick = {10, 100, 1000, 10000, 100000},
                    yticklabels={$10^{1}$, $10^{2}$, $10^{3}$, $10^{4}$, \textbf{INF}},
				ymode = log,
                    ymax=100050,
				width=0.43\textwidth,
    			height=.33\textwidth,
				mark size=6.0pt, 
				xlabel={\Huge \bf $\epsilon$}, 
				ticklabel style={font=\Huge},
				every axis plot/.append style={line width = 2.5pt},
				every axis/.append style={line width = 2.5pt},
				]
            \addplot [mark=o,color=c8] table[x=k,y=DFW]{\approxBA};
            \addplot [mark=square,color=c3] table[x=k,y=FW]{\approxBA};
            \addplot [mark=star,color=c1] table[x=k,y=MWU]{\approxBA};
            \addplot [mark=triangle,color=c2] table[x=k,y=FISTA]{\approxBA};
			\end{axis}
		\end{tikzpicture}
	}
	%
\hspace{-1pt}
    \subfigure[WE]{
        \begin{tikzpicture}[scale=0.35]
            \begin{axis}[
                    legend style = {
                    legend columns=-1,
                    draw=none,
                },
                xtick = {2, 4, 6, 8},
                xticklabels={0.1, 0.01, 0.001, 0.0001},
                ytick = {100, 1000, 10000, 100000},
                    yticklabels={$10^{2}$, $10^{3}$, $10^{4}$, \textbf{INF}},
                ymode = log,
                ymin=100, ymax=100050,
                mark size=6.0pt, 
                width=0.43\textwidth,
                height=.33\textwidth,
                xlabel={\Huge \bf $\epsilon$}, 
                ticklabel style={font=\Huge},
                every axis plot/.append style={line width = 2.5pt},
                every axis/.append style={line width = 2.5pt},
                ]
            \addplot [mark=o,color=c8] table[x=k,y=DFW]{\approxWE};
            \addplot [mark=square,color=c3] table[x=k,y=FW]{\approxWE};
            \addplot [mark=star,color=c1] table[x=k,y=MWU]{\approxWE};
            \addplot [mark=triangle,color=c2] table[x=k,y=FISTA]{\approxWE};
            \end{axis}
        \end{tikzpicture}
    }
	%
\hspace{-1pt}
    \subfigure[SK]{
		\begin{tikzpicture}[scale=0.35]
			\begin{axis}[
				 legend style = {
				    legend columns=-1,
				    draw=none,
				},
              xtick = {2, 4, 6, 8},
                xticklabels={0.1, 0.01, 0.001, 0.0001},
                ytick = {1000, 10000, 100000},
                    yticklabels={$10^{3}$, $10^{4}$, \textbf{INF}},
                    ymax = 100050,
				ymode = log,
				width=0.43\textwidth,
    			height=.33\textwidth,
				mark size=6.0pt, 
				xlabel={\Huge \bf $\epsilon$}, 
				ticklabel style={font=\Huge},
				every axis plot/.append style={line width = 2.5pt},
				every axis/.append style={line width = 2.5pt},
				]
             \addplot [mark=o,color=c8] table[x=k,y=DFW]{\approxSK};
            \addplot [mark=square,color=c3] table[x=k,y=FW]{\approxSK};
            \addplot [mark=star,color=c1] table[x=k,y=MWU]{\approxSK};
            \addplot [mark=triangle,color=c2] table[x=k,y=FISTA]{\approxSK};
			\end{axis}
		\end{tikzpicture}
	}
	\caption{Efficiency results of ($1+\epsilon$)-approximation algorithms on directed graphs.}
	\label{fig:dds:approx}
\end{figure*}

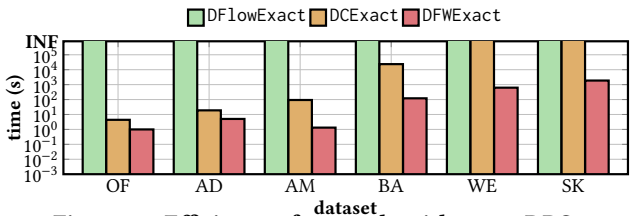
\begin{figure}[h]
    \centering
     \setlength{\abovecaptionskip}{-0.1cm}
       	\begin{tikzpicture}[scale=0.52]
        		\begin{axis}[
                        grid = major,
        			ybar=0.11pt,
        			bar width=0.6cm,
        			width=0.9\textwidth,
    				height=0.28\textwidth,
        			xlabel={\huge \bf dataset}, 
        			xtick={6, 8, 10, 12, 14, 16},	xticklabels={OF, AD, AM, BA, WE, SK},
                     legend style={at={(0.5,1.30)}, anchor=north,legend columns=-1,draw=none},
                           legend image code/.code={
            \draw [#1] (0cm,-0.263cm) rectangle (0.4cm,0.15cm); },
        			xmin=4.8,xmax=17.2,
    					ymin=0.001,ymax=800000,
                         ytick = {0.001, 0.01, 0.1, 1, 10, 100, 1000, 10000, 100000,800105},
    	        yticklabels = {$10^{-3}$,$10^{-2}$,$10^{-1}$, $10^0$,$10^1$, $10^2$, $10^3$, $10^4$, $10^5$, \textbf{INF}},
                    ymode = log,    
                log origin=infty,
        			tick align=inside,
        			ticklabel style={font=\huge},
        			every axis plot/.append style={line width = 1.6pt},
        			every axis/.append style={line width = 1.6pt},
                        ylabel={\textbf{\huge time (s) }},
        			]
        			\addplot[fill=p1] table[x=datasets,y=Flow]{\exactDDS};
        			\addplot[fill=p2] table[x=datasets,y=Core]{\exactDDS};
           			\addplot[fill=p4] table[x=datasets,y=FW]{\exactDDS};
                \legend{\huge {\tt DFlowExact},\huge {\tt DCExact}, \huge {\tt DFWExact}}
            		\end{axis}
        	\end{tikzpicture}
    	\caption{Efficiency  of exact algorithms on DDS.}
    \label{fig:dds:exact_time}
\end{figure}

\subsection{Evaluation of DDS algorithms}
\label{sec:exp:dds}
\noindent \textbf{1. Overall performance.} 
Similar to the UDS part, we first evaluate the overall performance of various DDS algorithms. To be specific, 
(1) we compare the efficiency and accuracy of all approximation algorithms in Table \ref{tab:dds:2_approx}. Here, we also set $\epsilon = 1$ for \texttt{DFWApp} and \texttt{VWApp};
(2) we test the performance of \texttt{DFWApp} and \texttt{VWApp} over different $\epsilon$ values, from $0.1$ to $0.0001$ in  Figure \ref{fig:dds:approx};
and (3) we record the running time of \texttt{DFlowExact}, \texttt{DCExact} and \texttt{DFWExact} in Figure \ref{fig:dds:exact_time}.
Based on the above results, we make the following observations:
(1) \texttt{WCoreApp} is the most efficient one for larger graphs, but \texttt{DFWApp} usually yields a more accurate solution.
(2) Although we have incorporated the SOTA UDS algorithm into {\tt VWApp}, it is still slower than \texttt{DFWApp}. 
The main reason is that, although \texttt{VWApp} reduces the number of different trials of $c$ from $O(n^2)$ to $O(\log_{1+\epsilon}n)$, this number is still too many compared to that of \texttt{DFWApp}.
(3) For exact algorithms, \texttt{DFWExact} always outperforms the rest, because it significantly reduces the time cost incurred for computing maximum flow, and it can compute the minimum cut on the smaller flow network.
\begin{table*}[h]
\renewcommand\arraystretch{0.9}
\centering
\small
 \setlength{\abovecaptionskip}{-0.001cm}
\begin{tabular}{c|r|lrrl|lrrl|lrrl|r}
\toprule
\multicolumn{1}{c|}{\multirow{2}{*}{Dataset}} & \multicolumn{1}{c|}{\multirow{2}{*}{$\epsilon$}} &  & \multicolumn{2}{c}{\#Vertices}                             &  &  & \multicolumn{2}{c}{\#Edges}                                &  &  & \multicolumn{2}{c}{Time (s)}                               &  & \multicolumn{1}{l}{\multirow{2}{*}{Speedup}} \\ \cline{3-14}
\multicolumn{1}{c|}{}                         & \multicolumn{1}{c|}{}                   &  & \multicolumn{1}{c}{Original} & \multicolumn{1}{c}{Reduced} &  &  & \multicolumn{1}{c}{Original} & \multicolumn{1}{c}{Reduced} &  &  & \multicolumn{1}{c}{Original} & \multicolumn{1}{c}{Reduced} &  & \multicolumn{1}{l}{}                         \\ \hline
\multirow{4}{*}{AM} &0.01 &  & 403,394 & 11,045 &  &  & 3,387,388 & 20,058 &  &  & 216.0 & 1.5 &  & 143.6\\
~ & 0.001 &  & 403,394 & 9,730 &  &  & 3,387,388 & 15,046 &  &  & 1,791.2 & 2.9 &  & 612.1\\
~ & 0.0001 &  & 403,394 & 9,066 &  &  & 3,387,388 & 12,513 &  &  & 7,885.0 & 17.3 &  & 455.7\\
~ & 0 &  & 403,394 & 10,632 &  &  & 3,387,388 & 18,483 &  &  & 119.5 & 1.7 &  & 71.1\\ \hline
\multirow{4}{*}{BA} &0.01 &  & 2,141,300 & 137,181 &  &  & 17,794,839 & 698,422 &  &  & 1,505.5 & 35.0 &  & 43.0\\
~ & 0.001 &  & 2,141,300 & 137,181 &  &  & 17,794,839 & 698,422 &  &  & 4,244.5 & 52.3 &  & 81.2\\
~ & 0.0001 &  & 2,141,300 & 156,486 &  &  & 17,794,839 & 652,086 &  &  & 13,870.6 & 156.6 &  & 88.6\\
~ & 0 &  & 2,141,300 & 151,901 &  &  & 17,794,839 & 711,706 &  &  & 1,319.3 & 153.5 &  & 8.6\\ \bottomrule
\end{tabular}
\caption{\textbf{The effect of graph reduction on directed graphs.}}
\label{tab:dds_reduction}
\end{table*}

\begin{table*}[h]
\renewcommand\arraystretch{0.89}
 \setlength{\abovecaptionskip}{-0.001cm}
    \centering
    \small
        \begin{tabular}{l|l|rr|rr|rr|rr}
            \toprule
             \multicolumn{2}{c|}{\multirow{2}*{Dataset}} & \multicolumn{2}{c|}{$\epsilon = 0.01$} & \multicolumn{2}{c|}{$\epsilon=0.001$} & \multicolumn{2}{c|}{$\epsilon=0.0001$} & \multicolumn{2}{c}{$\epsilon=0$} \\
             \cline{3-10}
            \multicolumn{2}{c|}{~} & \texttt{DFW} & \texttt{DFWReN} &\texttt{DFW} & \texttt{DFWReN} &\texttt{DFW} & \texttt{DFWReN} &\texttt{DFW} & \texttt{DFWReN} \\
            \hline
            
            \multirow{5}*{WE}&\#Ratios&23 & 21 & 29 & 32 & 32 & 35 & 20 & 19\\
            &\#Edges&9,458,420 & 41,014,857 & 9,597,565 & 30,606,451 & 9,221,802 & 28,859,994 & 10,767,492 & 44,912,585\\
            &\#Iterations&174 & 152 & 486 & 603 & 2,156 & 2,580 & 100 & 105\\
            &Product&$3.78\times 10^{10}$ & $1.31\times 10^{11}$ & $1.35\times 10^{11}$ & $5.91\times 10^{11}$ & $6.36\times 10^{11}$ & $2.61\times 10^{12}$ & $2.15\times 10^{10}$ & $8.98\times 10^{10}$\\
            &Time (s)&1,033.3 & 3,723.7 & 2,266.8 & 13,658.4 & 7,869.2 & 23,916.9 & 643.0 & 2,865.3\\\hline
            \multirow{5}*{SK}&\#Ratios&17 & 15 & 24 & 18 & 28 & --- & 21 & 15\\
            &\#Edges&44,552,779 & 375,979,205 & 44,837,242 & 313,662,857 & 44,362,182 & --- & 45,844,129 & 365,743,496\\
            &\#Iterations&124 & 120 & 258 & 344 & 1,914 & --- & 124 & 147\\
            &Product&$9.36\times 10^{10}$ & $6.77\times 10^{11}$ & $2.78\times 10^{11}$ & $1.94\times 10^{12}$ & $2.38\times 10^{12}$ & --- & $1.19\times 10^{11}$ & $8.05\times 10^{11}$\\
            &Time (s)&1,419.7 & 15,856.6 & 3,425.8 & 48,106.6 & 24,922.1 & --- & 2,425.7 & 15,909.8\\
            \bottomrule
        \end{tabular}
      \caption{Effectiveness of graph reduction technique on directed graphs.}
      \label{tab:dds:interval_adjstment}
\end{table*}

 \begin{figure*}[h]
 \small
  \setlength{\abovecaptionskip}{-0.1cm}
  \setlength{\belowcaptionskip}{-0.3cm}
    \centering
         %
     	\subfigure[AD (time)]{
       	\begin{tikzpicture}[scale=0.36]
        		\begin{axis}[
                        grid=major,
        			ybar=0.11pt,
        			bar width=0.5cm,
        			width=0.44\textwidth,
    				height=0.33\textwidth,
        			xlabel={\Huge $\epsilon$}, 
        			xtick=data,	xticklabels={$10^{-1}$, $10^{-2}$,$10^{-3}$,$10^{-4}$,0.0},
                     legend style={legend columns=-1,draw=none},
                           legend image code/.code={
            \draw [#1] (0cm,-0.263cm) rectangle (0.4cm,0.15cm); },
        			xmin=0.8,xmax=11.2,
                        ymax=1001,
                        ymode = log,
                        log origin=infty,
        			tick align=inside,
        			ticklabel style={font=\Huge},
        			every axis plot/.append style={line width = 2.5pt},
        			every axis/.append style={line width = 2.5pt},
                        ylabel={\textbf{\Huge time (s) }}
        			]
        			\addplot[fill=p1]table[x=datasets,y=FWSim]{\updateAD};
        			\addplot[fill=p4] table[x=datasets,y= FWSeq]{\updateAD};
           \legend{{\huge \tt DFWSeq},{\huge \tt DFWSim}};
        		\end{axis}
        	\end{tikzpicture}
         }
         \hspace{1pt}
		 \subfigure[AM (time)]{
       	\begin{tikzpicture}[scale=0.36]
        		\begin{axis}[
                        grid=major,
        			ybar=0.11pt,
        			bar width=0.5cm,
        			width=0.44\textwidth,
    				height=0.33\textwidth,
        			xlabel={\Huge $\epsilon$}, 
        			xtick=data,	xticklabels={$10^{-1}$, $10^{-2}$,$10^{-3}$,$10^{-4}$,0.0},
                     legend style={legend columns=-1,draw=none},
                           legend image code/.code={
            \draw [#1] (0cm,-0.263cm) rectangle (0.4cm,0.15cm); },
        			xmin=0.8,xmax=11.2,
                        ymin=0.09, ymax=110,
                        ymode = log,
                        log origin=infty,
        			tick align=inside,
        			ticklabel style={font=\Huge},
        			every axis plot/.append style={line width = 2.5pt},
        			every axis/.append style={line width = 2.5pt},
        			]
        			\addplot[fill=p1]table[x=datasets,y=FWSim]{\updateAM};
        			\addplot[fill=p4] table[x=datasets,y= FWSeq]{\updateAM};
           \legend{{\huge \tt DFWSeq},{\huge \tt DFWSim}};
        		\end{axis}
        	\end{tikzpicture}
         }
         \hspace{1pt}
     	\subfigure[BA (time)]{
       	\begin{tikzpicture}[scale=0.36]
        		\begin{axis}[
                        grid=major,
        			ybar=0.11pt,
        			bar width=0.5cm,
        			width=0.44\textwidth,
    				height=0.33\textwidth,
        			xlabel={\Huge $\epsilon$}, 
        			xtick=data,	xticklabels={$10^{-1}$, $10^{-2}$,$10^{-3}$,$10^{-4}$,0.0},
                     legend style={legend columns=-1,draw=none},
                           legend image code/.code={
            \draw [#1] (0cm,-0.263cm) rectangle (0.3cm,0.15cm); },
        			xmin=0.8,xmax=11.2,
                        ymin=0.9, ymax=1700,
                        ytick={1, 10, 100, 1000},
                        ymode = log,
                        log origin=infty,
        			tick align=inside,
        			ticklabel style={font=\Huge},
        			every axis plot/.append style={line width = 2.5pt},
        			every axis/.append style={line width = 2.5pt},
        			]
        			\addplot[fill=p1]table[x=datasets,y=FWSim]{\updateBA};
        			\addplot[fill=p4] table[x=datasets,y= FWSeq]{\updateBA};
           \legend{{\huge \tt DFWSeq},{\huge \tt DFWSim}};
        		\end{axis}
        	\end{tikzpicture}
         }
         %
	%
 \hspace{1pt}
	\subfigure[AD (\#c)]{
		\begin{tikzpicture}[scale=0.36]
			\begin{axis}[ 
				width=0.44\textwidth,
    			height=.33\textwidth,
			    legend style = {
				    legend columns=-1,
				    draw=none,
				},
			    xtick = {2, 4, 6, 8, 10},
                xticklabels={$10^{-1}$, $10^{-2}$,$10^{-3}$,$10^{-4}$,0.0},
				mark size=6.0pt, 
                    ymax=110,
                    ytick={0, 20, 40, 60, 80, 100},
				ylabel={\Huge \bf \#$c$},
				xlabel={\Huge \bf $\epsilon$}, 
				ticklabel style={font=\Huge},
				every axis plot/.append style={line width = 2.5pt},
				every axis/.append style={line width = 2.5pt},
				]
            \addplot [mark=o,color=c8] table[x=datasets,y=FWSim]{\updateADc};
            \addplot [mark=square,color=c2] table[x=datasets,y= FWSeq]{\updateADc};
            \legend{ \huge \texttt{DFWSeq},  \huge \texttt{DFWSim}};
			\end{axis}
		\end{tikzpicture}
	}
        \hspace{1pt}
	\subfigure[AM (\#c)]{
		\begin{tikzpicture}[scale=0.36]
			\begin{axis}[
			legend style = {
				    legend columns=-1,
				    draw=none,
				},
			width=0.44\textwidth,
			height=.33\textwidth,
	          xtick = {2, 4, 6, 8, 10},
                xticklabels={$10^{-1}$, $10^{-2}$,$10^{-3}$,$10^{-4}$,0.0},
				mark size=6.0pt, 
                    ymax=45,
                    ytick={0, 10, 20, 30, 40},
				xlabel={\Huge \bf $\epsilon$}, 
				ticklabel style={font=\Huge},
			every axis plot/.append style={line width = 2.5pt},
			every axis/.append style={line width = 2.5pt},
			]
            \addplot [mark=o,color=c8] table[x=datasets,y=FWSim]{\updateAMc};
            \addplot [mark=square,color=c2] table[x=datasets,y= FWSeq]{\updateAMc};
            \legend{ \huge \texttt{DFWSeq},  \huge \texttt{DFWSim}};
			\end{axis}
		\end{tikzpicture}
	}
	%
	\subfigure[BA (\#c)]{
		\begin{tikzpicture}[scale=0.36]
			\begin{axis}[
		        legend style = {
				    legend columns=-1,
				    draw=none,
				},
                xtick = {2, 4, 6, 8, 10},
                xticklabels={$10^{-1}$, $10^{-2}$,$10^{-3}$,$10^{-4}$,0.0},
				width=0.44\textwidth,
    			height=.33\textwidth,
				mark size=6.0pt, 
                    ymax=40,
				xlabel={\Huge \bf $\epsilon$}, 
				ticklabel style={font=\Huge},
				every axis plot/.append style={line width = 2.5pt},
				every axis/.append style={line width = 2.5pt},
				]
            \addplot [mark=o,color=c8] table[x=datasets,y=FWSim]{\updateBAc};
            \addplot [mark=square,color=c2] table[x=datasets,y= FWSeq]{\updateBAc};
           \legend{ \huge \texttt{DFWSeq},  \huge \texttt{DFWSim}};
			\end{axis}
		\end{tikzpicture}
	}
    	\caption{The effect of update strategies on directed graphs.}
    \label{fig:dds:update_strategy}
\end{figure*}

\begin{figure}[h]
  \setlength{\abovecaptionskip}{-0.1cm}
  \setlength{\belowcaptionskip}{-0.3cm}
            \subfigure[Running time]{
		\begin{tikzpicture}[scale = 0.41]
			\begin{axis}[
			   legend style={at={(0.5,1.30)}, anchor=north,legend columns=-1,draw=none},
                xtick = {2, 4, 6, 8, 10, 12},
                xticklabels={0, 0.2, 0.4, 0.6, 0.8, 1},
                ymin=1,ymax=800000,
                         ytick = {1, 10, 100, 1000, 10000, 100000,800105},
    	        yticklabels = {$10^0$,$10^1$, $10^2$, $10^3$, $10^4$, $10^5$, \textbf{INF}},
				ymode = log,
				mark size=6.0pt, 
				width=0.53\textwidth,
    			height=.33\textwidth,
           	    ylabel={\Huge \bf time (s)},
				xlabel={\Huge \bf $\gamma$}, 
				ticklabel style={font=\Huge},
				every axis plot/.append style={line width = 2.5pt},
				every axis/.append style={line width = 2.5pt},
				]
            \addplot [mark=o,color=c8] table[x=k,y=OF]{\learningrate};
            \addplot [mark=square,color=c3] table[x=k,y=AD]{\learningrate};
            \addplot [mark=star,color=c1] table[x=k,y=AM]{\learningrate};
            \addplot [mark=triangle,color=c2] table[x=k,y=BA]{\learningrate};
			\legend{ \huge{OF}, \huge{AD}, \huge{AM}, \huge{BA}};
			\end{axis}
		\end{tikzpicture}
            }
            \subfigure[Distribution of $c$ on OF]{
            \includegraphics[scale=0.41]{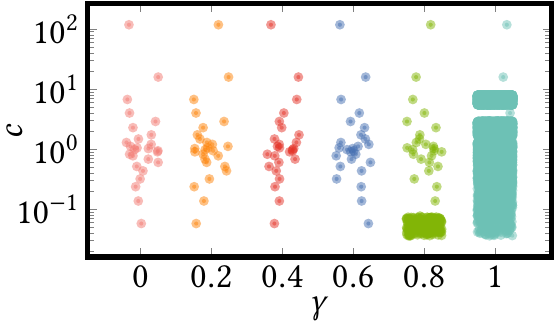}
            }
  \caption{The effect of setting $\underline{\rho}$ for \texttt{DCExact}.}
  \label{fig:dds:gamma}
  \end{figure}
  
\noindent \textbf{2. Effect of the lower bound of binary search.}
Recall that for {\tt DFlowExact}, it utilizes binary search to find the maximum density for each $c$. 
Intuitively, a tighter binary search space can speed up the algorithm, which suggests we should use $\underline{\rho}$ as the search lower bound. 
However, as shown in our experiment, we find that a higher lower bound may affect the effectiveness of the divide-and-conquer strategy.
{\tt DCExact} sets the lower bound of the binary search as $0$ for each $c$, instead of directly using $\underline{\rho}$.
%
To investigate the effect of the lower bound, we introduce a hyper-parameter $\gamma$, which controls the lower bound of binary search by setting $\underline{\rho} = \gamma \cdot \underline{\rho}$. 
We evaluate {\tt DCExact} by varying $\gamma$ from $0$ to $1$ on four datasets, and report their running time and the actual visited $c$ values in Figure \ref{fig:dds:gamma}.
We can see that a higher lower bound may increase the number of $c$ values examined and take more time.  
Although we cannot predict the best $\gamma$ in advance, smaller values of $\gamma$ usually perform better. We conjecture that this is because smaller $\gamma$ leads to more edges remaining, more feasible values of $c$ for the reduced graph, and then a larger interval of $c$ to be pruned. 
%
%
Hence, we follow \cite{ma2020efficient} to set $\underline{\rho}=0$ for each $c$ to keep the effectiveness of the divide and conquer strategy and improve the overall performance.

\noindent \textbf{3. Evaluation of graph reduction.}
{\tt DFWApp} \cite{ma2022convex} already adopts  multi-reduction to improve its efficiency.
To evaluate its usefulness, we conduct an ablation study by presenting the running time, the average numbers of vertices, and edges before and after the graph reduction for different values of $\epsilon$, across all datasets
in Table \ref{tab:dds_reduction}. For lack of space, we only present the results on two datasets here, and for other datasets, the conclusions are similar.
Clearly, the $[x,y]$-core graph reduction method can substantially reduce the size of the original graph and speed up the overall efficiency. 

Given a lower bound $\underline{\rho}$ for the optimal density, and the required interval of $c$ values ($c_l$, $c_r$) to be examined,
{\tt DFWExact} and {\tt DFWApp} do not use
$ \left[  \frac{\underline{\rho}}{2\sqrt{c_r}}, \frac{\sqrt{c_l}\underline{\rho}}{2}\right ]$-core to reduce the search space.
The main reason is that for a wide interval of $c$, (i.e., smaller $c_l$, and larger $c_r$), only a small fraction of vertices and edges will be removed due to the small values of $x$ and $y$.
To further harness the power of graph reduction, the CP-based algorithms adjust the interval length of $c$ values to enhance the effectiveness of $[x, y]$-core reduction. 
Specifically, a larger $c_l$ and a smaller $c_r$ lead to higher $x$ and $y$ values, which significantly reduce the size of the graph.
To test the effectiveness of this adjustment strategy, we evaluate the CP-based algorithms by employing and not employing this strategy, which are denoted by {\tt DFW} and {\tt DFWReN}, across different $\epsilon$ values on the two largest datasets in Table \ref{tab:dds:interval_adjstment}.
Specifically, we record the number of $c$ values to be checked (marked as ``\#c''), the average number of edges remaining after applying the $[x, y]$-core reduction (marked as `` \#edges''), the average number of iterations processed by the {\tt Frank-Wolfe} algorithm  (marked as ``\#iterations''), the product of these three items (marked as ``product''), and the running time. 
We can observe that: (1) the running time is generally proportional to the value of the ``product''. 
(2) While this strategy may increase the number of $c$ values that need to be examined, it significantly reduces the scale of the graph and reduces the value of the ``product'', indicating a better performance.
We find that this adjustment strategy is not useful for \texttt{DCExact}, since for each $c$, it sets $\underline{\rho}$ = 0 which means that it needs to calculate the maximum flow on a bigger graph.

\noindent \textbf{4. Evaluation of weight update strategies.}
We also compare the simultaneous and sequential update strategies on directed graphs in Figure \ref{fig:dds:update_strategy}.
%
Unlike the UDS problem, we discover that the simultaneous update strategy is not always more effective than the sequential one.
The main reason is that these two strategies have different effects on the divide-and-conquer strategy, and the performance gap mostly stems from the different number of $c$ values that need to be examined.

\noindent \textbf{5. Memory usage.}
In this experiment, we evaluate the memory usage of all exact algorithms. 
As shown in Figure \ref{fig:dds:exact_memory}, we can observe that \texttt{DCExact} uses less memory than \texttt{DFWExact} due to the effectiveness of its graph reduction.
For those approximation algorithms, they are around the same scale (i.e., $O(m)$), so we omit the details.

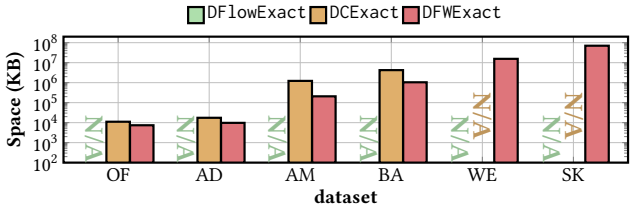
\begin{figure}[h]
    \centering
      \setlength{\abovecaptionskip}{-0.03cm}
  \setlength{\belowcaptionskip}{-0.2cm}
       	\begin{tikzpicture}[scale=0.52]
        		\begin{axis}[
                        grid = major,
        			ybar=0.11pt,
        			bar width=0.6cm,
        			width=0.9\textwidth,
    				height=0.27\textwidth,
        			xlabel={\huge \bf dataset}, 
        			xtick={2, 4, 6, 8, 10, 12, 14, 16},	
                        xticklabels={ML, MF, OF, AD, AM, BA, WE, SK},
                     legend style={at={(0.5,1.30)}, anchor=north,legend columns=-1,draw=none},
                           legend image code/.code={
            \draw [#1] (0cm,-0.263cm) rectangle (0.4cm,0.15cm); },
        			xmin=4.8,xmax=17.2,
    					ymin=100,ymax=200000050,
                    ytick = {100, 1000, 10000, 100000, 1000000, 10000000, 100000000},
                    yticklabels = {$10^2$,$10^3$, $10^4$, $10^5$, $10^6$, $10^7$, $10^8$},
                    ymode = log,    
                log origin=infty,
        			tick align=inside,
        			ticklabel style={font=\huge},
        			every axis plot/.append style={line width = 1.6pt},
        			every axis/.append style={line width = 1.6pt},
                        ylabel={\textbf{\huge Space (KB)}}
        			]	
                \addplot[fill=p1] table[x=datasets,y=DFlow]{\DDSmem};
                \addplot[fill=p2] table[x=datasets,y=DC]{\DDSmem};
                \addplot[fill=p4] table[x=datasets,y=DFW]{\DDSmem};
                \legend{\huge {\tt DFlowExact}, \huge {\tt DCExact}, \huge {\tt DFWExact}};
        		\end{axis}
                \draw[t1] (0.78,0.60) node[rotate=-90] {\large \bf N/A};
                \draw[t1] (3.12,0.60) node[rotate=-90] {\large \bf N/A};
                \draw[t1] (5.46,0.60) node[rotate=-90] {\large \bf N/A};
                \draw[t1] (7.80,0.60) node[rotate=-90] {\large \bf N/A};
                \draw[t2] (10.7,1.2) node[rotate=-90] {\large \bf N/A};
                \draw[t1] (10.14,0.60) node[rotate=-90] {\large \bf N/A};
                \draw[t2] (13.04,1.2) node[rotate=-90] {\large \bf N/A};
                \draw[t1] (12.48,0.60) node[rotate=-90] {\large \bf N/A};
        	\end{tikzpicture}
    	\caption{Memory usage on directed graphs.}
    \label{fig:dds:exact_memory}
\end{figure}

\section{Lessons and Opportunities}
\label{sec:lessons_opp}

We summarize the lessons (L) for practitioners and propose practical research opportunities (O) based on our observations.

\noindent 
\textbf{Lessons}:  

\begin{figure}[h]
	\centering
	\includegraphics[width=0.97     \linewidth]{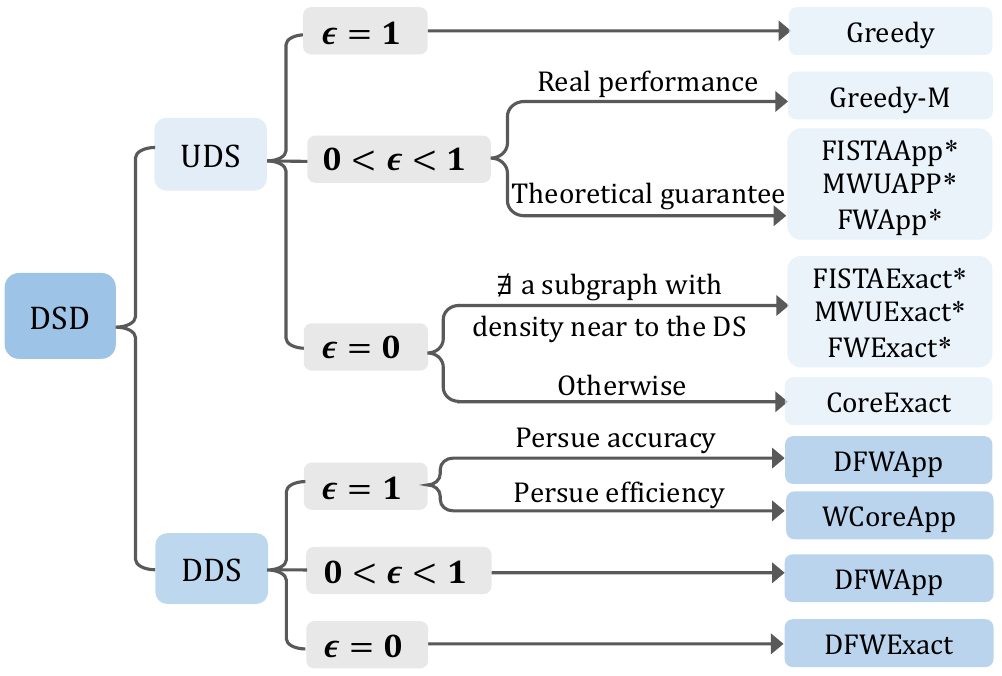}
	\caption{The taxonomy tree of DSD algorithms.}
	\label{fig:tree_DSD}
\end{figure}

\noindent\textbf{\underline{L1.}} In \Cref{fig:tree_DSD}, we depict a roadmap of the recommended DSD algorithms, highlighting which algorithms are best suited for different scenarios.  
%

\noindent\textbf{\underline{L2.}}  Graph reduction is very useful for all algorithms and using graph reduction multiple times is better than using only once.

\noindent\textbf{\underline{L3.}} 
Furthermore, no single CP-based UDS algorithm dominates other algorithms in all tests, as their performance highly depends on the structure of the graph.

\noindent\textbf{\underline{L4.}} For all iteration-based algorithms, the number of iterations required in practice is much less than the theoretical estimate.


%

\noindent\textbf{\underline{L5.}} It is not a good idea to transfer the DDS problem to the UDS problem, since it cannot utilize the divide and conquer strategy to reduce the number of $c$ values that need to be examined.

\noindent\textbf{\underline{L6.}} For the DDS problem, it is essential to reduce the number of $c$ values examined.
%
Besides, when attempting to speed up the efficiency for a specific $c$, it is important to evaluate if it affects the number of $c$ values that need to be examined subsequently.

\noindent 
\textbf{Opportunities}:  \\ 
\noindent\textbf{\underline{O1.}} For all UDS algorithms,  simultaneous weight updates tend to result in better efficiency than sequential weight updates. Providing a solid theoretical explanation of this phenomenon is an important open problem.
\eat{A theoretical explanation for this phenomenon is still missing, and providing such proof is very useful.} 

\noindent\underline{\textbf{O2.}} All existing DSD algorithms (both UDS and DDS) assume the setting of a single machine. What if the graph is too large to fit on a single machine? 
For example, the Facebook social network contains 1.32 billion nodes and 140 billion edges (\url{http://newsroom.fb.com/companyinfo}).
Can we design I/O-efficient, distributed, or sub-linear time algorithms for the DSD problem? 
Given the advantages of GPUs in executing multiple tasks concurrently due to their high-bandwidth parallel processing capabilities, designing GPU-friendly DSD algorithms \cite{ma2023accelerating} is another interesting opportunity.

\noindent\underline{\textbf{O3.}} In many domains, the network is private, and returning the DSD can reveal information about the network.
Some existing UDS algorithms \cite{dhulipala2022differential,dinitz2023improved} have considered the framework of local differential privacy. 
Designing a similar extension for DDS is important. 
\eat{However, for the DDS problem, no prior work has considered the privacy issue, so it is a very interesting direction to do it.} 
%

\noindent\underline{\textbf{O4.}} \eat{As shown in Table \ref{tab:overall_table}, the DSD problems on undirected and directed graphs have been widely studied. } 
Heterogeneous graphs are prevalent in various domains such as knowledge graphs, bibliographic networks, and biological networks.
A promising future research direction is to study DSD for  heterogeneous graphs. 


%
%

\section{Related works}
\label{sec:related}

In this section, we review the related works, including the variants of the UDS and DDS solutions and dense subgraph discovery.


$\bullet$ \textbf{Other variants of DSD.} Many variants of DSD have been studied \cite{andersen2009finding,dai2022anchored,feige1997densest,qin2015locally,xu2023efficient,ma2022finding,sawlani2020near}.
The clique densest subgraph (CDS) problem is proposed to better detect ``near-clique'' subgraphs \cite{mitzenmacher2015scalable,tsourakakis2015k,fang2019efficient,fang2022densest,he2023scaling,luo2023survey}.
Note that when $k$ = 2, it reduces to the UDS problem.
The network flow-based \cite{fang2019efficient,mitzenmacher2015scalable,tsourakakis2015k} and convex programming-based \cite{sun2020kclist++,he2023scaling} algorithms are developed to solve this problem.
The size-constrained DSD problems are studied in undirected and directed graphs~\cite{asahiro2000greedily,bhaskara2010detecting,bhaskara2012polynomial,bourgeois2013exact,nonner2016ptas,kawase2018densest,gonzales2019densest,bonchi2021finding}.
Another version of DSD called optimal quasi-clique \cite{tsourakakis2013denser,tsourakakis2014mathematical} extracts a subgraph that is more compact, with a smaller diameter than the DSD.
To identify locally dense regions, Qin et al.\cite{qin2015locally} and Ma et al. \cite{ma2022finding} studied the top-$k$ locally DS problem.
Veldt et al. \cite{veldt2021generalized} studied the $P$-mean DSD problem and proposed a generalized peeling algorithm.
To personalize results, the anchored DSD problem \cite{dai2022anchored} aims to maximize $R$-subgraph density of the subgraphs containing an anchored node set.
Besides, the DSD problems in bipartite, multilayer, and uncertain graphs were studied~\cite{andersen2010local,mitzenmacher2015scalable,hooi2016fraudar,jethava2015finding,galimberti2017core,galimberti2020core,miyauchi2018robust}.
Recently, the fair DS problem and diverse DS problems \cite{anagnostopoulos2020spectral, miyauchi2023densest} have been explored to achieve equitable outcomes and overcome algorithmic bias.

$\bullet$ \textbf{Dense subgraph discovery.} Another group of works highly related to DSD are about dense subgraph discovery.
Many cohesive subgraph models like $k$-core~\cite{batagelj2003m,seidman1983network}, $k$-truss~\cite{cohen2008trusses,saito2008extracting,zhang2012extracting}, $k$-ECC~\cite{hu2016querying,yuan2017efficient}, $k$-clique~\cite{danisch2018listing}, quasi-clique~\cite{abello2002massive}, and $k$-plex~\cite{balasundaram2011clique,zhou2021improving} have been studied. 
$k$-core is one of the most widely used dense subgraphs, in which all vertices have a degree of at least $k$.
$k$-truss is a dense subgraph based on the constraint of the number of triangles, in which each edge is contained by at least $k$-2 triangles.
$k$-ECC is a subgraph based on edge connectivity, and the edge connectivity of two vertices $u$ and $v$ is the minimum value of the number of edges that need to be removed to make $u$ and $v$ disconnected. 
$k$-clique is a complete graph of size $k$, while $k$-plex allows for up to $k$ missing connections in a clique, reflecting a quasi-clique model.
These models are often used in community search as they are cornerstone of the community~\cite{fang2020survey}.
Besides, these models are extended to bipartite graphs, such as $(\alpha,\beta)$-core~\cite{ding2017efficient,liu2020efficient}, bitruss~\cite{zou2016bitruss,wang2022towards}, biclique~\cite{lyu2020maximum}, and biplex~\cite{luo2022maximum,YuLL022}.
The directed dense subgraph models such as D-core~\cite{giatsidis2013d,fang2018effective,LiaoLJHXC22} and D-truss~\cite{liu2020truss} have also been studied.
Nevertheless, these works are different from DSD since they do not use the density definition as a key metric.

To the best of our knowledge, our work is the first study that provides a unified framework for all existing DSD algorithms and compares existing solutions comprehensively via empirical results.
%

\section{Conclusions}
\label{sec:conclusions}

In this paper, we provide an in-depth experimental evaluation and comparison of existing Densest Subgraph Discovery (DSD) algorithms. 
We first provide a unified framework, which can cover all the existing DSD algorithms, including exact and approximation algorithms, using an abstraction of a few key operations.
\eat{from a high-level perspective.}
We then thoroughly analyze and compare different DSD algorithms under our framework for both undirected and directed graphs, respectively.
We further systematically evaluate these algorithms from different angles using various datasets, and also develop variations by combining existing techniques, which often outperform state-of-the-art methods.
From extensive empirical results and analysis, we have identified several important findings and analyzed the critical components that affect the performance.
In addition, we have summarized the lessons learned and proposed practical research opportunities that can facilitate future studies.
We believe that this is a comprehensive work that thoroughly evaluates and analyzes the state-of-the-art DSD algorithms.     


\clearpage
\bibliographystyle{ACM-Reference-Format}
\bibliography{ref}


\begin{thebibliography}{91}


\ifx \showCODEN    \undefined \def \showCODEN     #1{\unskip}     \fi
\ifx \showDOI      \undefined \def \showDOI       #1{#1}\fi
\ifx \showISBNx    \undefined \def \showISBNx     #1{\unskip}     \fi
\ifx \showISBNxiii \undefined \def \showISBNxiii  #1{\unskip}     \fi
\ifx \showISSN     \undefined \def \showISSN      #1{\unskip}     \fi
\ifx \showLCCN     \undefined \def \showLCCN      #1{\unskip}     \fi
\ifx \shownote     \undefined \def \shownote      #1{#1}          \fi
\ifx \showarticletitle \undefined \def \showarticletitle #1{#1}   \fi
\ifx \showURL      \undefined \def \showURL       {\relax}        \fi
\providecommand\bibfield[2]{#2}
\providecommand\bibinfo[2]{#2}
\providecommand\natexlab[1]{#1}
\providecommand\showeprint[2][]{arXiv:#2}

\bibitem[Abello et~al\mbox{.}(2002)]%
        {abello2002massive}
\bibfield{author}{\bibinfo{person}{James Abello}, \bibinfo{person}{Mauricio~GC Resende}, {and} \bibinfo{person}{Sandra Sudarsky}.} \bibinfo{year}{2002}\natexlab{}.
\newblock \showarticletitle{Massive quasi-clique detection}. In \bibinfo{booktitle}{\emph{LATIN}}. Springer, \bibinfo{pages}{598--612}.
\newblock


\bibitem[Albert et~al\mbox{.}(1999)]%
        {albert1999diameter}
\bibfield{author}{\bibinfo{person}{R{\'e}ka Albert}, \bibinfo{person}{Hawoong Jeong}, {and} \bibinfo{person}{Albert-L{\'a}szl{\'o} Barab{\'a}si}.} \bibinfo{year}{1999}\natexlab{}.
\newblock \showarticletitle{Diameter of the world-wide web}.
\newblock \bibinfo{journal}{\emph{nature}} \bibinfo{volume}{401}, \bibinfo{number}{6749} (\bibinfo{year}{1999}), \bibinfo{pages}{130--131}.
\newblock


\bibitem[Anagnostopoulos et~al\mbox{.}(2020)]%
        {anagnostopoulos2020spectral}
\bibfield{author}{\bibinfo{person}{Aris Anagnostopoulos}, \bibinfo{person}{Luca Becchetti}, \bibinfo{person}{Adriano Fazzone}, \bibinfo{person}{Cristina Menghini}, {and} \bibinfo{person}{Chris Schwiegelshohn}.} \bibinfo{year}{2020}\natexlab{}.
\newblock \showarticletitle{Spectral relaxations and fair densest subgraphs}. In \bibinfo{booktitle}{\emph{Proceedings of the 29th ACM International Conference on Information \& Knowledge Management}}. \bibinfo{pages}{35--44}.
\newblock


\bibitem[Andersen(2010)]%
        {andersen2010local}
\bibfield{author}{\bibinfo{person}{Reid Andersen}.} \bibinfo{year}{2010}\natexlab{}.
\newblock \showarticletitle{A local algorithm for finding dense subgraphs}.
\newblock \bibinfo{journal}{\emph{ACM TALG}} \bibinfo{volume}{6}, \bibinfo{number}{4} (\bibinfo{year}{2010}), \bibinfo{pages}{1--12}.
\newblock


\bibitem[Andersen and Chellapilla(2009)]%
        {andersen2009finding}
\bibfield{author}{\bibinfo{person}{Reid Andersen} {and} \bibinfo{person}{Kumar Chellapilla}.} \bibinfo{year}{2009}\natexlab{}.
\newblock \showarticletitle{Finding dense subgraphs with size bounds}. In \bibinfo{booktitle}{\emph{{WAW}}}. Springer, \bibinfo{pages}{25--37}.
\newblock


\bibitem[Asahiro et~al\mbox{.}(2000)]%
        {asahiro2000greedily}
\bibfield{author}{\bibinfo{person}{Yuichi Asahiro}, \bibinfo{person}{Kazuo Iwama}, \bibinfo{person}{Hisao Tamaki}, {and} \bibinfo{person}{Takeshi Tokuyama}.} \bibinfo{year}{2000}\natexlab{}.
\newblock \showarticletitle{Greedily finding a dense subgraph}.
\newblock \bibinfo{journal}{\emph{Journal of Algorithms}} \bibinfo{volume}{34}, \bibinfo{number}{2} (\bibinfo{year}{2000}), \bibinfo{pages}{203--221}.
\newblock


\bibitem[Bahmani et~al\mbox{.}(2014)]%
        {bahmani2014efficient}
\bibfield{author}{\bibinfo{person}{Bahman Bahmani}, \bibinfo{person}{Ashish Goel}, {and} \bibinfo{person}{Kamesh Munagala}.} \bibinfo{year}{2014}\natexlab{}.
\newblock \showarticletitle{Efficient primal-dual graph algorithms for mapreduce}. In \bibinfo{booktitle}{\emph{WAW}}. Springer, \bibinfo{pages}{59--78}.
\newblock


\bibitem[Bahmani et~al\mbox{.}(2012)]%
        {bahmani2012densest}
\bibfield{author}{\bibinfo{person}{Bahman Bahmani}, \bibinfo{person}{Ravi Kumar}, {and} \bibinfo{person}{Sergei Vassilvitskii}.} \bibinfo{year}{2012}\natexlab{}.
\newblock \showarticletitle{Densest Subgraph in Streaming and MapReduce}.
\newblock \bibinfo{journal}{\emph{PVLDB}} \bibinfo{volume}{5}, \bibinfo{number}{5} (\bibinfo{year}{2012}).
\newblock


\bibitem[Balasundaram et~al\mbox{.}(2011)]%
        {balasundaram2011clique}
\bibfield{author}{\bibinfo{person}{Balabhaskar Balasundaram}, \bibinfo{person}{Sergiy Butenko}, {and} \bibinfo{person}{Illya~V Hicks}.} \bibinfo{year}{2011}\natexlab{}.
\newblock \showarticletitle{Clique relaxations in social network analysis: The maximum k-plex problem}.
\newblock \bibinfo{journal}{\emph{Operations Research}} \bibinfo{volume}{59}, \bibinfo{number}{1} (\bibinfo{year}{2011}), \bibinfo{pages}{133--142}.
\newblock


\bibitem[Batagelj and Zaversnik(2003)]%
        {batagelj2003m}
\bibfield{author}{\bibinfo{person}{Vladimir Batagelj} {and} \bibinfo{person}{Matjaz Zaversnik}.} \bibinfo{year}{2003}\natexlab{}.
\newblock \showarticletitle{An O (m) algorithm for cores decomposition of networks}.
\newblock \bibinfo{journal}{\emph{arXiv preprint cs/0310049}} (\bibinfo{year}{2003}).
\newblock


\bibitem[Beutel et~al\mbox{.}(2013)]%
        {beutel2013copycatch}
\bibfield{author}{\bibinfo{person}{Alex Beutel}, \bibinfo{person}{Wanhong Xu}, \bibinfo{person}{Venkatesan Guruswami}, \bibinfo{person}{Christopher Palow}, {and} \bibinfo{person}{Christos Faloutsos}.} \bibinfo{year}{2013}\natexlab{}.
\newblock \showarticletitle{Copycatch: stopping group attacks by spotting lockstep behavior in social networks}. In \bibinfo{booktitle}{\emph{WWW}}. \bibinfo{pages}{119--130}.
\newblock


\bibitem[Bhaskara et~al\mbox{.}(2010)]%
        {bhaskara2010detecting}
\bibfield{author}{\bibinfo{person}{Aditya Bhaskara}, \bibinfo{person}{Moses Charikar}, \bibinfo{person}{Eden Chlamtac}, \bibinfo{person}{Uriel Feige}, {and} \bibinfo{person}{Aravindan Vijayaraghavan}.} \bibinfo{year}{2010}\natexlab{}.
\newblock \showarticletitle{Detecting high log-densities: an O (n $1/4$) approximation for densest k-subgraph}. In \bibinfo{booktitle}{\emph{STOC}}. \bibinfo{pages}{201--210}.
\newblock


\bibitem[Bhaskara et~al\mbox{.}(2012)]%
        {bhaskara2012polynomial}
\bibfield{author}{\bibinfo{person}{Aditya Bhaskara}, \bibinfo{person}{Moses Charikar}, \bibinfo{person}{Venkatesan Guruswami}, \bibinfo{person}{Aravindan Vijayaraghavan}, {and} \bibinfo{person}{Yuan Zhou}.} \bibinfo{year}{2012}\natexlab{}.
\newblock \showarticletitle{Polynomial integrality gaps for strong sdp relaxations of densest k-subgraph}. In \bibinfo{booktitle}{\emph{{SODA}}}. SIAM, \bibinfo{pages}{388--405}.
\newblock


\bibitem[Bonchi et~al\mbox{.}(2021)]%
        {bonchi2021finding}
\bibfield{author}{\bibinfo{person}{Francesco Bonchi}, \bibinfo{person}{David Garc{\'\i}a-Soriano}, \bibinfo{person}{Atsushi Miyauchi}, {and} \bibinfo{person}{Charalampos~E Tsourakakis}.} \bibinfo{year}{2021}\natexlab{}.
\newblock \showarticletitle{Finding densest k-connected subgraphs}.
\newblock \bibinfo{journal}{\emph{Discrete Applied Mathematics}}  \bibinfo{volume}{305} (\bibinfo{year}{2021}), \bibinfo{pages}{34--47}.
\newblock


\bibitem[Boob et~al\mbox{.}(2020)]%
        {boob2020flowless}
\bibfield{author}{\bibinfo{person}{Digvijay Boob}, \bibinfo{person}{Yu Gao}, \bibinfo{person}{Richard Peng}, \bibinfo{person}{Saurabh Sawlani}, \bibinfo{person}{Charalampos Tsourakakis}, \bibinfo{person}{Di Wang}, {and} \bibinfo{person}{Junxing Wang}.} \bibinfo{year}{2020}\natexlab{}.
\newblock \showarticletitle{Flowless: Extracting densest subgraphs without flow computations}. In \bibinfo{booktitle}{\emph{{WWW}}}.
\newblock


\bibitem[Bourgeois et~al\mbox{.}(2013)]%
        {bourgeois2013exact}
\bibfield{author}{\bibinfo{person}{Nicolas Bourgeois}, \bibinfo{person}{Aristotelis Giannakos}, \bibinfo{person}{Giorgio Lucarelli}, \bibinfo{person}{Ioannis Milis}, {and} \bibinfo{person}{Vangelis~Th Paschos}.} \bibinfo{year}{2013}\natexlab{}.
\newblock \showarticletitle{Exact and approximation algorithms for densest k-subgraph}. In \bibinfo{booktitle}{\emph{{WALCOM}}}. Springer, \bibinfo{pages}{114--125}.
\newblock


\bibitem[Charikar(2000)]%
        {charikar2000greedy}
\bibfield{author}{\bibinfo{person}{Moses Charikar}.} \bibinfo{year}{2000}\natexlab{}.
\newblock \showarticletitle{Greedy approximation algorithms for finding dense components in a graph}. In \bibinfo{booktitle}{\emph{{APPROX}}}. Springer, \bibinfo{pages}{84--95}.
\newblock


\bibitem[Chekuri et~al\mbox{.}(2022)]%
        {chekuri2022densest}
\bibfield{author}{\bibinfo{person}{Chandra Chekuri}, \bibinfo{person}{Kent Quanrud}, {and} \bibinfo{person}{Manuel~R Torres}.} \bibinfo{year}{2022}\natexlab{}.
\newblock \showarticletitle{Densest Subgraph: Supermodularity, Iterative Peeling, and Flow}. In \bibinfo{booktitle}{\emph{SODA}}. SIAM, \bibinfo{pages}{1531--1555}.
\newblock


\bibitem[Chen and Saad(2010)]%
        {chen2010dense}
\bibfield{author}{\bibinfo{person}{Jie Chen} {and} \bibinfo{person}{Yousef Saad}.} \bibinfo{year}{2010}\natexlab{}.
\newblock \showarticletitle{Dense subgraph extraction with application to community detection}.
\newblock \bibinfo{journal}{\emph{TKDE}} \bibinfo{volume}{24}, \bibinfo{number}{7} (\bibinfo{year}{2010}), \bibinfo{pages}{1216--1230}.
\newblock


\bibitem[Ching et~al\mbox{.}(2015)]%
        {ching2015one}
\bibfield{author}{\bibinfo{person}{Avery Ching}, \bibinfo{person}{Sergey Edunov}, \bibinfo{person}{Maja Kabiljo}, \bibinfo{person}{Dionysios Logothetis}, {and} \bibinfo{person}{Sambavi Muthukrishnan}.} \bibinfo{year}{2015}\natexlab{}.
\newblock \showarticletitle{One trillion edges: Graph processing at facebook-scale}.
\newblock \bibinfo{journal}{\emph{PVLDB}} \bibinfo{volume}{8}, \bibinfo{number}{12} (\bibinfo{year}{2015}), \bibinfo{pages}{1804--1815}.
\newblock


\bibitem[Cohen et~al\mbox{.}(2003)]%
        {cohen2003reachability}
\bibfield{author}{\bibinfo{person}{Edith Cohen}, \bibinfo{person}{Eran Halperin}, \bibinfo{person}{Haim Kaplan}, {and} \bibinfo{person}{Uri Zwick}.} \bibinfo{year}{2003}\natexlab{}.
\newblock \showarticletitle{Reachability and distance queries via 2-hop labels}.
\newblock \bibinfo{journal}{\emph{{SIAM} J. Comput.}} \bibinfo{volume}{32}, \bibinfo{number}{5} (\bibinfo{year}{2003}), \bibinfo{pages}{1338--1355}.
\newblock


\bibitem[Cohen(2008)]%
        {cohen2008trusses}
\bibfield{author}{\bibinfo{person}{Jonathan Cohen}.} \bibinfo{year}{2008}\natexlab{}.
\newblock \showarticletitle{Trusses: Cohesive subgraphs for social network analysis}.
\newblock \bibinfo{journal}{\emph{National security agency technical report}} \bibinfo{volume}{16}, \bibinfo{number}{3.1} (\bibinfo{year}{2008}).
\newblock


\bibitem[Dai et~al\mbox{.}(2022)]%
        {dai2022anchored}
\bibfield{author}{\bibinfo{person}{Yizhou Dai}, \bibinfo{person}{Miao Qiao}, {and} \bibinfo{person}{Lijun Chang}.} \bibinfo{year}{2022}\natexlab{}.
\newblock \showarticletitle{Anchored Densest Subgraph}. In \bibinfo{booktitle}{\emph{SIGMOD}}. \bibinfo{pages}{1200--1213}.
\newblock


\bibitem[Danisch et~al\mbox{.}(2018)]%
        {danisch2018listing}
\bibfield{author}{\bibinfo{person}{Maximilien Danisch}, \bibinfo{person}{Oana Balalau}, {and} \bibinfo{person}{Mauro Sozio}.} \bibinfo{year}{2018}\natexlab{}.
\newblock \showarticletitle{Listing k-cliques in sparse real-world graphs}. In \bibinfo{booktitle}{\emph{WWW}}. \bibinfo{pages}{589--598}.
\newblock


\bibitem[Danisch et~al\mbox{.}(2017)]%
        {danisch2017large}
\bibfield{author}{\bibinfo{person}{Maximilien Danisch}, \bibinfo{person}{T-H~Hubert Chan}, {and} \bibinfo{person}{Mauro Sozio}.} \bibinfo{year}{2017}\natexlab{}.
\newblock \showarticletitle{Large scale density-friendly graph decomposition via convex programming}. In \bibinfo{booktitle}{\emph{{WWW}}}. \bibinfo{pages}{233--242}.
\newblock


\bibitem[Dhulipala et~al\mbox{.}(2022)]%
        {dhulipala2022differential}
\bibfield{author}{\bibinfo{person}{Laxman Dhulipala}, \bibinfo{person}{Quanquan~C Liu}, \bibinfo{person}{Sofya Raskhodnikova}, \bibinfo{person}{Jessica Shi}, \bibinfo{person}{Julian Shun}, {and} \bibinfo{person}{Shangdi Yu}.} \bibinfo{year}{2022}\natexlab{}.
\newblock \showarticletitle{Differential privacy from locally adjustable graph algorithms: k-core decomposition, low out-degree ordering, and densest subgraphs}. In \bibinfo{booktitle}{\emph{2022 IEEE 63rd Annual Symposium on Foundations of Computer Science (FOCS)}}. IEEE, \bibinfo{pages}{754--765}.
\newblock


\bibitem[Ding et~al\mbox{.}(2017)]%
        {ding2017efficient}
\bibfield{author}{\bibinfo{person}{Danhao Ding}, \bibinfo{person}{Hui Li}, \bibinfo{person}{Zhipeng Huang}, {and} \bibinfo{person}{Nikos Mamoulis}.} \bibinfo{year}{2017}\natexlab{}.
\newblock \showarticletitle{Efficient fault-tolerant group recommendation using alpha-beta-core}. In \bibinfo{booktitle}{\emph{CIKM}}. \bibinfo{pages}{2047--2050}.
\newblock


\bibitem[Dinitz et~al\mbox{.}(2023)]%
        {dinitz2023improved}
\bibfield{author}{\bibinfo{person}{Michael Dinitz}, \bibinfo{person}{Satyen Kale}, \bibinfo{person}{Silvio Lattanzi}, {and} \bibinfo{person}{Sergei Vassilvitskii}.} \bibinfo{year}{2023}\natexlab{}.
\newblock \showarticletitle{Improved Differentially Private Densest Subgraph: Local and Purely Additive}.
\newblock \bibinfo{journal}{\emph{arXiv preprint arXiv:2308.10316}} (\bibinfo{year}{2023}).
\newblock


\bibitem[Fang et~al\mbox{.}(2017)]%
        {fang2017effective}
\bibfield{author}{\bibinfo{person}{Yixiang Fang}, \bibinfo{person}{Reynold Cheng}, \bibinfo{person}{Xiaodong Li}, \bibinfo{person}{Siqiang Luo}, {and} \bibinfo{person}{Jiafeng Hu}.} \bibinfo{year}{2017}\natexlab{}.
\newblock \showarticletitle{Effective community search over large spatial graphs}.
\newblock \bibinfo{journal}{\emph{PVLDB}} \bibinfo{volume}{10}, \bibinfo{number}{6} (\bibinfo{year}{2017}), \bibinfo{pages}{709--720}.
\newblock


\bibitem[Fang et~al\mbox{.}(2020)]%
        {fang2020survey}
\bibfield{author}{\bibinfo{person}{Yixiang Fang}, \bibinfo{person}{Xin Huang}, \bibinfo{person}{Lu Qin}, \bibinfo{person}{Ying Zhang}, \bibinfo{person}{Wenjie Zhang}, \bibinfo{person}{Reynold Cheng}, {and} \bibinfo{person}{Xuemin Lin}.} \bibinfo{year}{2020}\natexlab{}.
\newblock \showarticletitle{A survey of community search over big graphs}.
\newblock \bibinfo{journal}{\emph{VLDBJ}} \bibinfo{volume}{29}, \bibinfo{number}{1} (\bibinfo{year}{2020}), \bibinfo{pages}{353--392}.
\newblock


\bibitem[Fang et~al\mbox{.}(2022)]%
        {fang2022densest}
\bibfield{author}{\bibinfo{person}{Yixiang Fang}, \bibinfo{person}{Wensheng Luo}, {and} \bibinfo{person}{Chenhao Ma}.} \bibinfo{year}{2022}\natexlab{}.
\newblock \showarticletitle{Densest subgraph discovery on large graphs: Applications, challenges, and techniques}.
\newblock \bibinfo{journal}{\emph{Proceedings of the VLDB Endowment}} \bibinfo{volume}{15}, \bibinfo{number}{12} (\bibinfo{year}{2022}), \bibinfo{pages}{3766--3769}.
\newblock


\bibitem[Fang et~al\mbox{.}(2018)]%
        {fang2018effective}
\bibfield{author}{\bibinfo{person}{Yixiang Fang}, \bibinfo{person}{Zhongran Wang}, \bibinfo{person}{Reynold Cheng}, \bibinfo{person}{Hongzhi Wang}, {and} \bibinfo{person}{Jiafeng Hu}.} \bibinfo{year}{2018}\natexlab{}.
\newblock \showarticletitle{Effective and efficient community search over large directed graphs}.
\newblock \bibinfo{journal}{\emph{TKDE}} \bibinfo{volume}{31}, \bibinfo{number}{11} (\bibinfo{year}{2018}), \bibinfo{pages}{2093--2107}.
\newblock


\bibitem[Fang et~al\mbox{.}(2019)]%
        {fang2019efficient}
\bibfield{author}{\bibinfo{person}{Yixiang Fang}, \bibinfo{person}{Kaiqiang Yu}, \bibinfo{person}{Reynold Cheng}, \bibinfo{person}{Laks~VS Lakshmanan}, {and} \bibinfo{person}{Xuemin Lin}.} \bibinfo{year}{2019}\natexlab{}.
\newblock \showarticletitle{Efficient algorithms for densest subgraph discovery}.
\newblock \bibinfo{journal}{\emph{PVLDB}} \bibinfo{volume}{12}, \bibinfo{number}{11} (\bibinfo{year}{2019}), \bibinfo{pages}{1719--1732}.
\newblock


\bibitem[Feige et~al\mbox{.}(1997)]%
        {feige1997densest}
\bibfield{author}{\bibinfo{person}{Uriel Feige}, \bibinfo{person}{Michael Seltser}, {et~al\mbox{.}}} \bibinfo{year}{1997}\natexlab{}.
\newblock \bibinfo{booktitle}{\emph{On the densest k-subgraph problem}}.
\newblock \bibinfo{publisher}{Citeseer}.
\newblock


\bibitem[Fratkin et~al\mbox{.}(2006)]%
        {fratkin2006motifcut}
\bibfield{author}{\bibinfo{person}{Eugene Fratkin}, \bibinfo{person}{Brian~T Naughton}, \bibinfo{person}{Douglas~L Brutlag}, {and} \bibinfo{person}{Serafim Batzoglou}.} \bibinfo{year}{2006}\natexlab{}.
\newblock \showarticletitle{MotifCut: regulatory motifs finding with maximum density subgraphs}.
\newblock \bibinfo{journal}{\emph{Bioinformatics}} \bibinfo{volume}{22}, \bibinfo{number}{14} (\bibinfo{year}{2006}), \bibinfo{pages}{e150--e157}.
\newblock


\bibitem[Galimberti et~al\mbox{.}(2017)]%
        {galimberti2017core}
\bibfield{author}{\bibinfo{person}{Edoardo Galimberti}, \bibinfo{person}{Francesco Bonchi}, {and} \bibinfo{person}{Francesco Gullo}.} \bibinfo{year}{2017}\natexlab{}.
\newblock \showarticletitle{Core decomposition and densest subgraph in multilayer networks}. In \bibinfo{booktitle}{\emph{{CIKM}}}. \bibinfo{pages}{1807--1816}.
\newblock


\bibitem[Galimberti et~al\mbox{.}(2020)]%
        {galimberti2020core}
\bibfield{author}{\bibinfo{person}{Edoardo Galimberti}, \bibinfo{person}{Francesco Bonchi}, \bibinfo{person}{Francesco Gullo}, {and} \bibinfo{person}{Tommaso Lanciano}.} \bibinfo{year}{2020}\natexlab{}.
\newblock \showarticletitle{Core decomposition in multilayer networks: theory, algorithms, and applications}.
\newblock \bibinfo{journal}{\emph{TKDD}} \bibinfo{volume}{14}, \bibinfo{number}{1} (\bibinfo{year}{2020}), \bibinfo{pages}{1--40}.
\newblock


\bibitem[Giatsidis et~al\mbox{.}(2013)]%
        {giatsidis2013d}
\bibfield{author}{\bibinfo{person}{Christos Giatsidis}, \bibinfo{person}{Dimitrios~M Thilikos}, {and} \bibinfo{person}{Michalis Vazirgiannis}.} \bibinfo{year}{2013}\natexlab{}.
\newblock \showarticletitle{D-cores: measuring collaboration of directed graphs based on degeneracy}.
\newblock \bibinfo{journal}{\emph{KAIS}} \bibinfo{volume}{35}, \bibinfo{number}{2} (\bibinfo{year}{2013}), \bibinfo{pages}{311--343}.
\newblock


\bibitem[Gionis and Tsourakakis(2015)]%
        {gionis2015dense}
\bibfield{author}{\bibinfo{person}{Aristides Gionis} {and} \bibinfo{person}{Charalampos~E Tsourakakis}.} \bibinfo{year}{2015}\natexlab{}.
\newblock \showarticletitle{Dense subgraph discovery: Kdd 2015 tutorial}. In \bibinfo{booktitle}{\emph{SIGKDD}}. \bibinfo{pages}{2313--2314}.
\newblock


\bibitem[Goldberg(1984)]%
        {goldberg1984finding}
\bibfield{author}{\bibinfo{person}{Andrew~V Goldberg}.} \bibinfo{year}{1984}\natexlab{}.
\newblock \bibinfo{booktitle}{\emph{Finding a maximum density subgraph}}.
\newblock \bibinfo{publisher}{University of California Berkeley}.
\newblock


\bibitem[Gonzales and Migler(2019)]%
        {gonzales2019densest}
\bibfield{author}{\bibinfo{person}{Sean Gonzales} {and} \bibinfo{person}{Theresa Migler}.} \bibinfo{year}{2019}\natexlab{}.
\newblock \showarticletitle{The Densest k Subgraph Problem in b-Outerplanar Graphs}. In \bibinfo{booktitle}{\emph{{COMPLEX} {NETWORKS}}}. Springer, \bibinfo{pages}{116--127}.
\newblock


\bibitem[Harb et~al\mbox{.}(2022)]%
        {harb2022faster}
\bibfield{author}{\bibinfo{person}{Elfarouk Harb}, \bibinfo{person}{Kent Quanrud}, {and} \bibinfo{person}{Chandra Chekuri}.} \bibinfo{year}{2022}\natexlab{}.
\newblock \showarticletitle{Faster and Scalable Algorithms for Densest Subgraph and Decomposition}. In \bibinfo{booktitle}{\emph{NIPS}}.
\newblock


\bibitem[He et~al\mbox{.}(2023)]%
        {he2023scaling}
\bibfield{author}{\bibinfo{person}{Yizhang He}, \bibinfo{person}{Kai Wang}, \bibinfo{person}{Wenjie Zhang}, \bibinfo{person}{Xuemin Lin}, {and} \bibinfo{person}{Ying Zhang}.} \bibinfo{year}{2023}\natexlab{}.
\newblock \showarticletitle{Scaling Up k-Clique Densest Subgraph Detection}.
\newblock \bibinfo{journal}{\emph{Proceedings of the ACM on Management of Data}} \bibinfo{volume}{1}, \bibinfo{number}{1} (\bibinfo{year}{2023}), \bibinfo{pages}{1--26}.
\newblock


\bibitem[Hooi et~al\mbox{.}(2016)]%
        {hooi2016fraudar}
\bibfield{author}{\bibinfo{person}{Bryan Hooi}, \bibinfo{person}{Hyun~Ah Song}, \bibinfo{person}{Alex Beutel}, \bibinfo{person}{Neil Shah}, \bibinfo{person}{Kijung Shin}, {and} \bibinfo{person}{Christos Faloutsos}.} \bibinfo{year}{2016}\natexlab{}.
\newblock \showarticletitle{Fraudar: Bounding graph fraud in the face of camouflage}. In \bibinfo{booktitle}{\emph{SIGKDD}}. \bibinfo{pages}{895--904}.
\newblock


\bibitem[Hu et~al\mbox{.}(2016)]%
        {hu2016querying}
\bibfield{author}{\bibinfo{person}{Jiafeng Hu}, \bibinfo{person}{Xiaowei Wu}, \bibinfo{person}{Reynold Cheng}, \bibinfo{person}{Siqiang Luo}, {and} \bibinfo{person}{Yixiang Fang}.} \bibinfo{year}{2016}\natexlab{}.
\newblock \showarticletitle{Querying minimal steiner maximum-connected subgraphs in large graphs}. In \bibinfo{booktitle}{\emph{CIKM}}. \bibinfo{pages}{1241--1250}.
\newblock


\bibitem[Jaggi(2013)]%
        {jaggi2013revisiting}
\bibfield{author}{\bibinfo{person}{Martin Jaggi}.} \bibinfo{year}{2013}\natexlab{}.
\newblock \showarticletitle{Revisiting Frank-Wolfe: Projection-free sparse convex optimization}. In \bibinfo{booktitle}{\emph{ICML}}. PMLR, \bibinfo{pages}{427--435}.
\newblock


\bibitem[Java et~al\mbox{.}(2007)]%
        {java2007we}
\bibfield{author}{\bibinfo{person}{Akshay Java}, \bibinfo{person}{Xiaodan Song}, \bibinfo{person}{Tim Finin}, {and} \bibinfo{person}{Belle Tseng}.} \bibinfo{year}{2007}\natexlab{}.
\newblock \showarticletitle{Why we twitter: understanding microblogging usage and communities}. In \bibinfo{booktitle}{\emph{WebKDD/SNA-KDD}}. \bibinfo{pages}{56--65}.
\newblock


\bibitem[Jethava and Beerenwinkel(2015)]%
        {jethava2015finding}
\bibfield{author}{\bibinfo{person}{Vinay Jethava} {and} \bibinfo{person}{Niko Beerenwinkel}.} \bibinfo{year}{2015}\natexlab{}.
\newblock \showarticletitle{Finding dense subgraphs in relational graphs}. In \bibinfo{booktitle}{\emph{{ECML} {PKDD}}}. Springer, \bibinfo{pages}{641--654}.
\newblock


\bibitem[Kannan and Vinay(1999)]%
        {kannan1999analyzing}
\bibfield{author}{\bibinfo{person}{Ravindran Kannan} {and} \bibinfo{person}{V Vinay}.} \bibinfo{year}{1999}\natexlab{}.
\newblock \bibinfo{booktitle}{\emph{Analyzing the structure of large graphs}}.
\newblock \bibinfo{publisher}{Forschungsinst. f{\"u}r Diskrete Mathematik}.
\newblock


\bibitem[Karlebach and Shamir(2008)]%
        {karlebach2008modelling}
\bibfield{author}{\bibinfo{person}{Guy Karlebach} {and} \bibinfo{person}{Ron Shamir}.} \bibinfo{year}{2008}\natexlab{}.
\newblock \showarticletitle{Modelling and analysis of gene regulatory networks}.
\newblock \bibinfo{journal}{\emph{Nature reviews Molecular cell biology}} \bibinfo{volume}{9}, \bibinfo{number}{10} (\bibinfo{year}{2008}), \bibinfo{pages}{770--780}.
\newblock


\bibitem[Kawase and Miyauchi(2018)]%
        {kawase2018densest}
\bibfield{author}{\bibinfo{person}{Yasushi Kawase} {and} \bibinfo{person}{Atsushi Miyauchi}.} \bibinfo{year}{2018}\natexlab{}.
\newblock \showarticletitle{The densest subgraph problem with a convex/concave size function}.
\newblock \bibinfo{journal}{\emph{Algorithmica}} \bibinfo{volume}{80}, \bibinfo{number}{12} (\bibinfo{year}{2018}), \bibinfo{pages}{3461--3480}.
\newblock


\bibitem[Khuller and Saha(2009)]%
        {khuller2009finding}
\bibfield{author}{\bibinfo{person}{Samir Khuller} {and} \bibinfo{person}{Barna Saha}.} \bibinfo{year}{2009}\natexlab{}.
\newblock \showarticletitle{On finding dense subgraphs}. In \bibinfo{booktitle}{\emph{{ICALP}}}. Springer, \bibinfo{pages}{597--608}.
\newblock


\bibitem[Lakshmanan(2022)]%
        {lakshmanan2022quest}
\bibfield{author}{\bibinfo{person}{Laks~VS Lakshmanan}.} \bibinfo{year}{2022}\natexlab{}.
\newblock \showarticletitle{On a Quest for Combating Filter Bubbles and Misinformation}. In \bibinfo{booktitle}{\emph{SIGMOD}}. \bibinfo{pages}{2--2}.
\newblock


\bibitem[Lanciano et~al\mbox{.}(2023)]%
        {lanciano2023survey}
\bibfield{author}{\bibinfo{person}{Tommaso Lanciano}, \bibinfo{person}{Atsushi Miyauchi}, \bibinfo{person}{Adriano Fazzone}, {and} \bibinfo{person}{Francesco Bonchi}.} \bibinfo{year}{2023}\natexlab{}.
\newblock \showarticletitle{A Survey on the Densest Subgraph Problem and its Variants}.
\newblock \bibinfo{journal}{\emph{arXiv preprint arXiv:2303.14467}} (\bibinfo{year}{2023}).
\newblock


\bibitem[Liao et~al\mbox{.}(2022)]%
        {LiaoLJHXC22}
\bibfield{author}{\bibinfo{person}{Xuankun Liao}, \bibinfo{person}{Qing Liu}, \bibinfo{person}{Jiaxin Jiang}, \bibinfo{person}{Xin Huang}, \bibinfo{person}{Jianliang Xu}, {and} \bibinfo{person}{Byron Choi}.} \bibinfo{year}{2022}\natexlab{}.
\newblock \showarticletitle{Distributed D-core Decomposition over Large Directed Graphs}.
\newblock \bibinfo{journal}{\emph{PVLDB}} \bibinfo{volume}{15}, \bibinfo{number}{8} (\bibinfo{year}{2022}), \bibinfo{pages}{1546--1558}.
\newblock


\bibitem[Liu et~al\mbox{.}(2020a)]%
        {liu2020efficient}
\bibfield{author}{\bibinfo{person}{Boge Liu}, \bibinfo{person}{Long Yuan}, \bibinfo{person}{Xuemin Lin}, \bibinfo{person}{Lu Qin}, \bibinfo{person}{Wenjie Zhang}, {and} \bibinfo{person}{Jingren Zhou}.} \bibinfo{year}{2020}\natexlab{a}.
\newblock \showarticletitle{Efficient ({\(\alpha\)}, {\(\beta\)})-core computation in bipartite graphs}.
\newblock \bibinfo{journal}{\emph{The VLDB Journal}} \bibinfo{volume}{29}, \bibinfo{number}{5} (\bibinfo{year}{2020}), \bibinfo{pages}{1075--1099}.
\newblock


\bibitem[Liu et~al\mbox{.}(2020b)]%
        {liu2020truss}
\bibfield{author}{\bibinfo{person}{Qing Liu}, \bibinfo{person}{Minjun Zhao}, \bibinfo{person}{Xin Huang}, \bibinfo{person}{Jianliang Xu}, {and} \bibinfo{person}{Yunjun Gao}.} \bibinfo{year}{2020}\natexlab{b}.
\newblock \showarticletitle{Truss-based community search over large directed graphs}. In \bibinfo{booktitle}{\emph{SIGMOD}}. \bibinfo{pages}{2183--2197}.
\newblock


\bibitem[Luo et~al\mbox{.}(2022)]%
        {luo2022maximum}
\bibfield{author}{\bibinfo{person}{Wensheng Luo}, \bibinfo{person}{Kenli Li}, \bibinfo{person}{Xu Zhou}, \bibinfo{person}{Yunjun Gao}, {and} \bibinfo{person}{Keqin Li}.} \bibinfo{year}{2022}\natexlab{}.
\newblock \showarticletitle{Maximum Biplex Search over Bipartite Graphs}. In \bibinfo{booktitle}{\emph{ICDE}}. IEEE, \bibinfo{pages}{898--910}.
\newblock


\bibitem[Luo et~al\mbox{.}(2023a)]%
        {luo2023survey}
\bibfield{author}{\bibinfo{person}{Wensheng Luo}, \bibinfo{person}{Chenhao Ma}, \bibinfo{person}{Yixiang Fang}, {and} \bibinfo{person}{Laks~VS Lakshman}.} \bibinfo{year}{2023}\natexlab{a}.
\newblock \showarticletitle{A Survey of Densest Subgraph Discovery on Large Graphs}.
\newblock \bibinfo{journal}{\emph{arXiv preprint arXiv:2306.07927}} (\bibinfo{year}{2023}).
\newblock


\bibitem[Luo et~al\mbox{.}(2023b)]%
        {luo2023scalable}
\bibfield{author}{\bibinfo{person}{Wensheng Luo}, \bibinfo{person}{Zhuo Tang}, \bibinfo{person}{Yixiang Fang}, \bibinfo{person}{Chenhao Ma}, {and} \bibinfo{person}{Xu Zhou}.} \bibinfo{year}{2023}\natexlab{b}.
\newblock \showarticletitle{Scalable Algorithms for Densest Subgraph Discovery}. In \bibinfo{booktitle}{\emph{ICDE}}. IEEE.
\newblock


\bibitem[Lyu et~al\mbox{.}(2020)]%
        {lyu2020maximum}
\bibfield{author}{\bibinfo{person}{Bingqing Lyu}, \bibinfo{person}{Lu Qin}, \bibinfo{person}{Xuemin Lin}, \bibinfo{person}{Ying Zhang}, \bibinfo{person}{Zhengping Qian}, {and} \bibinfo{person}{Jingren Zhou}.} \bibinfo{year}{2020}\natexlab{}.
\newblock \showarticletitle{Maximum biclique search at billion scale}.
\newblock \bibinfo{journal}{\emph{PVLDB}} \bibinfo{volume}{13}, \bibinfo{number}{9} (\bibinfo{year}{2020}), \bibinfo{pages}{1359--1372}.
\newblock


\bibitem[Ma et~al\mbox{.}(2019)]%
        {ma2019linc}
\bibfield{author}{\bibinfo{person}{Chenhao Ma}, \bibinfo{person}{Reynold Cheng}, \bibinfo{person}{Laks~VS Lakshmanan}, \bibinfo{person}{Tobias Grubenmann}, \bibinfo{person}{Yixiang Fang}, {and} \bibinfo{person}{Xiaodong Li}.} \bibinfo{year}{2019}\natexlab{}.
\newblock \showarticletitle{Linc: a motif counting algorithm for uncertain graphs}.
\newblock \bibinfo{journal}{\emph{PVLDB}} \bibinfo{volume}{13}, \bibinfo{number}{2} (\bibinfo{year}{2019}), \bibinfo{pages}{155--168}.
\newblock


\bibitem[Ma et~al\mbox{.}(2022a)]%
        {ma2022finding}
\bibfield{author}{\bibinfo{person}{Chenhao Ma}, \bibinfo{person}{Reynold Cheng}, \bibinfo{person}{Laks~VS Lakshmanan}, {and} \bibinfo{person}{Xiaolin Han}.} \bibinfo{year}{2022}\natexlab{a}.
\newblock \showarticletitle{Finding locally densest subgraphs: a convex programming approach}.
\newblock \bibinfo{journal}{\emph{PVLDB}} \bibinfo{volume}{15}, \bibinfo{number}{11} (\bibinfo{year}{2022}), \bibinfo{pages}{2719--2732}.
\newblock


\bibitem[Ma et~al\mbox{.}(2022b)]%
        {ma2022convex}
\bibfield{author}{\bibinfo{person}{Chenhao Ma}, \bibinfo{person}{Yixiang Fang}, \bibinfo{person}{Reynold Cheng}, \bibinfo{person}{Laks~VS Lakshmanan}, {and} \bibinfo{person}{Xiaolin Han}.} \bibinfo{year}{2022}\natexlab{b}.
\newblock \showarticletitle{A Convex-Programming Approach for Efficient Directed Densest Subgraph Discovery}. In \bibinfo{booktitle}{\emph{SIGMOD}}. \bibinfo{pages}{845--859}.
\newblock


\bibitem[Ma et~al\mbox{.}(2023)]%
        {ma2023accelerating}
\bibfield{author}{\bibinfo{person}{Chenhao Ma}, \bibinfo{person}{Yixiang Fang}, \bibinfo{person}{Reynold Cheng}, \bibinfo{person}{Laks~VS Lakshmanan}, \bibinfo{person}{Xiaolin Han}, {and} \bibinfo{person}{Xiaodong Li}.} \bibinfo{year}{2023}\natexlab{}.
\newblock \showarticletitle{Accelerating directed densest subgraph queries with software and hardware approaches}.
\newblock \bibinfo{journal}{\emph{The VLDB Journal}} (\bibinfo{year}{2023}), \bibinfo{pages}{1--24}.
\newblock


\bibitem[Ma et~al\mbox{.}(2020)]%
        {ma2020efficient}
\bibfield{author}{\bibinfo{person}{Chenhao Ma}, \bibinfo{person}{Yixiang Fang}, \bibinfo{person}{Reynold Cheng}, \bibinfo{person}{Laks~VS Lakshmanan}, \bibinfo{person}{Wenjie Zhang}, {and} \bibinfo{person}{Xuemin Lin}.} \bibinfo{year}{2020}\natexlab{}.
\newblock \showarticletitle{Efficient algorithms for densest subgraph discovery on large directed graphs}. In \bibinfo{booktitle}{\emph{SIGMOD}}. \bibinfo{pages}{1051--1066}.
\newblock


\bibitem[Ma et~al\mbox{.}(2021a)]%
        {ma2021efficient}
\bibfield{author}{\bibinfo{person}{Chenhao Ma}, \bibinfo{person}{Yixiang Fang}, \bibinfo{person}{Reynold Cheng}, \bibinfo{person}{Laks~VS Lakshmanan}, \bibinfo{person}{Wenjie Zhang}, {and} \bibinfo{person}{Xuemin Lin}.} \bibinfo{year}{2021}\natexlab{a}.
\newblock \showarticletitle{Efficient Directed Densest Subgraph Discovery}.
\newblock \bibinfo{journal}{\emph{ACM SIGMOD Record}} \bibinfo{volume}{50}, \bibinfo{number}{1} (\bibinfo{year}{2021}), \bibinfo{pages}{33--40}.
\newblock


\bibitem[Ma et~al\mbox{.}(2021b)]%
        {ma2021directed}
\bibfield{author}{\bibinfo{person}{Chenhao Ma}, \bibinfo{person}{Yixiang Fang}, \bibinfo{person}{Reynold Cheng}, \bibinfo{person}{Laks~VS Lakshmanan}, \bibinfo{person}{Wenjie Zhang}, {and} \bibinfo{person}{Xuemin Lin}.} \bibinfo{year}{2021}\natexlab{b}.
\newblock \showarticletitle{On Directed Densest Subgraph Discovery}.
\newblock \bibinfo{journal}{\emph{TODS}} \bibinfo{volume}{46}, \bibinfo{number}{4} (\bibinfo{year}{2021}), \bibinfo{pages}{1--45}.
\newblock


\bibitem[Mitzenmacher et~al\mbox{.}(2015)]%
        {mitzenmacher2015scalable}
\bibfield{author}{\bibinfo{person}{Michael Mitzenmacher}, \bibinfo{person}{Jakub Pachocki}, \bibinfo{person}{Richard Peng}, \bibinfo{person}{Charalampos Tsourakakis}, {and} \bibinfo{person}{Shen~Chen Xu}.} \bibinfo{year}{2015}\natexlab{}.
\newblock \showarticletitle{Scalable large near-clique detection in large-scale networks via sampling}. In \bibinfo{booktitle}{\emph{{SIGKDD}}}. \bibinfo{pages}{815--824}.
\newblock


\bibitem[Miyauchi et~al\mbox{.}(2023)]%
        {miyauchi2023densest}
\bibfield{author}{\bibinfo{person}{Atsushi Miyauchi}, \bibinfo{person}{Tianyi Chen}, \bibinfo{person}{Konstantinos Sotiropoulos}, {and} \bibinfo{person}{Charalampos~E Tsourakakis}.} \bibinfo{year}{2023}\natexlab{}.
\newblock \showarticletitle{Densest Diverse Subgraphs: How to Plan a Successful Cocktail Party with Diversity}. In \bibinfo{booktitle}{\emph{Proceedings of the 29th ACM SIGKDD Conference on Knowledge Discovery \& Data Mining}}. \bibinfo{pages}{1710--1721}.
\newblock


\bibitem[Miyauchi and Takeda(2018)]%
        {miyauchi2018robust}
\bibfield{author}{\bibinfo{person}{Atsushi Miyauchi} {and} \bibinfo{person}{Akiko Takeda}.} \bibinfo{year}{2018}\natexlab{}.
\newblock \showarticletitle{Robust densest subgraph discovery}. In \bibinfo{booktitle}{\emph{ICDM}}. IEEE, \bibinfo{pages}{1188--1193}.
\newblock


\bibitem[Nesterov(1983)]%
        {nesterov1983method}
\bibfield{author}{\bibinfo{person}{Yu~E Nesterov}.} \bibinfo{year}{1983}\natexlab{}.
\newblock \showarticletitle{A method for solving the convex programming problem with convergence rate {O}$(1/k^2)$}. In \bibinfo{booktitle}{\emph{Dokl. Akad. Nauk SSSR,}}, Vol.~\bibinfo{volume}{269}. \bibinfo{pages}{543--547}.
\newblock


\bibitem[Nonner(2016)]%
        {nonner2016ptas}
\bibfield{author}{\bibinfo{person}{Tim Nonner}.} \bibinfo{year}{2016}\natexlab{}.
\newblock \showarticletitle{PTAS for Densest k-Subgraph in Interval Graphs}.
\newblock \bibinfo{journal}{\emph{Algorithmica}} \bibinfo{volume}{74}, \bibinfo{number}{1} (\bibinfo{year}{2016}), \bibinfo{pages}{528--539}.
\newblock


\bibitem[Qin et~al\mbox{.}(2015)]%
        {qin2015locally}
\bibfield{author}{\bibinfo{person}{Lu Qin}, \bibinfo{person}{Rong-Hua Li}, \bibinfo{person}{Lijun Chang}, {and} \bibinfo{person}{Chengqi Zhang}.} \bibinfo{year}{2015}\natexlab{}.
\newblock \showarticletitle{Locally densest subgraph discovery}. In \bibinfo{booktitle}{\emph{KDD}}. \bibinfo{pages}{965--974}.
\newblock


\bibitem[Saha et~al\mbox{.}(2010)]%
        {saha2010dense}
\bibfield{author}{\bibinfo{person}{Barna Saha}, \bibinfo{person}{Allison Hoch}, \bibinfo{person}{Samir Khuller}, \bibinfo{person}{Louiqa Raschid}, {and} \bibinfo{person}{Xiao-Ning Zhang}.} \bibinfo{year}{2010}\natexlab{}.
\newblock \showarticletitle{Dense subgraphs with restrictions and applications to gene annotation graphs}. In \bibinfo{booktitle}{\emph{{RECOMB}}}. Springer, \bibinfo{pages}{456--472}.
\newblock


\bibitem[Saito et~al\mbox{.}(2008)]%
        {saito2008extracting}
\bibfield{author}{\bibinfo{person}{Kazumi Saito}, \bibinfo{person}{Takeshi Yamada}, {and} \bibinfo{person}{Kazuhiro Kazama}.} \bibinfo{year}{2008}\natexlab{}.
\newblock \showarticletitle{Extracting communities from complex networks by the k-dense method}.
\newblock \bibinfo{journal}{\emph{IEICE Transactions on Fundamentals of Electronics, Communications and Computer Sciences}} \bibinfo{volume}{91}, \bibinfo{number}{11} (\bibinfo{year}{2008}), \bibinfo{pages}{3304--3311}.
\newblock


\bibitem[Sawlani and Wang(2020)]%
        {sawlani2020near}
\bibfield{author}{\bibinfo{person}{Saurabh Sawlani} {and} \bibinfo{person}{Junxing Wang}.} \bibinfo{year}{2020}\natexlab{}.
\newblock \showarticletitle{Near-optimal fully dynamic densest subgraph}. In \bibinfo{booktitle}{\emph{{STOC}}}. \bibinfo{pages}{181--193}.
\newblock


\bibitem[Seidman(1983)]%
        {seidman1983network}
\bibfield{author}{\bibinfo{person}{Stephen~B Seidman}.} \bibinfo{year}{1983}\natexlab{}.
\newblock \showarticletitle{Network structure and minimum degree}.
\newblock \bibinfo{journal}{\emph{Social networks}} \bibinfo{volume}{5}, \bibinfo{number}{3} (\bibinfo{year}{1983}), \bibinfo{pages}{269--287}.
\newblock


\bibitem[Sukprasert et~al\mbox{.}(2024)]%
        {sukprasert2024practical}
\bibfield{author}{\bibinfo{person}{Pattara Sukprasert}, \bibinfo{person}{Quanquan~C Liu}, \bibinfo{person}{Laxman Dhulipala}, {and} \bibinfo{person}{Julian Shun}.} \bibinfo{year}{2024}\natexlab{}.
\newblock \showarticletitle{Practical Parallel Algorithms for Near-Optimal Densest Subgraphs on Massive Graphs}. In \bibinfo{booktitle}{\emph{2024 Proceedings of the Symposium on Algorithm Engineering and Experiments (ALENEX)}}. SIAM, \bibinfo{pages}{59--73}.
\newblock


\bibitem[Sun et~al\mbox{.}(2020)]%
        {sun2020kclist++}
\bibfield{author}{\bibinfo{person}{Bintao Sun}, \bibinfo{person}{Maximilien Danisch}, \bibinfo{person}{TH Chan}, {and} \bibinfo{person}{Mauro Sozio}.} \bibinfo{year}{2020}\natexlab{}.
\newblock \showarticletitle{KClist++: A Simple Algorithm for Finding k-Clique Densest Subgraphs in Large Graphs}.
\newblock \bibinfo{journal}{\emph{PVLDB}} (\bibinfo{year}{2020}).
\newblock


\bibitem[Tsourakakis(2015)]%
        {tsourakakis2015k}
\bibfield{author}{\bibinfo{person}{Charalampos Tsourakakis}.} \bibinfo{year}{2015}\natexlab{}.
\newblock \showarticletitle{The k-clique densest subgraph problem}. In \bibinfo{booktitle}{\emph{{WWW}}}. \bibinfo{pages}{1122--1132}.
\newblock


\bibitem[Tsourakakis et~al\mbox{.}(2013)]%
        {tsourakakis2013denser}
\bibfield{author}{\bibinfo{person}{Charalampos Tsourakakis}, \bibinfo{person}{Francesco Bonchi}, \bibinfo{person}{Aristides Gionis}, \bibinfo{person}{Francesco Gullo}, {and} \bibinfo{person}{Maria Tsiarli}.} \bibinfo{year}{2013}\natexlab{}.
\newblock \showarticletitle{Denser than the densest subgraph: extracting optimal quasi-cliques with quality guarantees}. In \bibinfo{booktitle}{\emph{SIGKDD}}. \bibinfo{pages}{104--112}.
\newblock


\bibitem[Tsourakakis(2014)]%
        {tsourakakis2014mathematical}
\bibfield{author}{\bibinfo{person}{Charalampos~E Tsourakakis}.} \bibinfo{year}{2014}\natexlab{}.
\newblock \showarticletitle{Mathematical and algorithmic analysis of network and biological data}.
\newblock \bibinfo{journal}{\emph{arXiv preprint arXiv:1407.0375}} (\bibinfo{year}{2014}).
\newblock


\bibitem[Veldt et~al\mbox{.}(2021)]%
        {veldt2021generalized}
\bibfield{author}{\bibinfo{person}{Nate Veldt}, \bibinfo{person}{Austin~R Benson}, {and} \bibinfo{person}{Jon Kleinberg}.} \bibinfo{year}{2021}\natexlab{}.
\newblock \showarticletitle{The generalized mean densest subgraph problem}. In \bibinfo{booktitle}{\emph{Proceedings of the 27th ACM SIGKDD Conference on Knowledge Discovery \& Data Mining}}. \bibinfo{pages}{1604--1614}.
\newblock


\bibitem[Wang et~al\mbox{.}(2022)]%
        {wang2022towards}
\bibfield{author}{\bibinfo{person}{Kai Wang}, \bibinfo{person}{Xuemin Lin}, \bibinfo{person}{Lu Qin}, \bibinfo{person}{Wenjie Zhang}, {and} \bibinfo{person}{Ying Zhang}.} \bibinfo{year}{2022}\natexlab{}.
\newblock \showarticletitle{Towards efficient solutions of bitruss decomposition for large-scale bipartite graphs}.
\newblock \bibinfo{journal}{\emph{The VLDB Journal}} \bibinfo{volume}{31}, \bibinfo{number}{2} (\bibinfo{year}{2022}), \bibinfo{pages}{203--226}.
\newblock


\bibitem[Xu et~al\mbox{.}(2023)]%
        {xu2023efficient}
\bibfield{author}{\bibinfo{person}{Yichen Xu}, \bibinfo{person}{Chenhao Ma}, \bibinfo{person}{Yixiang Fang}, {and} \bibinfo{person}{Zhifeng Bao}.} \bibinfo{year}{2023}\natexlab{}.
\newblock \showarticletitle{Efficient and Effective Algorithms for Generalized Densest Subgraph Discovery}.
\newblock \bibinfo{journal}{\emph{Proceedings of the ACM on Management of Data}} \bibinfo{volume}{1}, \bibinfo{number}{2} (\bibinfo{year}{2023}), \bibinfo{pages}{1--27}.
\newblock


\bibitem[Yu et~al\mbox{.}(2022)]%
        {YuLL022}
\bibfield{author}{\bibinfo{person}{Kaiqiang Yu}, \bibinfo{person}{Cheng Long}, \bibinfo{person}{Shengxin Liu}, {and} \bibinfo{person}{Da Yan}.} \bibinfo{year}{2022}\natexlab{}.
\newblock \showarticletitle{Efficient Algorithms for Maximal k-Biplex Enumeration}. In \bibinfo{booktitle}{\emph{SIGMOD}}. \bibinfo{publisher}{{ACM}}, \bibinfo{pages}{860--873}.
\newblock


\bibitem[Yuan et~al\mbox{.}(2017)]%
        {yuan2017efficient}
\bibfield{author}{\bibinfo{person}{Long Yuan}, \bibinfo{person}{Lu Qin}, \bibinfo{person}{Xuemin Lin}, \bibinfo{person}{Lijun Chang}, {and} \bibinfo{person}{Wenjie Zhang}.} \bibinfo{year}{2017}\natexlab{}.
\newblock \showarticletitle{I/O efficient ECC graph decomposition via graph reduction}.
\newblock \bibinfo{journal}{\emph{The VLDB Journal}} \bibinfo{volume}{26}, \bibinfo{number}{2} (\bibinfo{year}{2017}), \bibinfo{pages}{275--300}.
\newblock


\bibitem[Zhang and Parthasarathy(2012)]%
        {zhang2012extracting}
\bibfield{author}{\bibinfo{person}{Yang Zhang} {and} \bibinfo{person}{Srinivasan Parthasarathy}.} \bibinfo{year}{2012}\natexlab{}.
\newblock \showarticletitle{Extracting analyzing and visualizing triangle k-core motifs within networks}. In \bibinfo{booktitle}{\emph{ICDE}}. IEEE, \bibinfo{pages}{1049--1060}.
\newblock


\bibitem[Zhou et~al\mbox{.}(2021)]%
        {zhou2021improving}
\bibfield{author}{\bibinfo{person}{Yi Zhou}, \bibinfo{person}{Shan Hu}, \bibinfo{person}{Mingyu Xiao}, {and} \bibinfo{person}{Zhang-Hua Fu}.} \bibinfo{year}{2021}\natexlab{}.
\newblock \showarticletitle{Improving maximum k-plex solver via second-order reduction and graph color bounding}. In \bibinfo{booktitle}{\emph{AAAI}}, Vol.~\bibinfo{volume}{35}. \bibinfo{pages}{12453--12460}.
\newblock


\bibitem[Zou(2016)]%
        {zou2016bitruss}
\bibfield{author}{\bibinfo{person}{Zhaonian Zou}.} \bibinfo{year}{2016}\natexlab{}.
\newblock \showarticletitle{Bitruss decomposition of bipartite graphs}. In \bibinfo{booktitle}{\emph{DASFAA}}. Springer, \bibinfo{pages}{218--233}.
\newblock


\end{thebibliography}

\balance
\clearpage

\appendix
\nobalance

\section{Additional Definitions}
\subsection{ The $\mathsf{LP}$ formulation and Dual of UDS problem}
\label{sec:app_lp_formulation}
The following is a well-known LP formulation of the UDS problem, introduced in \cite{charikar2000greedy}.
Associate each vertex $v$ with a variable $x_v \in \{0, 1\}$, where $x_v$= 1 signifies $v$ being included in $\mathcal{D}(G)$.
Similarly, for each edge, let $y_e \in \{0, 1\}$ denote whether or not it is in $E(\mathcal{D}(G))$. 
Relaxing the variables to be real numbers, we get the following $\mathsf{LP}$s, which we denote by $\mathsf{LP}(G)$, whose optimal is known to be  $\rho^{*}_G$ \cite{charikar2000greedy}:
\begin{align}
\mathsf{LP}(G) \quad \max &\sum_{e \in E}{y_e} \\
\text{s.t.} \quad &y_e \leq x_u, x_v, && \forall e=(u, v) \in E \\ 
&\sum_{ v \in V} x_v \leq 1, \\
&y_e \geq 0, x_v \geq 0, && \forall e \in E, \forall v \in V
\end{align}

The Lagrangian dual $\mathsf{DP}(G)$ of the $\mathsf{LP}(G)$ is as follows \cite{bahmani2014efficient,danisch2017large}:
\begin{align}
\mathsf{DP}(G) \quad \min & \max_{v \in V }\mathbf{w}(v) \\
\text{s.t.} \quad &\mathbf{w}(v) = \sum_{e \in E } \alpha_e(v), && \forall v \in V   \\
&\alpha_e(u) + \alpha_e(v) = 1, && \forall e = (u,v) \in E \\
&\alpha_e(u) \geq 0,  \alpha_e(v) \geq 0 && \forall e = (u,v) \in E
\end{align}
Note that $\Vert \mathbf{w} \Vert_{\infty}$ =  $\max_{v \in V }\mathbf{w}(v)$, which means that the objective function of $\mathsf{DP}(G)$ is:
$\min \Vert \mathbf{w} \Vert_{\infty}$.
This $\mathsf{DP}$ can be visualized as follows. Each edge $e$ = $(u, v)$ has a weight of 1, which it wants to assign to its endpoints: $\alpha_e(u)$ and $\alpha_e(v)$ such that the total weight on each vertex is at most $\Vert \mathbf{w} \Vert_{\infty}$. The objective is to find the minimum $\Vert \mathbf{w} \Vert_{\infty}$ for which such a load assignment is feasible.
Here, we introduce the vertex weight vector $\mathbf{w}$:
\begin{equation*}
    \mathbf{w} =  \begin{bmatrix} \mathbf{w}(v_1) & \mathbf{w}(v_2) & \cdots & \mathbf{w}(v_n) \end{bmatrix} 
\end{equation*}

\subsection{More Definitions of DDS problem}
\begin{definition} [$c$-biased density\cite{ma2022convex}]
    Given a directed graph $D$=$(V,E)$, a fixed $c$ and two vertex sets $S$ and $T$, the $c$-biased density of an ($S$, $T$)-induced subgraph is proportional to its edge-density, and is defined as 
$\rho_c(S, T)= \frac{2\sqrt{c}\sqrt{c'}}{c+c'}\rho (S, T) = \frac{2\sqrt{c}\sqrt{c'}}{c+c'}\frac{{|E(S, T)|}}{{\sqrt{|S||T|}}}
$, where $c'=\frac{|S|}{|T|}$. Note when $c'=c$, $\rho_c(S, T)=\rho(S,T)$.
\end{definition}

\begin{definition} [$c$-biased directed densest subgraph (DDS)~\cite{ma2022convex}]
    Given a directed graph $D$ and a fixed $c$, the subgraph, whose corresponding $c$-biased density is the highest among all possible ones, is called the $c$-biased DDS.
\end{definition}
Charikar \cite{charikar2000greedy} proposed the first exact DDS solution, which is based on linear programming (LP). Because the DDS is related to two vertex subset $S$ and $T$, Charikar formulated the DS problem to a series of linear programs w.r.t. the possible values of $c=\frac{|S|}{|T|}$. Because the ratio $c=\frac{|S|}{|T|}$ is not known in advance, there are $O(n^{2})$ possible values, which result in $O(n^{2})$ different LPs. For each $c=\frac{|S|}{|T|}$, the corresponding LP is formulated by Equation (\ref{equ:charikar-lp}).
\begin{equation}
\label{equ:charikar-lp}
\begin{aligned}
\mathsf{LP}(c)&&  \max  &&    x_{\mathrm{sum}}&=\sum_{(u,v)\in E}x_{u,v}		 \\
	&& \text{s.t.}	    && 0\leq x_{u,v} &\leq s_{u},				&& \forall (u,v) \in E \\
 	&&					&& x_{u,v} &\leq t_{v},				&& \forall (u,v) \in E \\
 	&& 					&& \sum_{u \in V} s_{u} &= \sqrt{c},  \\
 	&&					&& \sum_{v \in V} t_{v} &= \frac{1}{\sqrt{c}}.    
\end{aligned}
\end{equation}
The variables in Eq. (\ref{equ:charikar-lp}) can be used to infer the DDS when $c=\frac{|S^{*}|}{|T^{*}|}$. Specifically, $s_{u}$, $t_{v}$, and $x_{u,v}$ indicate the inclusion of a vertex $u$/vertex $v$/edge $(u,v)$ in an optimal densest subgraph according to whether the variable value is larger than 0, when $c=\frac{|S^{*}|}{|T^{*}|}$. To find the DDS, Charikar's algorithm needs to solve $O(n^{2})$ LPs with LP solvers.

To reduce the number of LPs to be solved, Ma et al. \cite{ma2022convex} introduced a relaxation, $a+b=2$, to the LP formulation of the DDS problem, as shown in \Cref{equ:ma-lp}. 
\begin{equation}
\label{equ:ma-lp}
\begin{aligned}
\mathsf{LP}(c)&&  \max  &&    x_{\mathrm{sum}}&=\sum_{(u,v)\in E}x_{u,v}		 \\
	&& \text{s.t.}					&& x_{u,v} &\geq 0,					&& \forall (u,v) \in E  \\
	&& && x_{u,v} &\leq s_{u},				&& \forall (u,v) \in E \\
 	&&					&& x_{u,v} &\leq t_{v},				&& \forall (u,v) \in E \\
 	&& 					&& \sum_{u \in V} s_{u} &= a \sqrt{c},  \\
 	&&					&& \sum_{v \in V} t_{v} &= \frac{b}{\sqrt{c}}, \\
 	&&					&& a + b &= 2.
\end{aligned}
\end{equation}
Comparing \Cref{equ:ma-lp} with \Cref{equ:charikar-lp}, we can find that the two formulations are identical if we restrict $a=1$ and $b=1$. By introducing the relaxation, Ma et al. \cite{ma2022convex} managed to build the connection between each LP and the DDS. Based on the connection, they developed a divide-and-conquer strategy to reduce the number of LPs to be solved. 

Ma et al. \cite{ma2022convex} derived the convex program (CP) of the linear programming formulation of the DDS problem.
For each $c=\frac{|S|}{|T|}$, the corresponding CP is formulated by Equation (\ref{equ:ma-dual-lp}).
Besides, Ma et al \cite{ma2022convex}
presented the convergence of employing Frank-Wolfe algorithm to solve \Cref{equ:ma-dual-lp}, which is given by the following theorem:
\begin{theorem}[\cite{ma2022convex}]
Given a directed graph $D$ and $c$, suppose \(d_{\text{max}}^\text{+}\) (resp. \(d_{\text{max}}^\text{-}\)) is the maximum out-degree (resp. in-degree) of \(D\). For $t > 16\frac{\kappa m}{\varepsilon^2}$, we have $\| \max_{u \in V} \{|\mathbf{w}_{\alpha}(u)|,  |\mathbf{w}_{\beta}(u)|\}\|_\infty - \rho^* \le \varepsilon$, where $\kappa = \sum_{c\in C}(\sqrt{c} + \frac{1}{\sqrt{c}})\max\{\sqrt{c} d_{\text{max}}^{+}, \frac{1}{\sqrt{c}} d_{\text{max}}^-\}$, and $C$ = $\{\frac{a}{b} \mid 1\le a, b\le n\}$.
\end{theorem}


\begin{table*}[]
\centering
\small
\begin{tabularx}{\textwidth}{l|l|l|l}
\toprule
\textbf{Method} 
& \textbf{Stage (1):} {\tt ReduceGraph}
& \textbf{Stage (2):} {\tt VWU}
& \textbf{Stage (3):} {\tt CSV}
\\ \hline
\begin{tabular}[c]{@{}l@{}} {\tt FlowExact} \\   {\tt DCExact}\end{tabular} & \begin{tabular}[c]{@{}l@{}} no reduction  \\ locate graph into certain [$x,y$]-core \end{tabular} & compute the maximum flow & \begin{tabular}[c]{@{}l@{}} {\Large\ding{182}} extract the minimum cut;\\ {\Large\ding{183}}  verify if it is optimal;\end{tabular} \\ \hline
CP-based & locate graph into certain [$x,y$]-core & optimize \(\mathsf{CP}(c)\); & \begin{tabular}[c]{@{}l@{}}{\Large\ding{182}}  extract the maximum prefix sum set;\\ {\Large\ding{183}} verify if it is exact or satisfies the approximation ratio criteria;\end{tabular} \\ \hline
Peeling-based & no reduction & iteratively remove vertices &  {\Large\ding{182}} extract the subgraph with the highest density during the peeling;\\
\bottomrule
\end{tabularx}
\caption{Overview of the three stages of the existing DDS algorithms.}
\label{addtab:brief_overview}
\end{table*}
\section{More Discussions}
\label{appendix:sec:more_dis}

\subsection{The equality of the two divide-and-conquer strategies}
To reduce the number of  $\frac{|S|}{|T|}$ values that need to be examined,  {\tt DCExact} \cite{ma2020efficient} first proposed a divide and conquer method to reduce the number of  $\frac{|S|}{|T|}$ values examined from $n^2$ to $k$, where theoretically $k \leq n^2$, but practically $k \ll n^2$.
This method effectively omits searching for some values of $c$.
{\tt DFWExact} \cite{ma2022convex} introduced a new divide and conquer method, utilizing the relationship between the densest subgraph and the $c$-biased densest subgraph to skip searches for certain $c$ values in the DDS process.
In this study, we thoroughly analyze these two strategies and, surprisingly, find them to be theoretically identical. 
Specifically, for a given $c$, \texttt{DCExact} \cite{ma2020efficient} solves such optimization problem:
\begin{equation}
\label{eq:dcexact}
\begin{aligned}
    &\max_{S, T \in V} &&g\\
    &\text{s.t.} &&\frac{|S|}{\sqrt{c}}(g-\frac{|E(S, T)|}{|S|/\sqrt{c}})+|T|\sqrt{c}(g-\frac{|E(S, T)|}{|T|\sqrt{c}})\le 0.
\end{aligned}
\end{equation}
Assume that $S'$ and $T'$ are the optimal choices for Equation \ref{eq:dcexact}, we can derive that
\begin{equation}
\label{eq:dccbiased}
    g^*(c) \le \frac{2\rho(S', T')}{\frac{\sqrt{c'}}{\sqrt{c}} + \frac{\sqrt{c}}{\sqrt{c'}}} = \frac{2\sqrt{c}\sqrt{c'}}{c + c'}\rho(S', T')=\rho_c(S',T') 
\end{equation}
where $c'=\frac{|S'|}{|T'|}$. Equation (\ref{eq:dccbiased}) reveals that the divide and conquer strategy in {\tt DCExact} \cite{ma2020efficient} is also based on $c$-biased DDS. To be more specific, {\tt DCExact} \cite{ma2020efficient} utilizes network flow computation to find the $c$-biased DDS, while {\tt DFWExact}~\cite{ma2022convex} obtains it via {\tt Frank-Wolfe}.

\subsection{The verification of CP-based Exact algorithms}

\begin{algorithm}[h]
  \caption{Verification of CP-based Exact algorithms} 
  \label{alg:verify:lp_based}
   \SetKwInOut{Input}{input}\SetKwInOut{Output}{output}
    \Input{$S$=$(V(S), E(S))$, $\mathbf{w}$}
    \Output{Whether $S$ is a densest subgraph of $G$}
    \If{$V(S)$ is a stable set} {
        \lIf{the \Cref{theorem:test_lp_eaact} is satisfied} {\Return{\texttt{True}}}
        $n \gets |V(S)|$; $m \gets |E(S)|$\;
        build flow network $\mathcal{F}$ on $S$\;
        $f \gets$ maximum flow from $s$ to $t$\;
        \Return{$f$ = $nm$;}
    }
    \Return {\texttt{False};}
\end{algorithm}

\textbf{CP-based Exact Algorithms.} 
After a sufficient number of iterations, CP-based algorithms tend to converge to the optimal solution. We begin by explaining a stable set to verify this.
\begin{definition}[Stable Set \cite{danisch2017large,sun2020kclist++}]
A non-empty vertex set $S \subseteq V$ is a stable set if the following conditions hold:
\begin{enumerate}
    \item $\forall u \in S$ and $v \in V \setminus S$, $\mathbf{w}(u) > \mathbf{w}(v)$.
    \item $\forall v \in V \setminus S$, we have $\forall (u,v) \in E \cap (S \times \{v\})$, $\alpha_{u,v}$ = 0.
\end{enumerate}
    \label{def:stable_set}

\end{definition}

Next, we present a theorem for optimal testing.

\begin{theorem}[Improved Goldberg's Condition \cite{sun2020kclist++}]
\label{theorem:test_lp_eaact}
Given a vertex weight vector $\mathbf{w}$ on subgraph $S$ of $G$ and supposed $\mathbf{w}(u_1) \geq \mathbf{w}(u_2) \geq \cdots \geq \mathbf{w}(u_{|n|})$. Let $n$ = $|V(S)|$, and $m$ = $|E(S)|$, if $ $ $\forall i \in [0, n - 1]$,
   $\min \left\{\frac{1}{i} {i \choose 2}, \frac{1}{i} \sum_{j=1}^i \mathbf{w}(u_j) \right\} - \rho(S) < \max \left\{ \frac{1}{ni}, \frac{1}{i} \left ( \left\lceil \frac{im}{n} \right\rceil  - \frac{im}{n} \right ) \right \}$, then $S$ is densest subgraph of $G$.
\end{theorem}

Algorithm \ref{alg:verify:lp_based} presents the details of using maximum flow for optimal result testing.
Specifically, if $S$ is a stable set and either the condition of \Cref{theorem:test_lp_eaact} is satisfied or there is a feasible flow with value $nm$ exists in $\mathcal{F}$, then the Algorithm \ref{alg:verify:lp_based} will return true; otherwise it will return false (lines 1-9).

\subsection{Overview of the existing DDS algorithms}
We illustrate the details of these three stages for each category of DSD algorithms for directed graphs in Table \ref{addtab:brief_overview}, which is very similar to the algorithms for undirected graphs.

\subsection{More Lessons and Opportunities}
\noindent\textbf{\underline{L7.}} For the exact UDS algorithms, {\tt CoreExact} and those new CP-based exact algorithms achieve comparable performance.
For the DDS problem, \texttt{DFWExact} is always the best one.

\noindent\textbf{\underline{L8.}} For all the DSD algorithms, including exact and approximation, the memory usage is almost the same, scaling linearly with the number of edges in the graph. 

\noindent\underline{\textbf{O5.}} An interesting future research direction is to study the application-driven variants of the DSD problem, by carefully considering the requirements of real-life scenarios. For example, the DSD solutions can be used for detecting network communities \cite{chen2010dense}. However, in a geo-social network, a community often contains a group of users that are not only linked densely, but also have close physical distance \cite{fang2017effective}. Thus, it would be interesting to study how to incorporate distance into the DSD problem.

\noindent\underline{\textbf{O6.}} While {\tt Greedy++} algorithm achieves performs well in practice, theoretically it needs $\Omega(\frac{\Delta(G)}{\rho^{*}_G \epsilon^2})$ iterations to obtain a (1+$\epsilon$)-approximation ratio solution. 
Hence, designing an early-stop strategy for {\tt Greedy++}, similar to those used in CP-based algorithms, is very useful for real situations.
Besides, how to devise the {\tt Greedy++} like algorithm for the DDS problem is also an exciting research problem.

\section{Additional Experiments }

\subsection{Additional Datasets.}
We present the statistics of the additional four graphs on \Cref{tab:add_dataset}, where two undirect graphs, two direct graphs, and from the different domains.
They are available on the Stanford Network Analysis Platform \footnote{http://snap.stanford.edu/data/}, Laboratory of Web Algorithmics \footnote{http://law.di.unimi.it/datasets.php}, Network Repository \footnote{https://networkrepository.com/network-data.php}, and Konect \footnote{http://konect.cc/networks/}.

\begin{table}[]
\small
  \caption{Additional four graphs.}
  \label{tab:add_dataset}
\begin{tabular}{l|l|r|r}
    \toprule
     Dataset &Category & $|V|$ & $|E|$  \\
    \hline    bio-SC-GT (BG)& Biological & 1,716 & 31,564  \\
    UK-2002 (UK)  & Road & 18,483,186 &261,787,258\\ \hline
     maayan-lake (ML) & Foodweb & 183 & 2,494 \\
    maayan-figeys (MF)   &  Metabolic  &2,239 &6,452 \\
   \bottomrule
  \end{tabular}
\end{table}

\begin{table}[]
\caption{Comparison of 2-approximation algorithms on undirected graphs (\textcolor{red}{Red} denotes most efficient algorithm, and \textcolor{t3}{Blue} denotes the best accuracy).}
%
	}
	\caption{Efficiency results of ($1+\epsilon$)-approximation algorithms on undirected graph.}
	\label{addfig:uds:time_approx}
\end{figure}
\begin{figure}[]
    \centering
    \ref{plot_addbg}\\
       	\subfigure[BG]{
        %
         }
    	\caption{Efficiency results of exact algorithms for UDS.}
    \label{addfig:uds:exact_time}
\end{figure}

\begin{figure}[]
 \small
	\centering		
 \ref{poltNewadd}\\
\subfigure[BG]{
		%
	}
	\caption{Efficiency results of new ($1+\epsilon$)-approximation algorithms on undirected graphs.}
	\label{addfig:uds:new_approx}
\end{figure}

\begin{figure}[]
    \centering
  \ref{addplotMem}\\
\subfigure[BG]{       	
%
         }
    	\caption{Memory usage on undirected graphs.}
    \label{addfig:uds:exact_memory}
\end{figure}

\begin{table}[]
\begin{minipage}{0.5\textwidth}
\centering
\large
\caption{Comparison of 2-approximation algorithms on directed graphs (\textcolor{red}{Red} denotes most efficient algorithm, and \textcolor{t3}{Blue} denotes the best accuracy).}
%
         }
    	\caption{Running time and memory usage of Exact DDS algorithm.}
    \label{addfig:dds:exact_time}
\end{figure}

\begin{figure}[]
 \small
	\centering		
 \ref{addpolt_appro_dds}\\
\subfigure[ML]{
		%
	}

	\caption{Efficiency results of ($1+\epsilon$)-approximation algorithms on directed graphs.}
	\label{addfig:dds:approx}
\end{figure}

\begin{table*}[]
\begin{minipage}{\textwidth}
\centering
\begin{tabular}{c|r|lrrl|lrrl|lrrl|r}
\toprule
\multicolumn{1}{c|}{\multirow{2}{*}{Dataset}} & \multicolumn{1}{c|}{\multirow{2}{*}{$\epsilon$}} &  & \multicolumn{2}{c}{\#Vertices}                             &  &  & \multicolumn{2}{c}{\#Edges}                                &  &  & \multicolumn{2}{c}{Time (s)}                               &  & \multicolumn{1}{l}{\multirow{2}{*}{Speedup}} \\ \cline{3-14}
\multicolumn{1}{c|}{}                         & \multicolumn{1}{c|}{}                   &  & \multicolumn{1}{c}{Original} & \multicolumn{1}{c}{Reduced} &  &  & \multicolumn{1}{c}{Original} & \multicolumn{1}{c}{Reduced} &  &  & \multicolumn{1}{c}{Original} & \multicolumn{1}{c}{Reduced} &  & \multicolumn{1}{l}{}                         \\ \hline
\multirow{4}{*}{OF} &0.01 &  & 2,939 & 464 &  &  & 30,501 & 12,263 &  &  & 3.5 & 1.5 &  & 2.3\\
~ & 0.001 &  & 2,939 & 439 &  &  & 30,501 & 12,473 &  &  & 22.6 & 9.1 &  & 2.5\\
~ & 0.0001 &  & 2,939 & 426 &  &  & 30,501 & 12,458 &  &  & 140.3 & 34.5 &  & 4.1\\
~ & 0 &  & 2,939 & 443 &  &  & 30,501 & 12,364 &  &  & 3.0 & 1.0 &  & 2.9\\ \hline
\multirow{4}{*}{AD} &0.01 &  & 6,541 & 1,195 &  &  & 51,127 & 17,137 &  &  & 10.4 & 3.7 &  & 2.8\\
~ & 0.001 &  & 6,541 & 1,128 &  &  & 51,127 & 17,541 &  &  & 54.2 & 21.2 &  & 2.6\\
~ & 0.0001 &  & 6,541 & 1,064 &  &  & 51,127 & 17,708 &  &  & 296.6 & 138.8 &  & 2.1\\
~ & 0 &  & 6,541 & 1,161 &  &  & 51,127 & 16,986 &  &  & 17.4 & 5.9 &  & 2.9\\ 
\bottomrule
\end{tabular}
\caption{\textbf{The effect of graph reduction on directed graphs.}}
               \label{addtab:dds_reduction}
\end{minipage}
\end{table*}

\begin{table*}[]
\begin{minipage}{\textwidth}
    \centering
        \begin{tabular}{l|l|rr|rr|rr|rr}
            \toprule
             \multicolumn{2}{c|}{\multirow{2}*{Dataset}} & \multicolumn{2}{c|}{$\epsilon = 0.01$} & \multicolumn{2}{c|}{$\epsilon=0.001$} & \multicolumn{2}{c|}{$\epsilon=0.0001$} & \multicolumn{2}{c}{$\epsilon=0$} \\
             \cline{3-10}
            \multicolumn{2}{c|}{~} & \texttt{DFW} & \texttt{DFWReN} &\texttt{DFW} & \texttt{DFWReN} &\texttt{DFW} & \texttt{DFWReN} &\texttt{DFW} & \texttt{DFWReN} \\
            \hline
            \multirow{5}*{AM}&\#Ratios&6 & 6 & 11 & 10 & 19 & 17 & 7 & 6\\
            &\#Edges&20,058 & 31,439 & 15,046 & 22,476 & 12,513 & 17,009 & 18,483 & 31,439\\
            &\#Iterations&233 & 267 & 1,045 & 1,390 & 7,974 & 7,265 & 200 & 217\\
            &Product&$2.81\times 10^{7}$ & $5.03\times 10^{7}$ & $1.73\times 10^{8}$ & $3.12\times 10^{8}$ & $1.90\times 10^{9}$ & $2.10\times 10^{9}$ & $2.59\times 10^{7}$ & $4.09\times 10^{7}$\\
            &Time (s)&1.0 & 1.1 & 2.5 & 2.6 & 13.8 & 11.6 & 1.3 & 1.2\\\hline
            \multirow{5}*{BA}&\#Ratios&16 & 9 & 16 & 9 & 19 & 12 & 16 & 10\\
            &\#Edges&698,422 & 2,665,311 & 698,422 & 2,674,577 & 652,086 & 2,099,222 & 711,706 & 2,449,856\\
            &\#Iterations&112 & 122 & 281 & 322 & 1,242 & 2,450 & 100 & 100\\
            &Product&$1.26\times 10^{9}$ & $2.93\times 10^{9}$ & $3.14\times 10^{9}$ & $7.76\times 10^{9}$ & $1.54\times 10^{10}$ & $6.17\times 10^{10}$ & $1.14\times 10^{9}$ & $2.45\times 10^{9}$\\
            &Time (s)&28.2 & 56.4 & 38.5 & 61.7 & 117.7 & 166.3 & 119.7 & 133.0\\
            \bottomrule
        \end{tabular}
    \caption{Effectiveness of graph reduction technique on directed graphs.}
    \label{addtab:dds:interval_adjstment}
\end{minipage}
\end{table*}

\begin{figure}[]
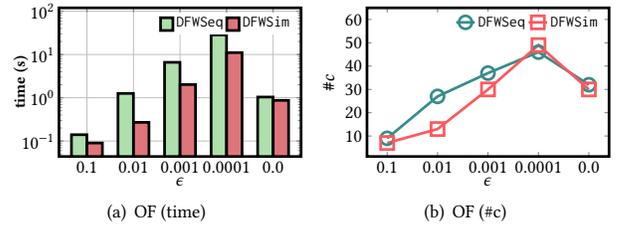

    \centering
    	\subfigure[OF (time)]{
       	%
	}
    	\caption{The effect of update strategies on directed graphs.}
    \label{addfig:dds:update_strategy}
\end{figure}

\subsection{Additional Experiments on UDS problem}
In this section, we present the additional experimental results on the BG and UK datasets in table \ref{addtab:uds:2_approx}, and Figure \ref{addfig:uds:time_approx} - \ref{addfig:uds:exact_memory}.
The results for these two datasets align with experimental results from other datasets presented in Section 6.2.

\subsection{Additional Experiments on DDS problem}
In this section, we present the additional experimental results on the ML and MF datasets in table \ref{addtab:dds:2_approx}, and Figure \ref{addfig:dds:exact_time} - \ref{addfig:dds:update_strategy}.
The results for these two datasets align with experimental results from other datasets presented in Section 6.3.

\end{document}